\documentclass[longauth]{aa} 

\usepackage{amsmath,amssymb}
\usepackage[breaklinks=true]{hyperref}

\usepackage{graphicx}
\usepackage{xcolor}
\usepackage[normalem]{ulem}
\usepackage{listings}
\usepackage{placeins}
\definecolor{dkgreen}{rgb}{0,0.6,0}
\definecolor{gray}{rgb}{0.5,0.5,0.5}
\definecolor{mauve}{rgb}{0.58,0,0.82}
\usepackage{txfonts}
\newcommand{\logg}{{\mathrm{logg}}}

\def\deg{\ensuremath{^\circ}}

\providecommand{\kms}{\ensuremath{\rm \,km\,s^{-1}}\xspace}
\providecommand{\masyr}{\ensuremath{\rm \,mas\,yr^{-1}}\xspace}

\providecommand{\teff}{\ensuremath{{\mathrm{T_{eff}}}}\xspace}

\newcommand{\bpminrp}{\ensuremath{G_\mathrm{BP}-G_\mathrm{RP}}\xspace}

\providecommand{\gmag}{\ensuremath{G}}
\providecommand{\bpmag}{\ensuremath{G_\mathrm{BP}}}
\providecommand{\rpmag}{\ensuremath{G_\mathrm{RP}}}

\providecommand{\kms}{\ensuremath{\textrm{km\,s}^{-1}}}

\providecommand{\masyr}{\ensuremath{\textrm{mas\,yr}^{-1}}}

\newcommand{\gbp}{{G_\mathrm{BP}}}
\newcommand{\grp}{{G_\mathrm{RP}}}

\providecommand{\modulename}[1]{#1\xspace}

\providecommand{\gspphot}{\modulename{GSP-Phot}}

\providecommand{\esphs}{\modulename{ESP-HS}}

\providecommand{\gaia}{\textit{Gaia}}

\newcommand\gdrtwo{\gaia~DR2}
\newcommand\gdrthree{\gaia~DR3 }
\newcommand\gedrthree{\gaia~EDR3}

\newcommand{\vrad}{$V_R$}
\newcommand{\vtan}{$V_\phi$}
\newcommand{\vz}{$V_z$}
\newcommand{\svrad}{$\sigma^*_R$}
\newcommand{\svtan}{$\sigma^*_\phi$}

\newcommand{\galah}{{\sl GALAH}}

\newcommand{\sdss}{{\sl SDSS}}
\newcommand{\apogee}{{\sl APOGEE}}
\newcommand{\rave}{{\sl RAVE}}
\newcommand{\lamost}{{\sl LAMOST}}

\providecommand{\rpmag}{\ensuremath{G_\mathrm{RP}}}
\usepackage{graphicx}
\usepackage{txfonts}
\usepackage{hyperref}
\newcommand{\orcit}[1]{\protect\href{https://orcid.org/#1}{\protect\includegraphics[width=8pt]{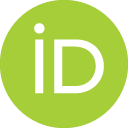}}}

\makeatletter
\renewcommand*\maketitle{%
  \thispagestyle{firstpage}
\begingroup
    \if@wideboxfn
    \setlength\bibindent{1.4\parindent}
    \else
    \setlength\bibindent{\parindent}
    \fi
    \renewcommand*\thefootnote{\@fnsymbol\c@footnote}%
    \renewcommand\@makefntext[1]{%
    \ifaa@longfn\hsize\textwidth\fi
    \noindent
    \hb@xt@\bibindent{\hss\@makefnmark\enspace}##1}
  \ifaa@twocolumn
  \begingroup
    \begin{aa@strip}
          \aa@maketitle
    \end{aa@strip}
    \@thanks	  	
  \endgroup
  \else
    \begingroup
      \let\thanks\footnote
      \aa@maketitle
    \endgroup
  \fi
\endgroup
  \setcounter{footnote}{0}%
}
\makeatother
\makeatletter
\DeclareRobustCommand*{\fieldName}[1]{%
  \begingroup\@fieldName\scantokens{\texttt{\small {#1}}\noexpand}\endgroup}
\begingroup\lccode`\~=`\_\relax
   \lowercase{\endgroup\def\@fieldName{\catcode`\_=\active \let~\_}}
\makeatother

\providecommand{\linktodoc}{https://geapre.esac.esa.int/archive/documentation/GDR3}  
\providecommand{\linktoparam}[2]{\href{\linktodoc/Gaia_archive/chap_datamodel/sec_dm_main_tables/ssec_dm_#1.html\##1-#2}{\fieldName{#2}\xspace}}
\providecommand{\linktotable}[1]{\href{\linktodoc/Gaia_archive/chap_datamodel/sec_dm_main_tables/ssec_dm_#1.html}{\fieldName{#1}\xspace}}

\begin{document}

   \title{\gaia{} Data Release 3:\\ Mapping the asymmetric disc of the Milky Way\thanks{The data for the following tables and figures can be retrieved in electronic format from CDS via anonymous ftp at cdsarc.u-strasbg.fr (130.79.128.5) or via http://cdsweb.u-strasbg.fr/cgi-bin/qcat?J/A+A/  
   : Tables \ref{tab:meanOcParameters} and \ref{tab:meanDCEPsParameters}, figures \ref{fig:velomapsgiants}, \ref{fig:vphiprof}, \ref{fig:velomaps_OBstars}, and \ref{fig:uncertaintyvelomaps}.
   }}

\author{
{\it Gaia} Collaboration
\and         R.~                       Drimmel\orcit{0000-0002-1777-5502}\inst{\ref{inst:0001}}
\and         M.~              Romero-G\'{o}mez\orcit{0000-0003-3936-1025}\inst{\ref{inst:0002}}
\and         L.~                        Chemin\orcit{0000-0002-3834-7937}\inst{\ref{inst:0003}}
\and         P.~                         Ramos\orcit{0000-0002-5080-7027}\inst{\ref{inst:0002},\ref{inst:0005}}
\and         E.~                        Poggio\orcit{0000-0003-3793-8505}\inst{\ref{inst:0006},\ref{inst:0001}}
\and         V.~                        Ripepi\orcit{0000-0003-1801-426X}\inst{\ref{inst:0008}}
\and         R.~                        Andrae\orcit{0000-0001-8006-6365}\inst{\ref{inst:0009}}
\and         R.~                        Blomme\orcit{0000-0002-2526-346X}\inst{\ref{inst:0010}}
\and         T.~                 Cantat-Gaudin\orcit{0000-0001-8726-2588}\inst{\ref{inst:0002},\ref{inst:0009}}
\and         A.~                 Castro-Ginard\orcit{0000-0002-9419-3725}\inst{\ref{inst:0013}}
\and         G.~                    Clementini\orcit{0000-0001-9206-9723}\inst{\ref{inst:0014}}
\and         F.~                      Figueras\orcit{0000-0002-3393-0007}\inst{\ref{inst:0002}}
\and         M.~                     Fouesneau\orcit{0000-0001-9256-5516}\inst{\ref{inst:0009}}
\and         Y.~                    Fr\'{e}mat\orcit{0000-0002-4645-6017}\inst{\ref{inst:0010}}
\and         K.~                       Jardine\inst{\ref{inst:0018}}
\and         S.~                        Khanna\orcit{0000-0002-2604-4277}\inst{\ref{inst:0019},\ref{inst:0001}}
\and         A.~                         Lobel\orcit{0000-0001-5030-019X}\inst{\ref{inst:0010}}
\and       D.J.~                      Marshall\orcit{0000-0003-3956-3524}\inst{\ref{inst:0022}}
\and         T.~                      Muraveva\orcit{0000-0002-0969-1915}\inst{\ref{inst:0014}}
\and     A.G.A.~                         Brown\orcit{0000-0002-7419-9679}\inst{\ref{inst:0013}}
\and         A.~                     Vallenari\orcit{0000-0003-0014-519X}\inst{\ref{inst:0025}}
\and         T.~                        Prusti\orcit{0000-0003-3120-7867}\inst{\ref{inst:0026}}
\and     J.H.J.~                    de Bruijne\orcit{0000-0001-6459-8599}\inst{\ref{inst:0026}}
\and         F.~                        Arenou\orcit{0000-0003-2837-3899}\inst{\ref{inst:0028}}
\and         C.~                     Babusiaux\orcit{0000-0002-7631-348X}\inst{\ref{inst:0029},\ref{inst:0028}}
\and         M.~                      Biermann\inst{\ref{inst:0031}}
\and       O.L.~                       Creevey\orcit{0000-0003-1853-6631}\inst{\ref{inst:0006}}
\and         C.~                     Ducourant\orcit{0000-0003-4843-8979}\inst{\ref{inst:0033}}
\and       D.W.~                         Evans\orcit{0000-0002-6685-5998}\inst{\ref{inst:0034}}
\and         L.~                          Eyer\orcit{0000-0002-0182-8040}\inst{\ref{inst:0035}}
\and         R.~                        Guerra\orcit{0000-0002-9850-8982}\inst{\ref{inst:0036}}
\and         A.~                        Hutton\inst{\ref{inst:0037}}
\and         C.~                         Jordi\orcit{0000-0001-5495-9602}\inst{\ref{inst:0002}}
\and       S.A.~                       Klioner\orcit{0000-0003-4682-7831}\inst{\ref{inst:0039}}
\and       U.L.~                       Lammers\orcit{0000-0001-8309-3801}\inst{\ref{inst:0036}}
\and         L.~                     Lindegren\orcit{0000-0002-5443-3026}\inst{\ref{inst:0041}}
\and         X.~                          Luri\orcit{0000-0001-5428-9397}\inst{\ref{inst:0002}}
\and         F.~                       Mignard\inst{\ref{inst:0006}}
\and         C.~                         Panem\inst{\ref{inst:0044}}
\and         D.~            Pourbaix$^\dagger$\orcit{0000-0002-3020-1837}\inst{\ref{inst:0045},\ref{inst:0046}}
\and         S.~                       Randich\orcit{0000-0003-2438-0899}\inst{\ref{inst:0047}}
\and         P.~                    Sartoretti\inst{\ref{inst:0028}}
\and         C.~                      Soubiran\orcit{0000-0003-3304-8134}\inst{\ref{inst:0033}}
\and         P.~                         Tanga\orcit{0000-0002-2718-997X}\inst{\ref{inst:0006}}
\and       N.A.~                        Walton\orcit{0000-0003-3983-8778}\inst{\ref{inst:0034}}
\and     C.A.L.~                  Bailer-Jones\inst{\ref{inst:0009}}
\and         U.~                       Bastian\orcit{0000-0002-8667-1715}\inst{\ref{inst:0031}}
\and         F.~                        Jansen\inst{\ref{inst:0054}}
\and         D.~                          Katz\orcit{0000-0001-7986-3164}\inst{\ref{inst:0028}}
\and       M.G.~                      Lattanzi\orcit{0000-0003-0429-7748}\inst{\ref{inst:0001},\ref{inst:0057}}
\and         F.~                   van Leeuwen\inst{\ref{inst:0034}}
\and         J.~                        Bakker\inst{\ref{inst:0036}}
\and         C.~                      Cacciari\orcit{0000-0001-5174-3179}\inst{\ref{inst:0014}}
\and         J.~                 Casta\~{n}eda\orcit{0000-0001-7820-946X}\inst{\ref{inst:0061}}
\and         F.~                     De Angeli\orcit{0000-0003-1879-0488}\inst{\ref{inst:0034}}
\and         C.~                     Fabricius\orcit{0000-0003-2639-1372}\inst{\ref{inst:0002}}
\and         L.~                     Galluccio\orcit{0000-0002-8541-0476}\inst{\ref{inst:0006}}
\and         A.~                      Guerrier\inst{\ref{inst:0044}}
\and         U.~                        Heiter\orcit{0000-0001-6825-1066}\inst{\ref{inst:0066}}
\and         E.~                        Masana\orcit{0000-0002-4819-329X}\inst{\ref{inst:0002}}
\and         R.~                      Messineo\inst{\ref{inst:0068}}
\and         N.~                       Mowlavi\orcit{0000-0003-1578-6993}\inst{\ref{inst:0035}}
\and         C.~                       Nicolas\inst{\ref{inst:0044}}
\and         K.~                  Nienartowicz\orcit{0000-0001-5415-0547}\inst{\ref{inst:0071},\ref{inst:0072}}
\and         F.~                       Pailler\orcit{0000-0002-4834-481X}\inst{\ref{inst:0044}}
\and         P.~                       Panuzzo\orcit{0000-0002-0016-8271}\inst{\ref{inst:0028}}
\and         F.~                        Riclet\inst{\ref{inst:0044}}
\and         W.~                          Roux\orcit{0000-0002-7816-1950}\inst{\ref{inst:0044}}
\and       G.M.~                      Seabroke\orcit{0000-0003-4072-9536}\inst{\ref{inst:0077}}
\and         R.~                         Sordo\orcit{0000-0003-4979-0659}\inst{\ref{inst:0025}}
\and         F.~                  Th\'{e}venin\inst{\ref{inst:0006}}
\and         G.~                  Gracia-Abril\inst{\ref{inst:0080},\ref{inst:0031}}
\and         J.~                       Portell\orcit{0000-0002-8886-8925}\inst{\ref{inst:0002}}
\and         D.~                      Teyssier\orcit{0000-0002-6261-5292}\inst{\ref{inst:0083}}
\and         M.~                       Altmann\orcit{0000-0002-0530-0913}\inst{\ref{inst:0031},\ref{inst:0085}}
\and         M.~                        Audard\orcit{0000-0003-4721-034X}\inst{\ref{inst:0035},\ref{inst:0072}}
\and         I.~                Bellas-Velidis\inst{\ref{inst:0088}}
\and         K.~                        Benson\inst{\ref{inst:0077}}
\and         J.~                      Berthier\orcit{0000-0003-1846-6485}\inst{\ref{inst:0090}}
\and       P.W.~                       Burgess\inst{\ref{inst:0034}}
\and         D.~                      Busonero\orcit{0000-0002-3903-7076}\inst{\ref{inst:0001}}
\and         G.~                         Busso\orcit{0000-0003-0937-9849}\inst{\ref{inst:0034}}
\and         H.~                   C\'{a}novas\orcit{0000-0001-7668-8022}\inst{\ref{inst:0083}}
\and         B.~                         Carry\orcit{0000-0001-5242-3089}\inst{\ref{inst:0006}}
\and         A.~                       Cellino\orcit{0000-0002-6645-334X}\inst{\ref{inst:0001}}
\and         N.~                         Cheek\inst{\ref{inst:0097}}
\and         Y.~                      Damerdji\orcit{0000-0002-3107-4024}\inst{\ref{inst:0098},\ref{inst:0099}}
\and         M.~                      Davidson\inst{\ref{inst:0100}}
\and         P.~                    de Teodoro\inst{\ref{inst:0036}}
\and         M.~              Nu\~{n}ez Campos\inst{\ref{inst:0037}}
\and         L.~                    Delchambre\orcit{0000-0003-2559-408X}\inst{\ref{inst:0098}}
\and         A.~                      Dell'Oro\orcit{0000-0003-1561-9685}\inst{\ref{inst:0047}}
\and         P.~                        Esquej\orcit{0000-0001-8195-628X}\inst{\ref{inst:0105}}
\and         J.~   Fern\'{a}ndez-Hern\'{a}ndez\inst{\ref{inst:0106}}
\and         E.~                        Fraile\inst{\ref{inst:0105}}
\and         D.~                      Garabato\orcit{0000-0002-7133-6623}\inst{\ref{inst:0108}}
\and         P.~              Garc\'{i}a-Lario\orcit{0000-0003-4039-8212}\inst{\ref{inst:0036}}
\and         E.~                        Gosset\inst{\ref{inst:0098},\ref{inst:0046}}
\and         R.~                       Haigron\inst{\ref{inst:0028}}
\and      J.-L.~                     Halbwachs\orcit{0000-0003-2968-6395}\inst{\ref{inst:0005}}
\and       N.C.~                        Hambly\orcit{0000-0002-9901-9064}\inst{\ref{inst:0100}}
\and       D.L.~                      Harrison\orcit{0000-0001-8687-6588}\inst{\ref{inst:0034},\ref{inst:0116}}
\and         J.~                 Hern\'{a}ndez\orcit{0000-0002-0361-4994}\inst{\ref{inst:0036}}
\and         D.~                    Hestroffer\orcit{0000-0003-0472-9459}\inst{\ref{inst:0090}}
\and       S.T.~                       Hodgkin\orcit{0000-0002-5470-3962}\inst{\ref{inst:0034}}
\and         B.~                          Holl\orcit{0000-0001-6220-3266}\inst{\ref{inst:0035},\ref{inst:0072}}
\and         K.~                    Jan{\ss}en\orcit{0000-0002-8163-2493}\inst{\ref{inst:0122}}
\and         G.~          Jevardat de Fombelle\inst{\ref{inst:0035}}
\and         S.~                        Jordan\orcit{0000-0001-6316-6831}\inst{\ref{inst:0031}}
\and         A.~                 Krone-Martins\orcit{0000-0002-2308-6623}\inst{\ref{inst:0125},\ref{inst:0126}}
\and       A.C.~                     Lanzafame\orcit{0000-0002-2697-3607}\inst{\ref{inst:0127},\ref{inst:0128}}
\and         W.~                  L\"{ o}ffler\inst{\ref{inst:0031}}
\and         O.~                       Marchal\orcit{ 0000-0001-7461-892}\inst{\ref{inst:0005}}
\and       P.M.~                       Marrese\orcit{0000-0002-8162-3810}\inst{\ref{inst:0131},\ref{inst:0132}}
\and         A.~                      Moitinho\orcit{0000-0003-0822-5995}\inst{\ref{inst:0125}}
\and         K.~                      Muinonen\orcit{0000-0001-8058-2642}\inst{\ref{inst:0134},\ref{inst:0135}}
\and         P.~                       Osborne\inst{\ref{inst:0034}}
\and         E.~                       Pancino\orcit{0000-0003-0788-5879}\inst{\ref{inst:0047},\ref{inst:0132}}
\and         T.~                       Pauwels\inst{\ref{inst:0010}}
\and         A.~                  Recio-Blanco\orcit{0000-0002-6550-7377}\inst{\ref{inst:0006}}
\and         C.~                     Reyl\'{e}\orcit{0000-0003-2258-2403}\inst{\ref{inst:0141}}
\and         M.~                        Riello\orcit{0000-0002-3134-0935}\inst{\ref{inst:0034}}
\and         L.~                     Rimoldini\orcit{0000-0002-0306-585X}\inst{\ref{inst:0072}}
\and         T.~                      Roegiers\orcit{0000-0002-1231-4440}\inst{\ref{inst:0144}}
\and         J.~                       Rybizki\orcit{0000-0002-0993-6089}\inst{\ref{inst:0009}}
\and       L.M.~                         Sarro\orcit{0000-0002-5622-5191}\inst{\ref{inst:0146}}
\and         C.~                        Siopis\orcit{0000-0002-6267-2924}\inst{\ref{inst:0045}}
\and         M.~                         Smith\inst{\ref{inst:0077}}
\and         A.~                      Sozzetti\orcit{0000-0002-7504-365X}\inst{\ref{inst:0001}}
\and         E.~                       Utrilla\inst{\ref{inst:0037}}
\and         M.~                   van Leeuwen\orcit{0000-0001-9698-2392}\inst{\ref{inst:0034}}
\and         U.~                         Abbas\orcit{0000-0002-5076-766X}\inst{\ref{inst:0001}}
\and         P.~               \'{A}brah\'{a}m\orcit{0000-0001-6015-646X}\inst{\ref{inst:0153},\ref{inst:0154}}
\and         A.~                Abreu Aramburu\inst{\ref{inst:0106}}
\and         C.~                         Aerts\orcit{0000-0003-1822-7126}\inst{\ref{inst:0156},\ref{inst:0157},\ref{inst:0009}}
\and       J.J.~                        Aguado\inst{\ref{inst:0146}}
\and         M.~                          Ajaj\inst{\ref{inst:0028}}
\and         F.~                 Aldea-Montero\inst{\ref{inst:0036}}
\and         G.~                     Altavilla\orcit{0000-0002-9934-1352}\inst{\ref{inst:0131},\ref{inst:0132}}
\and       M.A.~                   \'{A}lvarez\orcit{0000-0002-6786-2620}\inst{\ref{inst:0108}}
\and         J.~                         Alves\orcit{0000-0002-4355-0921}\inst{\ref{inst:0165}}
\and         F.~                        Anders\inst{\ref{inst:0002}}
\and       R.I.~                      Anderson\orcit{0000-0001-8089-4419}\inst{\ref{inst:0167}}
\and         E.~                Anglada Varela\orcit{0000-0001-7563-0689}\inst{\ref{inst:0106}}
\and         T.~                        Antoja\orcit{0000-0003-2595-5148}\inst{\ref{inst:0002}}
\and         D.~                        Baines\orcit{0000-0002-6923-3756}\inst{\ref{inst:0083}}
\and       S.G.~                         Baker\orcit{0000-0002-6436-1257}\inst{\ref{inst:0077}}
\and         L.~        Balaguer-N\'{u}\~{n}ez\orcit{0000-0001-9789-7069}\inst{\ref{inst:0002}}
\and         E.~                      Balbinot\orcit{0000-0002-1322-3153}\inst{\ref{inst:0019}}
\and         Z.~                         Balog\orcit{0000-0003-1748-2926}\inst{\ref{inst:0031},\ref{inst:0009}}
\and         C.~                       Barache\inst{\ref{inst:0085}}
\and         D.~                       Barbato\inst{\ref{inst:0035},\ref{inst:0001}}
\and         M.~                        Barros\orcit{0000-0002-9728-9618}\inst{\ref{inst:0125}}
\and       M.A.~                       Barstow\orcit{0000-0002-7116-3259}\inst{\ref{inst:0180}}
\and         S.~                 Bartolom\'{e}\orcit{0000-0002-6290-6030}\inst{\ref{inst:0002}}
\and      J.-L.~                     Bassilana\inst{\ref{inst:0182}}
\and         N.~                       Bauchet\inst{\ref{inst:0028}}
\and         U.~                      Becciani\orcit{0000-0002-4389-8688}\inst{\ref{inst:0127}}
\and         M.~                    Bellazzini\orcit{0000-0001-8200-810X}\inst{\ref{inst:0014}}
\and         A.~                     Berihuete\orcit{0000-0002-8589-4423}\inst{\ref{inst:0186}}
\and         M.~                        Bernet\orcit{0000-0001-7503-1010}\inst{\ref{inst:0002}}
\and         S.~                       Bertone\orcit{0000-0001-9885-8440}\inst{\ref{inst:0188},\ref{inst:0189},\ref{inst:0001}}
\and         L.~                       Bianchi\orcit{0000-0002-7999-4372}\inst{\ref{inst:0191}}
\and         A.~                    Binnenfeld\orcit{0000-0002-9319-3838}\inst{\ref{inst:0192}}
\and         S.~               Blanco-Cuaresma\orcit{0000-0002-1584-0171}\inst{\ref{inst:0193}}
\and         T.~                          Boch\orcit{0000-0001-5818-2781}\inst{\ref{inst:0005}}
\and         A.~                       Bombrun\inst{\ref{inst:0195}}
\and         D.~                       Bossini\orcit{0000-0002-9480-8400}\inst{\ref{inst:0196}}
\and         S.~                    Bouquillon\inst{\ref{inst:0085},\ref{inst:0198}}
\and         A.~                     Bragaglia\orcit{0000-0002-0338-7883}\inst{\ref{inst:0014}}
\and         L.~                      Bramante\inst{\ref{inst:0068}}
\and         E.~                        Breedt\orcit{0000-0001-6180-3438}\inst{\ref{inst:0034}}
\and         A.~                       Bressan\orcit{0000-0002-7922-8440}\inst{\ref{inst:0202}}
\and         N.~                     Brouillet\orcit{0000-0002-3274-7024}\inst{\ref{inst:0033}}
\and         E.~                    Brugaletta\orcit{0000-0003-2598-6737}\inst{\ref{inst:0127}}
\and         B.~                   Bucciarelli\orcit{0000-0002-5303-0268}\inst{\ref{inst:0001},\ref{inst:0057}}
\and         A.~                       Burlacu\inst{\ref{inst:0207}}
\and       A.G.~                     Butkevich\orcit{0000-0002-4098-3588}\inst{\ref{inst:0001}}
\and         R.~                         Buzzi\orcit{0000-0001-9389-5701}\inst{\ref{inst:0001}}
\and         E.~                        Caffau\orcit{0000-0001-6011-6134}\inst{\ref{inst:0028}}
\and         R.~                   Cancelliere\orcit{0000-0002-9120-3799}\inst{\ref{inst:0211}}
\and         R.~                      Carballo\orcit{0000-0001-7412-2498}\inst{\ref{inst:0212}}
\and         T.~                      Carlucci\inst{\ref{inst:0085}}
\and       M.I.~                     Carnerero\orcit{0000-0001-5843-5515}\inst{\ref{inst:0001}}
\and       J.M.~                      Carrasco\orcit{0000-0002-3029-5853}\inst{\ref{inst:0002}}
\and         L.~                   Casamiquela\orcit{0000-0001-5238-8674}\inst{\ref{inst:0033},\ref{inst:0028}}
\and         M.~                    Castellani\orcit{0000-0002-7650-7428}\inst{\ref{inst:0131}}
\and         L.~                        Chaoul\inst{\ref{inst:0044}}
\and         P.~                       Charlot\orcit{0000-0002-9142-716X}\inst{\ref{inst:0033}}
\and         V.~                    Chiaramida\inst{\ref{inst:0068}}
\and         A.~                     Chiavassa\orcit{0000-0003-3891-7554}\inst{\ref{inst:0006}}
\and         N.~                       Chornay\orcit{0000-0002-8767-3907}\inst{\ref{inst:0034}}
\and         G.~                     Comoretto\inst{\ref{inst:0083},\ref{inst:0225}}
\and         G.~                      Contursi\orcit{0000-0001-5370-1511}\inst{\ref{inst:0006}}
\and       W.J.~                        Cooper\orcit{0000-0003-3501-8967}\inst{\ref{inst:0227},\ref{inst:0001}}
\and         T.~                        Cornez\inst{\ref{inst:0182}}
\and         S.~                        Cowell\inst{\ref{inst:0034}}
\and         F.~                         Crifo\inst{\ref{inst:0028}}
\and         M.~                       Cropper\orcit{0000-0003-4571-9468}\inst{\ref{inst:0077}}
\and         M.~                        Crosta\orcit{0000-0003-4369-3786}\inst{\ref{inst:0001},\ref{inst:0234}}
\and         C.~                       Crowley\inst{\ref{inst:0195}}
\and         C.~                       Dafonte\orcit{0000-0003-4693-7555}\inst{\ref{inst:0108}}
\and         A.~                    Dapergolas\inst{\ref{inst:0088}}
\and         P.~                         David\inst{\ref{inst:0090}}
\and         P.~                    de Laverny\orcit{0000-0002-2817-4104}\inst{\ref{inst:0006}}
\and         F.~                      De Luise\orcit{0000-0002-6570-8208}\inst{\ref{inst:0240}}
\and         R.~                      De March\orcit{0000-0003-0567-842X}\inst{\ref{inst:0068}}
\and         J.~                     De Ridder\orcit{0000-0001-6726-2863}\inst{\ref{inst:0156}}
\and         R.~                      de Souza\inst{\ref{inst:0243}}
\and         A.~                     de Torres\inst{\ref{inst:0195}}
\and       E.F.~                    del Peloso\inst{\ref{inst:0031}}
\and         E.~                      del Pozo\inst{\ref{inst:0037}}
\and         M.~                         Delbo\orcit{0000-0002-8963-2404}\inst{\ref{inst:0006}}
\and         A.~                       Delgado\inst{\ref{inst:0105}}
\and      J.-B.~                       Delisle\orcit{0000-0001-5844-9888}\inst{\ref{inst:0035}}
\and         C.~                      Demouchy\inst{\ref{inst:0250}}
\and       T.E.~                 Dharmawardena\orcit{0000-0002-9583-5216}\inst{\ref{inst:0009}}
\and         P.~                     Di Matteo\inst{\ref{inst:0028}}
\and         S.~                       Diakite\inst{\ref{inst:0253}}
\and         C.~                        Diener\inst{\ref{inst:0034}}
\and         E.~                     Distefano\orcit{0000-0002-2448-2513}\inst{\ref{inst:0127}}
\and         C.~                       Dolding\inst{\ref{inst:0077}}
\and         H.~                          Enke\orcit{0000-0002-2366-8316}\inst{\ref{inst:0122}}
\and         C.~                         Fabre\inst{\ref{inst:0258}}
\and         M.~                      Fabrizio\orcit{0000-0001-5829-111X}\inst{\ref{inst:0131},\ref{inst:0132}}
\and         S.~                       Faigler\orcit{0000-0002-8368-5724}\inst{\ref{inst:0261}}
\and         G.~                      Fedorets\orcit{0000-0002-8418-4809}\inst{\ref{inst:0134},\ref{inst:0263}}
\and         P.~                      Fernique\orcit{0000-0002-3304-2923}\inst{\ref{inst:0005},\ref{inst:0265}}
\and         Y.~                      Fournier\orcit{0000-0002-6633-9088}\inst{\ref{inst:0122}}
\and         C.~                        Fouron\inst{\ref{inst:0207}}
\and         F.~                     Fragkoudi\orcit{0000-0002-0897-3013}\inst{\ref{inst:0268},\ref{inst:0269},\ref{inst:0270}}
\and         M.~                           Gai\orcit{0000-0001-9008-134X}\inst{\ref{inst:0001}}
\and         A.~              Garcia-Gutierrez\inst{\ref{inst:0002}}
\and         M.~              Garcia-Reinaldos\inst{\ref{inst:0036}}
\and         M.~             Garc\'{i}a-Torres\orcit{0000-0002-6867-7080}\inst{\ref{inst:0274}}
\and         A.~                      Garofalo\orcit{0000-0002-5907-0375}\inst{\ref{inst:0014}}
\and         A.~                         Gavel\orcit{0000-0002-2963-722X}\inst{\ref{inst:0066}}
\and         P.~                        Gavras\orcit{0000-0002-4383-4836}\inst{\ref{inst:0105}}
\and         E.~                       Gerlach\orcit{0000-0002-9533-2168}\inst{\ref{inst:0039}}
\and         R.~                         Geyer\orcit{0000-0001-6967-8707}\inst{\ref{inst:0039}}
\and         P.~                      Giacobbe\orcit{0000-0001-7034-7024}\inst{\ref{inst:0001}}
\and         G.~                       Gilmore\orcit{0000-0003-4632-0213}\inst{\ref{inst:0034}}
\and         S.~                        Girona\orcit{0000-0002-1975-1918}\inst{\ref{inst:0282}}
\and         G.~                     Giuffrida\inst{\ref{inst:0131}}
\and         R.~                         Gomel\inst{\ref{inst:0261}}
\and         A.~                         Gomez\orcit{0000-0002-3796-3690}\inst{\ref{inst:0108}}
\and         J.~    Gonz\'{a}lez-N\'{u}\~{n}ez\orcit{0000-0001-5311-5555}\inst{\ref{inst:0097},\ref{inst:0287}}
\and         I.~   Gonz\'{a}lez-Santamar\'{i}a\orcit{0000-0002-8537-9384}\inst{\ref{inst:0108}}
\and       J.J.~            Gonz\'{a}lez-Vidal\inst{\ref{inst:0002}}
\and         M.~                       Granvik\orcit{0000-0002-5624-1888}\inst{\ref{inst:0134},\ref{inst:0291}}
\and         P.~                      Guillout\inst{\ref{inst:0005}}
\and         J.~                       Guiraud\inst{\ref{inst:0044}}
\and         R.~     Guti\'{e}rrez-S\'{a}nchez\inst{\ref{inst:0083}}
\and       L.P.~                           Guy\orcit{0000-0003-0800-8755}\inst{\ref{inst:0072},\ref{inst:0296}}
\and         D.~                Hatzidimitriou\orcit{0000-0002-5415-0464}\inst{\ref{inst:0297},\ref{inst:0088}}
\and         M.~                        Hauser\inst{\ref{inst:0009},\ref{inst:0300}}
\and         M.~                       Haywood\orcit{0000-0003-0434-0400}\inst{\ref{inst:0028}}
\and         A.~                        Helmer\inst{\ref{inst:0182}}
\and         A.~                         Helmi\orcit{0000-0003-3937-7641}\inst{\ref{inst:0019}}
\and       M.H.~                     Sarmiento\orcit{0000-0003-4252-5115}\inst{\ref{inst:0037}}
\and       S.L.~                       Hidalgo\orcit{0000-0002-0002-9298}\inst{\ref{inst:0305},\ref{inst:0306}}
\and         N.~                   H\l{}adczuk\orcit{0000-0001-9163-4209}\inst{\ref{inst:0036},\ref{inst:0308}}
\and         D.~                         Hobbs\orcit{0000-0002-2696-1366}\inst{\ref{inst:0041}}
\and         G.~                       Holland\inst{\ref{inst:0034}}
\and       H.E.~                        Huckle\inst{\ref{inst:0077}}
\and         G.~                    Jasniewicz\inst{\ref{inst:0312}}
\and         A.~          Jean-Antoine Piccolo\orcit{0000-0001-8622-212X}\inst{\ref{inst:0044}}
\and     \'{O}.~            Jim\'{e}nez-Arranz\orcit{0000-0001-7434-5165}\inst{\ref{inst:0002}}
\and         J.~             Juaristi Campillo\inst{\ref{inst:0031}}
\and         F.~                         Julbe\inst{\ref{inst:0002}}
\and         L.~                     Karbevska\inst{\ref{inst:0072},\ref{inst:0318}}
\and         P.~                      Kervella\orcit{0000-0003-0626-1749}\inst{\ref{inst:0319}}
\and         G.~                    Kordopatis\orcit{0000-0002-9035-3920}\inst{\ref{inst:0006}}
\and       A.J.~                          Korn\orcit{0000-0002-3881-6756}\inst{\ref{inst:0066}}
\and      \'{A}~                K\'{o}sp\'{a}l\orcit{'{u}t 15-17, 1121 B}\inst{\ref{inst:0153},\ref{inst:0009},\ref{inst:0154}}
\and         Z.~           Kostrzewa-Rutkowska\inst{\ref{inst:0013},\ref{inst:0326}}
\and         K.~                Kruszy\'{n}ska\orcit{0000-0002-2729-5369}\inst{\ref{inst:0327}}
\and         M.~                           Kun\orcit{0000-0002-7538-5166}\inst{\ref{inst:0153}}
\and         P.~                       Laizeau\inst{\ref{inst:0329}}
\and         S.~                       Lambert\orcit{0000-0001-6759-5502}\inst{\ref{inst:0085}}
\and       A.F.~                         Lanza\orcit{0000-0001-5928-7251}\inst{\ref{inst:0127}}
\and         Y.~                         Lasne\inst{\ref{inst:0182}}
\and      J.-F.~                    Le Campion\inst{\ref{inst:0033}}
\and         Y.~                      Lebreton\orcit{0000-0002-4834-2144}\inst{\ref{inst:0319},\ref{inst:0335}}
\and         T.~                     Lebzelter\orcit{0000-0002-0702-7551}\inst{\ref{inst:0165}}
\and         S.~                        Leccia\orcit{0000-0001-5685-6930}\inst{\ref{inst:0008}}
\and         N.~                       Leclerc\inst{\ref{inst:0028}}
\and         I.~                 Lecoeur-Taibi\orcit{0000-0003-0029-8575}\inst{\ref{inst:0072}}
\and         S.~                          Liao\orcit{0000-0002-9346-0211}\inst{\ref{inst:0340},\ref{inst:0001},\ref{inst:0342}}
\and       E.L.~                        Licata\orcit{0000-0002-5203-0135}\inst{\ref{inst:0001}}
\and     H.E.P.~                  Lindstr{\o}m\inst{\ref{inst:0001},\ref{inst:0345},\ref{inst:0346}}
\and       T.A.~                        Lister\orcit{0000-0002-3818-7769}\inst{\ref{inst:0347}}
\and         E.~                       Livanou\orcit{0000-0003-0628-2347}\inst{\ref{inst:0297}}
\and         A.~                         Lorca\inst{\ref{inst:0037}}
\and         C.~                          Loup\inst{\ref{inst:0005}}
\and         P.~                 Madrero Pardo\inst{\ref{inst:0002}}
\and         A.~               Magdaleno Romeo\inst{\ref{inst:0207}}
\and         S.~                       Managau\inst{\ref{inst:0182}}
\and       R.G.~                          Mann\orcit{0000-0002-0194-325X}\inst{\ref{inst:0100}}
\and         M.~                      Manteiga\orcit{0000-0002-7711-5581}\inst{\ref{inst:0355}}
\and       J.M.~                      Marchant\orcit{0000-0002-3678-3145}\inst{\ref{inst:0356}}
\and         M.~                       Marconi\orcit{0000-0002-1330-2927}\inst{\ref{inst:0008}}
\and         J.~                        Marcos\inst{\ref{inst:0083}}
\and     M.M.S.~                 Marcos Santos\inst{\ref{inst:0097}}
\and         D.~                Mar\'{i}n Pina\orcit{0000-0001-6482-1842}\inst{\ref{inst:0002}}
\and         S.~                      Marinoni\orcit{0000-0001-7990-6849}\inst{\ref{inst:0131},\ref{inst:0132}}
\and         F.~                       Marocco\orcit{0000-0001-7519-1700}\inst{\ref{inst:0363}}
\and         L.~                   Martin Polo\inst{\ref{inst:0097}}
\and       J.M.~            Mart\'{i}n-Fleitas\orcit{0000-0002-8594-569X}\inst{\ref{inst:0037}}
\and         G.~                        Marton\orcit{0000-0002-1326-1686}\inst{\ref{inst:0153}}
\and         N.~                          Mary\inst{\ref{inst:0182}}
\and         A.~                         Masip\orcit{0000-0003-1419-0020}\inst{\ref{inst:0002}}
\and         D.~                       Massari\orcit{0000-0001-8892-4301}\inst{\ref{inst:0014}}
\and         A.~          Mastrobuono-Battisti\orcit{0000-0002-2386-9142}\inst{\ref{inst:0028}}
\and         T.~                         Mazeh\orcit{0000-0002-3569-3391}\inst{\ref{inst:0261}}
\and       P.J.~                      McMillan\orcit{0000-0002-8861-2620}\inst{\ref{inst:0041}}
\and         S.~                       Messina\orcit{0000-0002-2851-2468}\inst{\ref{inst:0127}}
\and         D.~                      Michalik\orcit{0000-0002-7618-6556}\inst{\ref{inst:0026}}
\and       N.R.~                        Millar\inst{\ref{inst:0034}}
\and         A.~                         Mints\orcit{0000-0002-8440-1455}\inst{\ref{inst:0122}}
\and         D.~                        Molina\orcit{0000-0003-4814-0275}\inst{\ref{inst:0002}}
\and         R.~                      Molinaro\orcit{0000-0003-3055-6002}\inst{\ref{inst:0008}}
\and         L.~                    Moln\'{a}r\orcit{0000-0002-8159-1599}\inst{\ref{inst:0153},\ref{inst:0380},\ref{inst:0154}}
\and         G.~                        Monari\orcit{0000-0002-6863-0661}\inst{\ref{inst:0005}}
\and         M.~                   Mongui\'{o}\orcit{0000-0002-4519-6700}\inst{\ref{inst:0002}}
\and         P.~                   Montegriffo\orcit{0000-0001-5013-5948}\inst{\ref{inst:0014}}
\and         A.~                       Montero\inst{\ref{inst:0037}}
\and         R.~                           Mor\orcit{0000-0002-8179-6527}\inst{\ref{inst:0002}}
\and         A.~                          Mora\inst{\ref{inst:0037}}
\and         R.~                    Morbidelli\orcit{0000-0001-7627-4946}\inst{\ref{inst:0001}}
\and         T.~                         Morel\orcit{0000-0002-8176-4816}\inst{\ref{inst:0098}}
\and         D.~                        Morris\inst{\ref{inst:0100}}
\and       C.P.~                        Murphy\inst{\ref{inst:0036}}
\and         I.~                       Musella\orcit{0000-0001-5909-6615}\inst{\ref{inst:0008}}
\and         Z.~                          Nagy\orcit{0000-0002-3632-1194}\inst{\ref{inst:0153}}
\and         L.~                         Noval\inst{\ref{inst:0182}}
\and         F.~                     Oca\~{n}a\inst{\ref{inst:0083},\ref{inst:0395}}
\and         A.~                         Ogden\inst{\ref{inst:0034}}
\and         C.~                     Ordenovic\inst{\ref{inst:0006}}
\and       J.O.~                        Osinde\inst{\ref{inst:0105}}
\and         C.~                        Pagani\orcit{0000-0001-5477-4720}\inst{\ref{inst:0180}}
\and         I.~                        Pagano\orcit{0000-0001-9573-4928}\inst{\ref{inst:0127}}
\and         L.~                     Palaversa\orcit{0000-0003-3710-0331}\inst{\ref{inst:0401},\ref{inst:0034}}
\and       P.A.~                       Palicio\orcit{0000-0002-7432-8709}\inst{\ref{inst:0006}}
\and         L.~               Pallas-Quintela\orcit{0000-0001-9296-3100}\inst{\ref{inst:0108}}
\and         A.~                        Panahi\orcit{0000-0001-5850-4373}\inst{\ref{inst:0261}}
\and         S.~               Payne-Wardenaar\inst{\ref{inst:0031}}
\and         X.~         Pe\~{n}alosa Esteller\inst{\ref{inst:0002}}
\and         A.~                 Penttil\"{ a}\orcit{0000-0001-7403-1721}\inst{\ref{inst:0134}}
\and         B.~                        Pichon\orcit{0000 0000 0062 1449}\inst{\ref{inst:0006}}
\and       A.M.~                    Piersimoni\orcit{0000-0002-8019-3708}\inst{\ref{inst:0240}}
\and      F.-X.~                        Pineau\orcit{0000-0002-2335-4499}\inst{\ref{inst:0005}}
\and         E.~                        Plachy\orcit{0000-0002-5481-3352}\inst{\ref{inst:0153},\ref{inst:0380},\ref{inst:0154}}
\and         G.~                          Plum\inst{\ref{inst:0028}}
\and         A.~                      Pr\v{s}a\orcit{0000-0002-1913-0281}\inst{\ref{inst:0415}}
\and         L.~                        Pulone\orcit{0000-0002-5285-998X}\inst{\ref{inst:0131}}
\and         E.~                        Racero\orcit{0000-0002-6101-9050}\inst{\ref{inst:0097},\ref{inst:0395}}
\and         S.~                       Ragaini\inst{\ref{inst:0014}}
\and         M.~                        Rainer\orcit{0000-0002-8786-2572}\inst{\ref{inst:0047},\ref{inst:0421}}
\and       C.M.~                       Raiteri\orcit{0000-0003-1784-2784}\inst{\ref{inst:0001}}
\and         M.~                  Ramos-Lerate\inst{\ref{inst:0083}}
\and         P.~                  Re Fiorentin\orcit{0000-0002-4995-0475}\inst{\ref{inst:0001}}
\and         S.~                        Regibo\inst{\ref{inst:0156}}
\and       P.J.~                      Richards\inst{\ref{inst:0426}}
\and         C.~                     Rios Diaz\inst{\ref{inst:0105}}
\and         A.~                          Riva\orcit{0000-0002-6928-8589}\inst{\ref{inst:0001}}
\and      H.-W.~                           Rix\orcit{0000-0003-4996-9069}\inst{\ref{inst:0009}}
\and         G.~                         Rixon\orcit{0000-0003-4399-6568}\inst{\ref{inst:0034}}
\and         N.~                      Robichon\orcit{0000-0003-4545-7517}\inst{\ref{inst:0028}}
\and       A.C.~                         Robin\orcit{0000-0001-8654-9499}\inst{\ref{inst:0141}}
\and         C.~                         Robin\inst{\ref{inst:0182}}
\and         M.~                       Roelens\orcit{0000-0003-0876-4673}\inst{\ref{inst:0035}}
\and     H.R.O.~                        Rogues\inst{\ref{inst:0250}}
\and         L.~                    Rohrbasser\inst{\ref{inst:0072}}
\and         N.~                        Rowell\orcit{0000-0003-3809-1895}\inst{\ref{inst:0100}}
\and         F.~                         Royer\orcit{0000-0002-9374-8645}\inst{\ref{inst:0028}}
\and         D.~                    Ruz Mieres\orcit{0000-0002-9455-157X}\inst{\ref{inst:0034}}
\and       K.A.~                       Rybicki\orcit{0000-0002-9326-9329}\inst{\ref{inst:0327}}
\and         G.~                      Sadowski\orcit{0000-0002-3411-1003}\inst{\ref{inst:0045}}
\and         A.~        S\'{a}ez N\'{u}\~{n}ez\inst{\ref{inst:0002}}
\and         A.~       Sagrist\`{a} Sell\'{e}s\orcit{0000-0001-6191-2028}\inst{\ref{inst:0031}}
\and         J.~                      Sahlmann\orcit{0000-0001-9525-3673}\inst{\ref{inst:0105}}
\and         E.~                      Salguero\inst{\ref{inst:0106}}
\and         N.~                       Samaras\orcit{0000-0001-8375-6652}\inst{\ref{inst:0010},\ref{inst:0447}}
\and         V.~               Sanchez Gimenez\orcit{0000-0003-1797-3557}\inst{\ref{inst:0002}}
\and         N.~                         Sanna\orcit{0000-0001-9275-9492}\inst{\ref{inst:0047}}
\and         R.~                 Santove\~{n}a\orcit{0000-0002-9257-2131}\inst{\ref{inst:0108}}
\and         M.~                       Sarasso\orcit{0000-0001-5121-0727}\inst{\ref{inst:0001}}
\and       M.S.~                    Schultheis\orcit{0000-0002-6590-1657}\inst{\ref{inst:0006}}
\and         E.~                       Sciacca\orcit{0000-0002-5574-2787}\inst{\ref{inst:0127}}
\and         M.~                         Segol\inst{\ref{inst:0250}}
\and       J.C.~                       Segovia\inst{\ref{inst:0097}}
\and         D.~                 S\'{e}gransan\orcit{0000-0003-2355-8034}\inst{\ref{inst:0035}}
\and         D.~                        Semeux\inst{\ref{inst:0258}}
\and         S.~                        Shahaf\orcit{0000-0001-9298-8068}\inst{\ref{inst:0458}}
\and       H.I.~                      Siddiqui\orcit{0000-0003-1853-6033}\inst{\ref{inst:0459}}
\and         A.~                       Siebert\orcit{0000-0001-8059-2840}\inst{\ref{inst:0005},\ref{inst:0265}}
\and         L.~                       Siltala\orcit{0000-0002-6938-794X}\inst{\ref{inst:0134}}
\and         A.~                       Silvelo\orcit{0000-0002-5126-6365}\inst{\ref{inst:0108}}
\and         E.~                        Slezak\inst{\ref{inst:0006}}
\and         I.~                        Slezak\inst{\ref{inst:0006}}
\and       R.L.~                         Smart\orcit{0000-0002-4424-4766}\inst{\ref{inst:0001}}
\and       O.N.~                        Snaith\inst{\ref{inst:0028}}
\and         E.~                        Solano\inst{\ref{inst:0468}}
\and         F.~                       Solitro\inst{\ref{inst:0068}}
\and         D.~                        Souami\orcit{0000-0003-4058-0815}\inst{\ref{inst:0319},\ref{inst:0471}}
\and         J.~                       Souchay\inst{\ref{inst:0085}}
\and         A.~                        Spagna\orcit{0000-0003-1732-2412}\inst{\ref{inst:0001}}
\and         L.~                         Spina\orcit{0000-0002-9760-6249}\inst{\ref{inst:0025}}
\and         F.~                         Spoto\orcit{0000-0001-7319-5847}\inst{\ref{inst:0193}}
\and       I.A.~                        Steele\orcit{0000-0001-8397-5759}\inst{\ref{inst:0356}}
\and         H.~            Steidelm\"{ u}ller\inst{\ref{inst:0039}}
\and       C.A.~                    Stephenson\inst{\ref{inst:0083},\ref{inst:0479}}
\and         M.~                  S\"{ u}veges\orcit{0000-0003-3017-5322}\inst{\ref{inst:0480}}
\and         J.~                        Surdej\orcit{0000-0002-7005-1976}\inst{\ref{inst:0098},\ref{inst:0482}}
\and         L.~                      Szabados\orcit{0000-0002-2046-4131}\inst{\ref{inst:0153}}
\and         E.~                  Szegedi-Elek\orcit{0000-0001-7807-6644}\inst{\ref{inst:0153}}
\and         F.~                         Taris\inst{\ref{inst:0085}}
\and       M.B.~                        Taylor\orcit{0000-0002-4209-1479}\inst{\ref{inst:0486}}
\and         R.~                      Teixeira\orcit{0000-0002-6806-6626}\inst{\ref{inst:0243}}
\and         L.~                       Tolomei\orcit{0000-0002-3541-3230}\inst{\ref{inst:0068}}
\and         N.~                       Tonello\orcit{0000-0003-0550-1667}\inst{\ref{inst:0282}}
\and         F.~                         Torra\orcit{0000-0002-8429-299X}\inst{\ref{inst:0061}}
\and         J.~               Torra$^\dagger$\inst{\ref{inst:0002}}
\and         G.~                Torralba Elipe\orcit{0000-0001-8738-194X}\inst{\ref{inst:0108}}
\and         M.~                     Trabucchi\orcit{0000-0002-1429-2388}\inst{\ref{inst:0493},\ref{inst:0035}}
\and       A.T.~                       Tsounis\inst{\ref{inst:0495}}
\and         C.~                         Turon\orcit{0000-0003-1236-5157}\inst{\ref{inst:0028}}
\and         A.~                          Ulla\orcit{0000-0001-6424-5005}\inst{\ref{inst:0497}}
\and         N.~                         Unger\orcit{0000-0003-3993-7127}\inst{\ref{inst:0035}}
\and       M.V.~                      Vaillant\inst{\ref{inst:0182}}
\and         E.~                    van Dillen\inst{\ref{inst:0250}}
\and         W.~                    van Reeven\inst{\ref{inst:0501}}
\and         O.~                         Vanel\orcit{0000-0002-7898-0454}\inst{\ref{inst:0028}}
\and         A.~                     Vecchiato\orcit{0000-0003-1399-5556}\inst{\ref{inst:0001}}
\and         Y.~                         Viala\inst{\ref{inst:0028}}
\and         D.~                       Vicente\orcit{0000-0002-1584-1182}\inst{\ref{inst:0282}}
\and         S.~                     Voutsinas\inst{\ref{inst:0100}}
\and         M.~                        Weiler\inst{\ref{inst:0002}}
\and         T.~                        Wevers\orcit{0000-0002-4043-9400}\inst{\ref{inst:0034},\ref{inst:0509}}
\and      \L{}.~                   Wyrzykowski\orcit{0000-0002-9658-6151}\inst{\ref{inst:0327}}
\and         A.~                        Yoldas\inst{\ref{inst:0034}}
\and         P.~                         Yvard\inst{\ref{inst:0250}}
\and         H.~                          Zhao\orcit{0000-0003-2645-6869}\inst{\ref{inst:0006}}
\and         J.~                         Zorec\inst{\ref{inst:0514}}
\and         S.~                        Zucker\orcit{0000-0003-3173-3138}\inst{\ref{inst:0192}}
\and         T.~                       Zwitter\orcit{0000-0002-2325-8763}\inst{\ref{inst:0516}}
}
\institute{
     INAF - Osservatorio Astrofisico di Torino, via Osservatorio 20, 10025 Pino Torinese (TO), Italy\relax                                                                                                                                                                                                                                                         \label{inst:0001}
\and Institut de Ci\`{e}ncies del Cosmos (ICCUB), Universitat  de  Barcelona  (IEEC-UB), Mart\'{i} i  Franqu\`{e}s  1, 08028 Barcelona, Spain\relax                                                                                                                                                                                                                \label{inst:0002}\vfill
\and Centro de Astronom\'{i}a - CITEVA, Universidad de Antofagasta, Avenida Angamos 601, Antofagasta 1270300, Chile\relax                                                                                                                                                                                                                                          \label{inst:0003}\vfill
\and Universit\'{e} de Strasbourg, CNRS, Observatoire astronomique de Strasbourg, UMR 7550,  11 rue de l'Universit\'{e}, 67000 Strasbourg, France\relax                                                                                                                                                                                                            \label{inst:0005}\vfill
\and Universit\'{e} C\^{o}te d'Azur, Observatoire de la C\^{o}te d'Azur, CNRS, Laboratoire Lagrange, Bd de l'Observatoire, CS 34229, 06304 Nice Cedex 4, France\relax                                                                                                                                                                                              \label{inst:0006}\vfill
\and INAF - Osservatorio Astronomico di Capodimonte, Via Moiariello 16, 80131, Napoli, Italy\relax                                                                                                                                                                                                                                                                 \label{inst:0008}\vfill
\and Max Planck Institute for Astronomy, K\"{ o}nigstuhl 17, 69117 Heidelberg, Germany\relax                                                                                                                                                                                                                                                                       \label{inst:0009}\vfill
\and Royal Observatory of Belgium, Ringlaan 3, 1180 Brussels, Belgium\relax                                                                                                                                                                                                                                                                                        \label{inst:0010}\vfill
\and Leiden Observatory, Leiden University, Niels Bohrweg 2, 2333 CA Leiden, The Netherlands\relax                                                                                                                                                                                                                                                                 \label{inst:0013}\vfill
\and INAF - Osservatorio di Astrofisica e Scienza dello Spazio di Bologna, via Piero Gobetti 93/3, 40129 Bologna, Italy\relax                                                                                                                                                                                                                                      \label{inst:0014}\vfill
\and Radagast Solutions\relax                                                                                                                                                                                                                                                                                                                                      \label{inst:0018}\vfill
\and Kapteyn Astronomical Institute, University of Groningen, Landleven 12, 9747 AD Groningen, The Netherlands\relax                                                                                                                                                                                                                                               \label{inst:0019}\vfill
\and IRAP, Universit\'{e} de Toulouse, CNRS, UPS, CNES, 9 Av. colonel Roche, BP 44346, 31028 Toulouse Cedex 4, France\relax                                                                                                                                                                                                                                        \label{inst:0022}\vfill
\and INAF - Osservatorio astronomico di Padova, Vicolo Osservatorio 5, 35122 Padova, Italy\relax                                                                                                                                                                                                                                                                   \label{inst:0025}\vfill
\and European Space Agency (ESA), European Space Research and Technology Centre (ESTEC), Keplerlaan 1, 2201AZ, Noordwijk, The Netherlands\relax                                                                                                                                                                                                                    \label{inst:0026}\vfill
\and GEPI, Observatoire de Paris, Universit\'{e} PSL, CNRS, 5 Place Jules Janssen, 92190 Meudon, France\relax                                                                                                                                                                                                                                                      \label{inst:0028}\vfill
\and Univ. Grenoble Alpes, CNRS, IPAG, 38000 Grenoble, France\relax                                                                                                                                                                                                                                                                                                \label{inst:0029}\vfill
\and Astronomisches Rechen-Institut, Zentrum f\"{ u}r Astronomie der Universit\"{ a}t Heidelberg, M\"{ o}nchhofstr. 12-14, 69120 Heidelberg, Germany\relax                                                                                                                                                                                                         \label{inst:0031}\vfill
\and Laboratoire d'astrophysique de Bordeaux, Univ. Bordeaux, CNRS, B18N, all{\'e}e Geoffroy Saint-Hilaire, 33615 Pessac, France\relax                                                                                                                                                                                                                             \label{inst:0033}\vfill
\and Institute of Astronomy, University of Cambridge, Madingley Road, Cambridge CB3 0HA, United Kingdom\relax                                                                                                                                                                                                                                                      \label{inst:0034}\vfill
\and Department of Astronomy, University of Geneva, Chemin Pegasi 51, 1290 Versoix, Switzerland\relax                                                                                                                                                                                                                                                              \label{inst:0035}\vfill
\and European Space Agency (ESA), European Space Astronomy Centre (ESAC), Camino bajo del Castillo, s/n, Urbanizacion Villafranca del Castillo, Villanueva de la Ca\~{n}ada, 28692 Madrid, Spain\relax                                                                                                                                                             \label{inst:0036}\vfill
\and Aurora Technology for European Space Agency (ESA), Camino bajo del Castillo, s/n, Urbanizacion Villafranca del Castillo, Villanueva de la Ca\~{n}ada, 28692 Madrid, Spain\relax                                                                                                                                                                               \label{inst:0037}\vfill
\and Lohrmann Observatory, Technische Universit\"{ a}t Dresden, Mommsenstra{\ss}e 13, 01062 Dresden, Germany\relax                                                                                                                                                                                                                                                 \label{inst:0039}\vfill
\and Lund Observatory, Department of Astronomy and Theoretical Physics, Lund University, Box 43, 22100 Lund, Sweden\relax                                                                                                                                                                                                                                          \label{inst:0041}\vfill
\and CNES Centre Spatial de Toulouse, 18 avenue Edouard Belin, 31401 Toulouse Cedex 9, France\relax                                                                                                                                                                                                                                                                \label{inst:0044}\vfill
\and Institut d'Astronomie et d'Astrophysique, Universit\'{e} Libre de Bruxelles CP 226, Boulevard du Triomphe, 1050 Brussels, Belgium\relax                                                                                                                                                                                                                       \label{inst:0045}\vfill
\and F.R.S.-FNRS, Rue d'Egmont 5, 1000 Brussels, Belgium\relax                                                                                                                                                                                                                                                                                                     \label{inst:0046}\vfill
\and INAF - Osservatorio Astrofisico di Arcetri, Largo Enrico Fermi 5, 50125 Firenze, Italy\relax                                                                                                                                                                                                                                                                  \label{inst:0047}\vfill
\and European Space Agency (ESA, retired)\relax                                                                                                                                                                                                                                                                                                                    \label{inst:0054}\vfill
\and University of Turin, Department of Physics, Via Pietro Giuria 1, 10125 Torino, Italy\relax                                                                                                                                                                                                                                                                    \label{inst:0057}\vfill
\and DAPCOM for Institut de Ci\`{e}ncies del Cosmos (ICCUB), Universitat  de  Barcelona  (IEEC-UB), Mart\'{i} i  Franqu\`{e}s  1, 08028 Barcelona, Spain\relax                                                                                                                                                                                                     \label{inst:0061}\vfill
\and Observational Astrophysics, Division of Astronomy and Space Physics, Department of Physics and Astronomy, Uppsala University, Box 516, 751 20 Uppsala, Sweden\relax                                                                                                                                                                                           \label{inst:0066}\vfill
\and ALTEC S.p.a, Corso Marche, 79,10146 Torino, Italy\relax                                                                                                                                                                                                                                                                                                       \label{inst:0068}\vfill
\and S\`{a}rl, Geneva, Switzerland\relax                                                                                                                                                                                                                                                                                                                           \label{inst:0071}\vfill
\and Department of Astronomy, University of Geneva, Chemin d'Ecogia 16, 1290 Versoix, Switzerland\relax                                                                                                                                                                                                                                                            \label{inst:0072}\vfill
\and Mullard Space Science Laboratory, University College London, Holmbury St Mary, Dorking, Surrey RH5 6NT, United Kingdom\relax                                                                                                                                                                                                                                  \label{inst:0077}\vfill
\and Gaia DPAC Project Office, ESAC, Camino bajo del Castillo, s/n, Urbanizacion Villafranca del Castillo, Villanueva de la Ca\~{n}ada, 28692 Madrid, Spain\relax                                                                                                                                                                                                  \label{inst:0080}\vfill
\and Telespazio UK S.L. for European Space Agency (ESA), Camino bajo del Castillo, s/n, Urbanizacion Villafranca del Castillo, Villanueva de la Ca\~{n}ada, 28692 Madrid, Spain\relax                                                                                                                                                                              \label{inst:0083}\vfill
\and SYRTE, Observatoire de Paris, Universit\'{e} PSL, CNRS,  Sorbonne Universit\'{e}, LNE, 61 avenue de l'Observatoire 75014 Paris, France\relax                                                                                                                                                                                                                  \label{inst:0085}\vfill
\and National Observatory of Athens, I. Metaxa and Vas. Pavlou, Palaia Penteli, 15236 Athens, Greece\relax                                                                                                                                                                                                                                                         \label{inst:0088}\vfill
\and IMCCE, Observatoire de Paris, Universit\'{e} PSL, CNRS, Sorbonne Universit{\'e}, Univ. Lille, 77 av. Denfert-Rochereau, 75014 Paris, France\relax                                                                                                                                                                                                             \label{inst:0090}\vfill
\and Serco Gesti\'{o}n de Negocios for European Space Agency (ESA), Camino bajo del Castillo, s/n, Urbanizacion Villafranca del Castillo, Villanueva de la Ca\~{n}ada, 28692 Madrid, Spain\relax                                                                                                                                                                   \label{inst:0097}\vfill
\and Institut d'Astrophysique et de G\'{e}ophysique, Universit\'{e} de Li\`{e}ge, 19c, All\'{e}e du 6 Ao\^{u}t, B-4000 Li\`{e}ge, Belgium\relax                                                                                                                                                                                                                    \label{inst:0098}\vfill
\and CRAAG - Centre de Recherche en Astronomie, Astrophysique et G\'{e}ophysique, Route de l'Observatoire Bp 63 Bouzareah 16340 Algiers, Algeria\relax                                                                                                                                                                                                             \label{inst:0099}\vfill
\and Institute for Astronomy, University of Edinburgh, Royal Observatory, Blackford Hill, Edinburgh EH9 3HJ, United Kingdom\relax                                                                                                                                                                                                                                  \label{inst:0100}\vfill
\and RHEA for European Space Agency (ESA), Camino bajo del Castillo, s/n, Urbanizacion Villafranca del Castillo, Villanueva de la Ca\~{n}ada, 28692 Madrid, Spain\relax                                                                                                                                                                                            \label{inst:0105}\vfill
\and ATG Europe for European Space Agency (ESA), Camino bajo del Castillo, s/n, Urbanizacion Villafranca del Castillo, Villanueva de la Ca\~{n}ada, 28692 Madrid, Spain\relax                                                                                                                                                                                      \label{inst:0106}\vfill
\and CIGUS CITIC - Department of Computer Science and Information Technologies, University of A Coru\~{n}a, Campus de Elvi\~{n}a s/n, A Coru\~{n}a, 15071, Spain\relax                                                                                                                                                                                             \label{inst:0108}\vfill
\and Kavli Institute for Cosmology Cambridge, Institute of Astronomy, Madingley Road, Cambridge, CB3 0HA\relax                                                                                                                                                                                                                                                     \label{inst:0116}\vfill
\and Leibniz Institute for Astrophysics Potsdam (AIP), An der Sternwarte 16, 14482 Potsdam, Germany\relax                                                                                                                                                                                                                                                          \label{inst:0122}\vfill
\and CENTRA, Faculdade de Ci\^{e}ncias, Universidade de Lisboa, Edif. C8, Campo Grande, 1749-016 Lisboa, Portugal\relax                                                                                                                                                                                                                                            \label{inst:0125}\vfill
\and Department of Informatics, Donald Bren School of Information and Computer Sciences, University of California, Irvine, 5226 Donald Bren Hall, 92697-3440 CA Irvine, United States\relax                                                                                                                                                                        \label{inst:0126}\vfill
\and INAF - Osservatorio Astrofisico di Catania, via S. Sofia 78, 95123 Catania, Italy\relax                                                                                                                                                                                                                                                                       \label{inst:0127}\vfill
\and Dipartimento di Fisica e Astronomia ""Ettore Majorana"", Universit\`{a} di Catania, Via S. Sofia 64, 95123 Catania, Italy\relax                                                                                                                                                                                                                               \label{inst:0128}\vfill
\and INAF - Osservatorio Astronomico di Roma, Via Frascati 33, 00078 Monte Porzio Catone (Roma), Italy\relax                                                                                                                                                                                                                                                       \label{inst:0131}\vfill
\and Space Science Data Center - ASI, Via del Politecnico SNC, 00133 Roma, Italy\relax                                                                                                                                                                                                                                                                             \label{inst:0132}\vfill
\and Department of Physics, University of Helsinki, P.O. Box 64, 00014 Helsinki, Finland\relax                                                                                                                                                                                                                                                                     \label{inst:0134}\vfill
\and Finnish Geospatial Research Institute FGI, Geodeetinrinne 2, 02430 Masala, Finland\relax                                                                                                                                                                                                                                                                      \label{inst:0135}\vfill
\and Institut UTINAM CNRS UMR6213, Universit\'{e} Bourgogne Franche-Comt\'{e}, OSU THETA Franche-Comt\'{e} Bourgogne, Observatoire de Besan\c{c}on, BP1615, 25010 Besan\c{c}on Cedex, France\relax                                                                                                                                                                 \label{inst:0141}\vfill
\and HE Space Operations BV for European Space Agency (ESA), Keplerlaan 1, 2201AZ, Noordwijk, The Netherlands\relax                                                                                                                                                                                                                                                \label{inst:0144}\vfill
\and Dpto. de Inteligencia Artificial, UNED, c/ Juan del Rosal 16, 28040 Madrid, Spain\relax                                                                                                                                                                                                                                                                       \label{inst:0146}\vfill
\and Konkoly Observatory, Research Centre for Astronomy and Earth Sciences, E\"{ o}tv\"{ o}s Lor{\'a}nd Research Network (ELKH), MTA Centre of Excellence, Konkoly Thege Mikl\'{o}s \'{u}t 15-17, 1121 Budapest, Hungary\relax                                                                                                                                     \label{inst:0153}\vfill
\and ELTE E\"{ o}tv\"{ o}s Lor\'{a}nd University, Institute of Physics, 1117, P\'{a}zm\'{a}ny P\'{e}ter s\'{e}t\'{a}ny 1A, Budapest, Hungary\relax                                                                                                                                                                                                                 \label{inst:0154}\vfill
\and Instituut voor Sterrenkunde, KU Leuven, Celestijnenlaan 200D, 3001 Leuven, Belgium\relax                                                                                                                                                                                                                                                                      \label{inst:0156}\vfill
\and Department of Astrophysics/IMAPP, Radboud University, P.O.Box 9010, 6500 GL Nijmegen, The Netherlands\relax                                                                                                                                                                                                                                                   \label{inst:0157}\vfill
\and University of Vienna, Department of Astrophysics, T\"{ u}rkenschanzstra{\ss}e 17, A1180 Vienna, Austria\relax                                                                                                                                                                                                                                                 \label{inst:0165}\vfill
\and Institute of Physics, Laboratory of Astrophysics, Ecole Polytechnique F\'ed\'erale de Lausanne (EPFL), Observatoire de Sauverny, 1290 Versoix, Switzerland\relax                                                                                                                                                                                              \label{inst:0167}\vfill
\and School of Physics and Astronomy / Space Park Leicester, University of Leicester, University Road, Leicester LE1 7RH, United Kingdom\relax                                                                                                                                                                                                                     \label{inst:0180}\vfill
\and Thales Services for CNES Centre Spatial de Toulouse, 18 avenue Edouard Belin, 31401 Toulouse Cedex 9, France\relax                                                                                                                                                                                                                                            \label{inst:0182}\vfill
\and Depto. Estad\'istica e Investigaci\'on Operativa. Universidad de C\'adiz, Avda. Rep\'ublica Saharaui s/n, 11510 Puerto Real, C\'adiz, Spain\relax                                                                                                                                                                                                             \label{inst:0186}\vfill
\and Center for Research and Exploration in Space Science and Technology, University of Maryland Baltimore County, 1000 Hilltop Circle, Baltimore MD, USA\relax                                                                                                                                                                                                    \label{inst:0188}\vfill
\and GSFC - Goddard Space Flight Center, Code 698, 8800 Greenbelt Rd, 20771 MD Greenbelt, United States\relax                                                                                                                                                                                                                                                      \label{inst:0189}\vfill
\and EURIX S.r.l., Corso Vittorio Emanuele II 61, 10128, Torino, Italy\relax                                                                                                                                                                                                                                                                                       \label{inst:0191}\vfill
\and Porter School of the Environment and Earth Sciences, Tel Aviv University, Tel Aviv 6997801, Israel\relax                                                                                                                                                                                                                                                      \label{inst:0192}\vfill
\and Harvard-Smithsonian Center for Astrophysics, 60 Garden St., MS 15, Cambridge, MA 02138, USA\relax                                                                                                                                                                                                                                                             \label{inst:0193}\vfill
\and HE Space Operations BV for European Space Agency (ESA), Camino bajo del Castillo, s/n, Urbanizacion Villafranca del Castillo, Villanueva de la Ca\~{n}ada, 28692 Madrid, Spain\relax                                                                                                                                                                          \label{inst:0195}\vfill
\and Instituto de Astrof\'{i}sica e Ci\^{e}ncias do Espa\c{c}o, Universidade do Porto, CAUP, Rua das Estrelas, PT4150-762 Porto, Portugal\relax                                                                                                                                                                                                                    \label{inst:0196}\vfill
\and LFCA/DAS,Universidad de Chile,CNRS,Casilla 36-D, Santiago, Chile\relax                                                                                                                                                                                                                                                                                        \label{inst:0198}\vfill
\and SISSA - Scuola Internazionale Superiore di Studi Avanzati, via Bonomea 265, 34136 Trieste, Italy\relax                                                                                                                                                                                                                                                        \label{inst:0202}\vfill
\and Telespazio for CNES Centre Spatial de Toulouse, 18 avenue Edouard Belin, 31401 Toulouse Cedex 9, France\relax                                                                                                                                                                                                                                                 \label{inst:0207}\vfill
\and University of Turin, Department of Computer Sciences, Corso Svizzera 185, 10149 Torino, Italy\relax                                                                                                                                                                                                                                                           \label{inst:0211}\vfill
\and Dpto. de Matem\'{a}tica Aplicada y Ciencias de la Computaci\'{o}n, Univ. de Cantabria, ETS Ingenieros de Caminos, Canales y Puertos, Avda. de los Castros s/n, 39005 Santander, Spain\relax                                                                                                                                                                   \label{inst:0212}\vfill
\and DLR Gesellschaft f\"{ u}r Raumfahrtanwendungen (GfR) mbH M\"{ u}nchener Stra{\ss}e 20 , 82234 We{\ss}ling\relax                                                                                                                                                                                                                                               \label{inst:0225}\vfill
\and Centre for Astrophysics Research, University of Hertfordshire, College Lane, AL10 9AB, Hatfield, United Kingdom\relax                                                                                                                                                                                                                                         \label{inst:0227}\vfill
\and University of Turin, Mathematical Department ""G.Peano"", Via Carlo Alberto 10, 10123 Torino, Italy\relax                                                                                                                                                                                                                                                     \label{inst:0234}\vfill
\and INAF - Osservatorio Astronomico d'Abruzzo, Via Mentore Maggini, 64100 Teramo, Italy\relax                                                                                                                                                                                                                                                                     \label{inst:0240}\vfill
\and Instituto de Astronomia, Geof\`{i}sica e Ci\^{e}ncias Atmosf\'{e}ricas, Universidade de S\~{a}o Paulo, Rua do Mat\~{a}o, 1226, Cidade Universitaria, 05508-900 S\~{a}o Paulo, SP, Brazil\relax                                                                                                                                                                \label{inst:0243}\vfill
\and APAVE SUDEUROPE SAS for CNES Centre Spatial de Toulouse, 18 avenue Edouard Belin, 31401 Toulouse Cedex 9, France\relax                                                                                                                                                                                                                                        \label{inst:0250}\vfill
\and M\'{e}socentre de calcul de Franche-Comt\'{e}, Universit\'{e} de Franche-Comt\'{e}, 16 route de Gray, 25030 Besan\c{c}on Cedex, France\relax                                                                                                                                                                                                                  \label{inst:0253}\vfill
\and ATOS for CNES Centre Spatial de Toulouse, 18 avenue Edouard Belin, 31401 Toulouse Cedex 9, France\relax                                                                                                                                                                                                                                                       \label{inst:0258}\vfill
\and School of Physics and Astronomy, Tel Aviv University, Tel Aviv 6997801, Israel\relax                                                                                                                                                                                                                                                                          \label{inst:0261}\vfill
\and Astrophysics Research Centre, School of Mathematics and Physics, Queen's University Belfast, Belfast BT7 1NN, UK\relax                                                                                                                                                                                                                                        \label{inst:0263}\vfill
\and Centre de Donn\'{e}es Astronomique de Strasbourg, Strasbourg, France\relax                                                                                                                                                                                                                                                                                    \label{inst:0265}\vfill
\and Institute for Computational Cosmology, Department of Physics, Durham University, Durham DH1 3LE, UK\relax                                                                                                                                                                                                                                                     \label{inst:0268}\vfill
\and European Southern Observatory, Karl-Schwarzschild-Str. 2, 85748 Garching, Germany\relax                                                                                                                                                                                                                                                                       \label{inst:0269}\vfill
\and Max-Planck-Institut f\"{ u}r Astrophysik, Karl-Schwarzschild-Stra{\ss}e 1, 85748 Garching, Germany\relax                                                                                                                                                                                                                                                      \label{inst:0270}\vfill
\and Data Science and Big Data Lab, Pablo de Olavide University, 41013, Seville, Spain\relax                                                                                                                                                                                                                                                                       \label{inst:0274}\vfill
\and Barcelona Supercomputing Center (BSC), Pla\c{c}a Eusebi G\"{ u}ell 1-3, 08034-Barcelona, Spain\relax                                                                                                                                                                                                                                                          \label{inst:0282}\vfill
\and ETSE Telecomunicaci\'{o}n, Universidade de Vigo, Campus Lagoas-Marcosende, 36310 Vigo, Galicia, Spain\relax                                                                                                                                                                                                                                                   \label{inst:0287}\vfill
\and Asteroid Engineering Laboratory, Space Systems, Lule\aa{} University of Technology, Box 848, S-981 28 Kiruna, Sweden\relax                                                                                                                                                                                                                                    \label{inst:0291}\vfill
\and Vera C Rubin Observatory,  950 N. Cherry Avenue, Tucson, AZ 85719, USA\relax                                                                                                                                                                                                                                                                                  \label{inst:0296}\vfill
\and Department of Astrophysics, Astronomy and Mechanics, National and Kapodistrian University of Athens, Panepistimiopolis, Zografos, 15783 Athens, Greece\relax                                                                                                                                                                                                  \label{inst:0297}\vfill
\and TRUMPF Photonic Components GmbH, Lise-Meitner-Stra{\ss}e 13,  89081 Ulm, Germany\relax                                                                                                                                                                                                                                                                        \label{inst:0300}\vfill
\and IAC - Instituto de Astrofisica de Canarias, Via L\'{a}ctea s/n, 38200 La Laguna S.C., Tenerife, Spain\relax                                                                                                                                                                                                                                                   \label{inst:0305}\vfill
\and Department of Astrophysics, University of La Laguna, Via L\'{a}ctea s/n, 38200 La Laguna S.C., Tenerife, Spain\relax                                                                                                                                                                                                                                          \label{inst:0306}\vfill
\and Faculty of Aerospace Engineering, Delft University of Technology, Kluyverweg 1, 2629 HS Delft, The Netherlands\relax                                                                                                                                                                                                                                          \label{inst:0308}\vfill
\and Laboratoire Univers et Particules de Montpellier, CNRS Universit\'{e} Montpellier, Place Eug\`{e}ne Bataillon, CC72, 34095 Montpellier Cedex 05, France\relax                                                                                                                                                                                                 \label{inst:0312}\vfill
\and Universit\'{e} de Caen Normandie, C\^{o}te de Nacre Boulevard Mar\'{e}chal Juin, 14032 Caen, France\relax                                                                                                                                                                                                                                                     \label{inst:0318}\vfill
\and LESIA, Observatoire de Paris, Universit\'{e} PSL, CNRS, Sorbonne Universit\'{e}, Universit\'{e} de Paris, 5 Place Jules Janssen, 92190 Meudon, France\relax                                                                                                                                                                                                   \label{inst:0319}\vfill
\and SRON Netherlands Institute for Space Research, Niels Bohrweg 4, 2333 CA Leiden, The Netherlands\relax                                                                                                                                                                                                                                                         \label{inst:0326}\vfill
\and Astronomical Observatory, University of Warsaw,  Al. Ujazdowskie 4, 00-478 Warszawa, Poland\relax                                                                                                                                                                                                                                                             \label{inst:0327}\vfill
\and Scalian for CNES Centre Spatial de Toulouse, 18 avenue Edouard Belin, 31401 Toulouse Cedex 9, France\relax                                                                                                                                                                                                                                                    \label{inst:0329}\vfill
\and Universit\'{e} Rennes, CNRS, IPR (Institut de Physique de Rennes) - UMR 6251, 35000 Rennes, France\relax                                                                                                                                                                                                                                                      \label{inst:0335}\vfill
\and Shanghai Astronomical Observatory, Chinese Academy of Sciences, 80 Nandan Road, Shanghai 200030, People's Republic of China\relax                                                                                                                                                                                                                             \label{inst:0340}\vfill
\and University of Chinese Academy of Sciences, No.19(A) Yuquan Road, Shijingshan District, Beijing 100049, People's Republic of China\relax                                                                                                                                                                                                                       \label{inst:0342}\vfill
\and Niels Bohr Institute, University of Copenhagen, Juliane Maries Vej 30, 2100 Copenhagen {\O}, Denmark\relax                                                                                                                                                                                                                                                    \label{inst:0345}\vfill
\and DXC Technology, Retortvej 8, 2500 Valby, Denmark\relax                                                                                                                                                                                                                                                                                                        \label{inst:0346}\vfill
\and Las Cumbres Observatory, 6740 Cortona Drive Suite 102, Goleta, CA 93117, USA\relax                                                                                                                                                                                                                                                                            \label{inst:0347}\vfill
\and CIGUS CITIC, Department of Nautical Sciences and Marine Engineering, University of A Coru\~{n}a, Paseo de Ronda 51, 15071, A Coru\~{n}a, Spain\relax                                                                                                                                                                                                          \label{inst:0355}\vfill
\and Astrophysics Research Institute, Liverpool John Moores University, 146 Brownlow Hill, Liverpool L3 5RF, United Kingdom\relax                                                                                                                                                                                                                                  \label{inst:0356}\vfill
\and IPAC, Mail Code 100-22, California Institute of Technology, 1200 E. California Blvd., Pasadena, CA 91125, USA\relax                                                                                                                                                                                                                                           \label{inst:0363}\vfill
\and MTA CSFK Lend\"{ u}let Near-Field Cosmology Research Group, Konkoly Observatory, MTA Research Centre for Astronomy and Earth Sciences, Konkoly Thege Mikl\'{o}s \'{u}t 15-17, 1121 Budapest, Hungary\relax                                                                                                                                                    \label{inst:0380}\vfill
\and Departmento de F\'{i}sica de la Tierra y Astrof\'{i}sica, Universidad Complutense de Madrid, 28040 Madrid, Spain\relax                                                                                                                                                                                                                                        \label{inst:0395}\vfill
\and Ru{\dj}er Bo\v{s}kovi\'{c} Institute, Bijeni\v{c}ka cesta 54, 10000 Zagreb, Croatia\relax                                                                                                                                                                                                                                                                     \label{inst:0401}\vfill
\and Villanova University, Department of Astrophysics and Planetary Science, 800 E Lancaster Avenue, Villanova PA 19085, USA\relax                                                                                                                                                                                                                                 \label{inst:0415}\vfill
\and INAF - Osservatorio Astronomico di Brera, via E. Bianchi, 46, 23807 Merate (LC), Italy\relax                                                                                                                                                                                                                                                                  \label{inst:0421}\vfill
\and STFC, Rutherford Appleton Laboratory, Harwell, Didcot, OX11 0QX, United Kingdom\relax                                                                                                                                                                                                                                                                         \label{inst:0426}\vfill
\and Charles University, Faculty of Mathematics and Physics, Astronomical Institute of Charles University, V Holesovickach 2, 18000 Prague, Czech Republic\relax                                                                                                                                                                                                   \label{inst:0447}\vfill
\and Department of Particle Physics and Astrophysics, Weizmann Institute of Science, Rehovot 7610001, Israel\relax                                                                                                                                                                                                                                                 \label{inst:0458}\vfill
\and Department of Astrophysical Sciences, 4 Ivy Lane, Princeton University, Princeton NJ 08544, USA\relax                                                                                                                                                                                                                                                         \label{inst:0459}\vfill
\and Departamento de Astrof\'{i}sica, Centro de Astrobiolog\'{i}a (CSIC-INTA), ESA-ESAC. Camino Bajo del Castillo s/n. 28692 Villanueva de la Ca\~{n}ada, Madrid, Spain\relax                                                                                                                                                                                      \label{inst:0468}\vfill
\and naXys, University of Namur, Rempart de la Vierge, 5000 Namur, Belgium\relax                                                                                                                                                                                                                                                                                   \label{inst:0471}\vfill
\and CGI Deutschland B.V. \& Co. KG, Mornewegstr. 30, 64293 Darmstadt, Germany\relax                                                                                                                                                                                                                                                                               \label{inst:0479}\vfill
\and Institute of Global Health, University of Geneva\relax                                                                                                                                                                                                                                                                                                        \label{inst:0480}\vfill
\and Astronomical Observatory Institute, Faculty of Physics, Adam Mickiewicz University, Pozna\'{n}, Poland\relax                                                                                                                                                                                                                                                  \label{inst:0482}\vfill
\and H H Wills Physics Laboratory, University of Bristol, Tyndall Avenue, Bristol BS8 1TL, United Kingdom\relax                                                                                                                                                                                                                                                    \label{inst:0486}\vfill
\and Department of Physics and Astronomy G. Galilei, University of Padova, Vicolo dell'Osservatorio 3, 35122, Padova, Italy\relax                                                                                                                                                                                                                                  \label{inst:0493}\vfill
\and CERN, Geneva, Switzerland\relax                                                                                                                                                                                                                                                                                                                               \label{inst:0495}\vfill
\and Applied Physics Department, Universidade de Vigo, 36310 Vigo, Spain\relax                                                                                                                                                                                                                                                                                     \label{inst:0497}\vfill
\and Association of Universities for Research in Astronomy, 1331 Pennsylvania Ave. NW, Washington, DC 20004, USA\relax                                                                                                                                                                                                                                             \label{inst:0501}\vfill
\and European Southern Observatory, Alonso de C\'ordova 3107, Casilla 19, Santiago, Chile\relax                                                                                                                                                                                                                                                                    \label{inst:0509}\vfill
\and Sorbonne Universit\'{e}, CNRS, UMR7095, Institut d'Astrophysique de Paris, 98bis bd. Arago, 75014 Paris, France\relax                                                                                                                                                                                                                                         \label{inst:0514}\vfill
\and Faculty of Mathematics and Physics, University of Ljubljana, Jadranska ulica 19, 1000 Ljubljana, Slovenia\relax                                                                                                                                                                                                                                               \label{inst:0516}\vfill
}

   \date{Received ; accepted }

 
  \abstract
   {With the most recent \gaia\ data release the number of sources with complete 6D phase space information (position and velocity) has increased to well over 33 million stars, while stellar astrophysical parameters are provided for more than 470 million sources, in addition to the identification of over 11 million variable stars.}
   {
   Using the astrophysical parameters and variability classifications provided in \gdrthree, we select various stellar populations to explore and identify non-axisymmetric features in the disc of the Milky Way in both configuration and velocity space.}
   {Using more about 580 thousand sources identified as hot OB stars, together with 988 known open clusters younger than 100 million years, we map the spiral structure associated with star formation 4-5 kpc from the Sun. We select over 2800 Classical Cepheids younger than 200 million years, which show spiral features extending as far as 10 kpc from the Sun in the outer disc.
   We also identify more than 8.7 million sources on the red giant branch (RGB), of which 5.7 million have line-of-sight velocities, allowing the velocity field of the Milky Way to be mapped as far as 8 kpc from the Sun, including the inner disc. 
}
   {The spiral structure revealed by the young populations is consistent with recent results using \gedrthree\ astrometry and source lists based on near infrared photometry, showing the Local (Orion) arm to be at least 8 kpc long, and an outer arm consistent with what is seen in HI surveys, which seems to be a continuation of the Perseus arm into the third quadrant. Meanwhile, the subset of RGB stars with velocities clearly reveals the large scale kinematic signature of the bar in the inner disc, as well as evidence of streaming motions in the outer disc that might be associated with spiral arms or bar resonances.  A local comparison of the velocity field of the OB stars reveals both similarities and differences with the RGB sample.}
   {This cursory study of \gdrthree\ data shows there is a rich bounty of kinematic information to be explored more deeply, which will undoubtedly lead us to an understanding of the dynamical nature of the Milky Way's non-axisymmetric structures.}  


   \maketitle
%

\section{Introduction}

The determination of the structure and kinematics of the Milky Way has been under investigation for more than a century. Researchers have been able to describe the morphology of external galaxies using deep photometric surveys, but the structure and evolution of our own Galaxy still remains a mystery in many aspects. Difficulties rise from the fact that we are observing it from the inside and cannot construct a complete picture as we can for other galaxies. Until recently it was necessary to infer large scale characteristics of the Milky Way from a limited amount of stars located in the Solar neighbourhood. Efforts to overcome this limitation over the past two decades have progressed thanks to a combination of large scale photometric and spectroscopic surveys, such as \sdss{} \citep{Juric:2008,York:2000}, \rave{} \citep{Steinmetz:2020}, \apogee{} \citep[][]{Majewski:2017, Jonsson:2020} , \lamost{} \citep{Cui:2012, Zhao:2012}, and \galah{} \citep{DeSilva:2015}.
\gaia\ has revolutionised this field starting with  first \gaia\ data release (DR1), providing new insights on the stability of the Galactic disc \citep[e.g.][]{DR2-DPACP-33,Antoja2018}, its merger history \citep[e.g.][]{Helmi2018,Belokurov2018}, its structure through the discovery of new open clusters \citep[e.g.][]{CantatGaudin2018,Castro-Ginard2019}, to name a few. These results largely used the unprecedented amount of about 7 million stars with 6-dimensional phase space information in \gaia's second data release (DR2). The sample with Radial Velocity Spectrometer (RVS) measurements of \citet{DR2-DPACP-33}, with full 6D phase space measurements, already showed that 
the disc of the Milky Way is not kinematically axisymmetric. Since then a large number of contributions have been published providing new results and characteristics of the Milky Way disc. We refer the reader to \citet{Brown:2021} for an updated review of the Milky Way with \gaia\, compared to a pre-\gaia\ view \citep[e.g.][]{JnO2016}. The purpose of this contribution is to highlight the new information that is contained in the most recent \gaia\ data release regarding the structure of the Milky Way's disc, as revealed in configuration and velocity space.


Being rich in gas and having a disc structure, it was immediately expected that the Milky Way would have non-axisymmetric structures, like other spiral galaxies. Clear evidence of spiral structure from the distribution of local OB associations has been known since the 1950s \citep{Morgan1953}. Since then, the location of spiral arms in density in the local neighbourhood has been studied using different tracers,
such as giant molecular clouds, high-mass star-formation region masers, H II regions, and young stars (OB stars, Cepheids, young open clusters).  Using parallaxes and proper motions of masers from the \textit{BESSEL} survey, \cite{Reid2019} built logarithmic models of the Galactic spiral arms. The main arms identified in the model are: Norma-Outer arm, Perseus, Sagittarius-Carina arm (Sag-Car hereafter), and Scutum-Centaurus arm. Another arm included is the Local Arm, which in the past has been mostly considered as a minor feature with respect to the other arms just mentioned, as the name suggests. 
\citet{Poggio:2021}, with a local OB sample, Cepheids and young open clusters, and similarly \citet{Hou:2021} with spectroscopically confirmed OB stars, made maps showing the spiral arm segments of the Perseus, Local and Sag-Car arms, extending the arm segments towards the third and fourth quadrant where the masers are mostly absent. While some progress has been made in detailing the large-scale spiral structure as evidenced by star formation products, the dynamical nature of these arms and the mechanisms causing their formation remains unknown. 

The Galactic bar in the inner disc is another long-known asymmetry of the Galaxy, whose kinematic signatures can be found from the inner to the outer disc. Like many external barred galaxies, the Galactic bar has a boxy-bulge shape, but its length, orientation angle and angular velocity are not yet well constrained. In this instance we now have a strong asymmetry in the stellar distribution, at least for the inner regions of the Milky Way. While some evidence of asymmetry in the form of spiral arms extending further out from the bar in the stellar disc is seen in the NIR \citep{Drimmel2000,Churchwell:2009}, it has been challenging to find confirmation that the Milky Way hosts a density-wave like structure in its kinematics. From earlier data releases, \gaia\ revealed that the velocity space of the stars is rich with structure. Most notably, the presence of arches and ridges in the $V_{\phi}-R$ space (galactocentric azimuthal velocity and radius) indicate large scale kinematic phenomena in the Galactic disc \citep{Antoja2018,Ramos:2018,Fragkoudi:2019,Khanna:2019}. Disentangling these into identified resonances with the bar \citep{Trick:2019,Fragkoudi:2019,Trick:2021,Monari:2019,Laporte:2020} and/or spiral arms and/or external perturbations is an ongoing process \citep{Hunt:2019,Khanna:2019,Khoperskov:2021}, and has proven difficult, mostly due to our detailed knowledge of stellar kinematics being contained within at most a few kiloparsecs of the Sun. 



To assist us in understanding the kinematics of the Milky Way, comparison between the observations and models will be important, and some of the previous works already mentioned have used this approach. Other recent works focused on how the spiral arms or a Galactic bar change the expected radial, tangential and vertical kinematic maps include \citet[][]{Faure:2014,Monari:2016,Hunt:2018,Monari:2019,Tepper-Garcia:2021}, based on either test particle simulations or pure N-body simulations. The observable used to compare with the data can either be directly mapping the average or dispersions of the velocity components on the galactic plane, or checking the known moving groups in the Solar Neighbourhood, or the ridges in the azimuthal velocity versus radius diagram. 
In any case, an appropriate comparison of models to data must take into account the selection effects and uncertainties in the data. It is important to check how these affect the prediction in contrast to ideal noise-free data. \citet{RomeroGomez2015} show the capabilities of the \gaia\ nominal mission to constrain the bar characteristic and construct \gaia\ mock catalogues based on the \gaia\ science performance prescriptions for disc Red Clump stars.


Aside from the bar, spiral arms and the galactic warp, there has been some kinematic evidence of additional asymmetries which may indicate disequilibrium on a larger scale. With the \rave{} survey, \cite{Williams:2013} showed the presence of large scale streaming motion in the disc, and revealed differences above and below the Galactic plane. With \sdss{} data, \citet{Widrow:2012} discovered similar wave-like compression/rarefaction features seen in both number density and bulk velocity, as well as towards the Galactic anticentre with \lamost{} data \citep{Carlin:2013}. The large scale velocity field has also been mapped using highly precise line-of-sight\footnote{We adopt this terminology to avoid confusion with galactocentric "radial velocity".} velocities and distance tracers such as Red Clump giants \citep{2015ApJ...800...83B,khanna1}. Their results hinted at streaming motion on scales much larger than about 2.5 kpc, but the analysis was likely limited by incompleteness in data coverage. As a demonstration of the enhanced astrometry and photometry in \gedrthree, \cite{gaia_anticenter2021} mapped the kinematics of the disc out to 14 kpc from the Galactic Center (GC). By selecting data in a narrow azimuthal range (20$^\circ$ about the Galactic anticentre), they studied the azimuthal and vertical velocity components, without requiring line-of-sight velocities. The large sample in their study allowed dissection of the stellar rotation curve in both the young, as well as the older population of stars. Additionally, they showed that kinematic features (such as ridges in $V_{\phi}-R$ space) seen in the inner disc with \gdrtwo{}, extended out to at least $R=14$ kpc. By separately considering the stars above and below the Galactic plane, they also revealed that the lower disc has predominantly higher rotational velocities than the upper disc.  

In this paper, we show the extraordinary capabilities of \gdrthree to shed light on the structure and kinematic issues mentioned above. We use similar tracers as other works, using only the new information provided in \gdrthree to select our samples, and then to map both the density and the kinematics over a large portion of the disc. The paper is organised as follows. Section~\ref{sec:samples} describe the selection of the four tracers used in this work, namely clusters, Cepheids, OB stars and red giant branch (RGB) stars, providing a description of their main properties. Section~\ref{sec:coord_maps} describes the derivation of the  positions, velocities and uncertainties, including a short study on possible systematic effects. Section~\ref{sec:confspace} maps the tracers into configuration space to show how they are distributed in relationship to each other. Section~\ref{sec:velomaps} focuses on the mapping the kinematics of the OB and RGB stars, and what information they contain about the bar and spiral arms. 
Section~\ref{sec:discussion} discusses our results in context with other works and highlights the caveats and short-comings that should be addressed in the future, followed by Section~\ref{sec:concl} summarising our conclusions. 

\section{Selection of tracers \label{sec:samples}}

To map the asymmetry of the Galactic disc with \gaia, we select both young and old stellar populations: the former as the traditional tracer for spiral structure used at optical wavelengths where the surface brightness of disk galaxies, like our Milky Way (MW), are dominated by star formation products; the later as being the component that determines the mass distribution. The latest \gdrthree release for the first time allows us to select samples based on stellar parameters, which we will use to select a sample of OB stars and giants. The subset of sources with line-of-sight velocities \citep{DR3-DPACP-159} will have full 6 dimensional phase space information, allowing us to map the velocity field for these samples. In addition we also investigate the distribution of open clusters and Classical Cepheids (DCEPs), both for which we can derive excellent distances as well as ages. In this section we describe how we construct each of these samples and how distances are derived for each. 

\subsection{Clusters}
\label{sec:OCdata}

We used the list of probable members of the 2017 clusters studied by \citet{CantatGaudin2020} with DR2 data. We obtained the DR3 \texttt{source\_id} of those sources via the available cross-match table, and removed stars whose EDR3 astrometry reveals them to be outliers by more than 3-$\sigma$. This list of members was supplemented with the stars from 628 clusters recently discovered by \citet{CastroGinard2021} in the EDR3 catalogue, and the members found by \citet{Tarricq2021} in the outskirts of 389 nearby clusters. Most of these clusters have associated ages estimated with an artificial neural network applied to the \gdrtwo{} data \citep[for those in][]{CantatGaudin2020} or EDR3 photometry, and 988 of them are younger than 100\,Myr.

The median astrometric parameters (parallax and proper motion) were computed for all clusters and are provided in Table~\ref{tab:meanOcParameters}. Before calculating the median parallax, the individual parallaxes were corrected following the recipe provided by \citet{EDR3-DPACP-132}. Given the statistical precision obtained from using a large number of (corrected) parallaxes, we estimate distances by inverting the median cluster parallax. The uncertainty on the bulk cluster astrometry is estimated as the quadratic sum of the statistical uncertainty and the uncertainty due to small-scale correlations, taken as 10\,$\mu$as in parallax and 25\,$\mu$as\,yr$^{-1}$ in proper motion \citep[][]{Vasiliev21}. 

\begin{figure}
    \centering
    \includegraphics[width=0.49\textwidth]{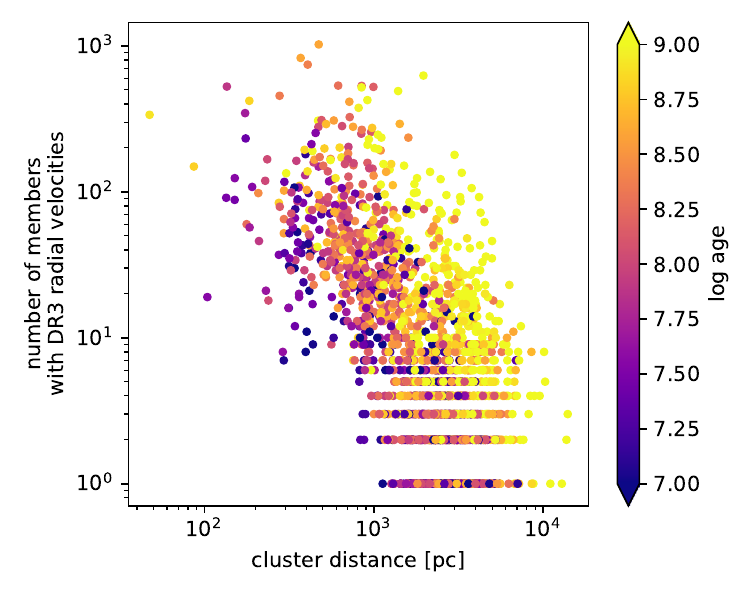}
    \caption{Number of cluster members with available line-of-sight velocities in \gdrthree\ as a function of cluster distance and age.
    }
    \label{fig:clusters_NbRV}
\end{figure}

We also computed the median line-of-sight velocity\footnote{See footnote 1} for the 2162 clusters in which at least one member has a DR3 line-of-sight velocity. Out of the 988 clusters younger than 100\,Myr, 698 have line-of-sight velocities from DR3. The bulk line-of-sight velocities were computed from an average of 48 members per cluster, although this number varies significantly with age and distance (Fig.~\ref{fig:clusters_NbRV}). For comparison, in \gdrtwo{} the line-of-sight velocities were only available for an average of 10 stars per cluster. We estimate the line-of-sight velocity uncertainty as:
\begin{equation}
\sigma_{los,cluster} = \sqrt{  \frac{ 1 }{ \sum_{i=1}^{N} \left( \frac{1}{\sigma_{v_{\mathrm{los},i}}} \right)^2 }    + 0.5^2 }
\label{eq:rv_errer_oc}
\end{equation}
\noindent where $\sigma_{v_{\mathrm{los},i}}$ are the nominal line-of-sight velocity uncertainties of the $N$ cluster members, and 0.5\,km\,s$^{-1}$ is a conservative estimate of the line-of-sight velocity accuracy estimated in DR2 \citep{Deepak18,Katz19}. Although future investigations of the DR3 line-of-sight velocities are likely to show improved systematics with respect to DR2, this conservative choice has no significant impact on the results of this paper.

\begin{table*}
\begin{center}
	\caption{ \label{tab:meanOcParameters} Summary of mean parameters for the cluster sample. The full table is available as an electronic table via CDS.}
	\small\addtolength{\tabcolsep}{-2pt}
	\begin{tabular}{ c  c  c  c  c  c  c  c  c  c  c  c  c  c  c }
	\hline
	\hline

	cluster & $\ell$ & $b$ & $N$ & $\mu_{\alpha}*$ & $\sigma_{\mu_{\alpha}*}$ & $\mu_{\delta}$ & $\sigma_{\mu_{\delta}}$ & $\varpi$ & $\sigma_{\varpi}$ & $v_\mathrm{los}$ & $\sigma_{v_\mathrm{los}}$ & $N_{v_\mathrm{los}}$ & $\log t$ & $n_0$ \\
  & [deg] & [deg] &   & [mas\,yr$^{-1}$] & [mas\,yr$^{-1}$] & [mas\,yr$^{-1}$] & [mas\,yr$^{-1}$] & [mas] & [mas] & [km\,s$^{-1}$] & [km\,s$^{-1}$] &   &   &   \\

	\hline

ASCC 10  &  155.558 & -17.801 & 147 & -1.840 & 0.197 & -1.396 & 0.157 & 1.530 & 0.047 & -17.07 & 0.54 & 84 & 8.42 & 5 \\
ASCC 101  &  67.978 & 11.608 & 106 & 0.940 & 0.228 & 1.200 & 0.250 & 2.548 & 0.052 & -19.04 & 0.53 & 56 & 8.69 & 1 \\
ASCC 105  &  62.860 & 2.025 & 148 & 1.429 & 0.147 & -1.626 & 0.147 & 1.817 & 0.039 & -16.83 & 0.62 & 71 & 7.87 & 3 \\

\multicolumn{15}{c}{...} \\

UBC 1628  &  319.829 & -4.227 & 10 & -2.554 & 0.032 & -2.840 & 0.043 & 0.552 & 0.013 & -44.87 & 18.35 & 1 & 8.47 & 1 \\
	\hline
	\hline
	\end{tabular}
\tablefoot{The full table is available as an electronic table via CDS. $N$: number of probable members kept to compute astrometric parameters. $\sigma_{\mu_{\alpha}*}$, $\sigma_{\mu_{\delta}}$, and  $\sigma_{\varpi}$ are the observed standard deviations of the members. $\sigma_{v_\mathrm{los}}$: computed line-of-sight velocity uncertainty. $N_{v_\mathrm{los}}$: number of members used to compute the cluster line-of-sight velocity. $n_0$: number of members with an absolute magnitude brighter than 0. }
\end{center}
\end{table*}

\subsection{Classical Cepheids}
\label{sect:dceps}

The sample of DCEPs adopted in this work is 
mainly based on the list of sources in the {\tt vari\_cepheid} 
table which is published in DR3 as a result of the processing by  the Specific Objects Study (SOS) pipeline specifically designed to validate and fully characterise DCEPs and RR Lyrae stars observed by \gaia\ (hereafter referred to as SOS Cep\&RRL pipeline) \citep[see][for full details]{Clementini2016,Clementini2019, DR3-DPACP-168, DR3-DPACP-169}. This sample is composed of 3286 DCEPs belonging to the MW, of which 1995 pulsate in the fundamental mode (F), 1097 in the first overtone (1O) and 194 are multi mode (MULTI) pulsators.
 For these DCEPs, the SOS Cep\&RRL pipeline provides pulsation periods, intensity-averaged magnitudes, peak-to-peak amplitudes, Fourier parameters and other quantities whose full description can be found in \citet{DR3-DPACP-169}. 
 The DR3 DCEPs sample was complemented with DCEPs taken by the recent compilations of \citet{Piet2021} and \citet{Inno2021}.
 After removing overlaps with the DR3 DCEPs sample,\footnote{We also neglected two additional multi-mode DCEPs pulsating in the second and third mode.} and retaining only objects with a valid measurements of the mean magnitude in all three \gaia\ passbands and reliable proper motions, we find an additional 564 objects from \citet{Piet2021} and 43 objects from \citet{Inno2021}, with 27 objects in common.  For these later we adopt the classifications and periods from  \citet{Piet2021}, giving a total of an additional 580 DCEPs from the literature.
 However, an additional 81 literature DCEPs were removed as suspect binaries from their position in the Period-Wesenheit diagram (see next section). We therefore are left with 486 and 13 DCEPs from the \citet{Piet2021} and \citet{Inno2021} catalogues, respectively.
The total sample is therefore composed of 3785 DCEPs. However, as we shall see below, we will further clip this sample.  

\subsubsection{Distances and cleaning of the sample.}

An estimation of the distance to each  DCEP in our sample was obtained directly from the definition of distance modulus $w-W$=$-5+5\log D$, where $w$ and $W$ are the apparent and absolute Wesenheit magnitudes\footnote{The Wesenheit magnitudes are reddening-free by construction, assuming that the extinction law is known \citep[][]{Madore1982}}, respectively. The coefficient of the $w$ magnitude has been derived in the \gaia\ bands on an empirical basis by \citet[][]{Ripepi2019} and is defined as $w=G-1.90 \times (G_{BP}-G_{RP})$. The absolute Wesenheit magnitude $W$ was calculated using the period-Wesenheit-metallicity ($PWZ$) relation recently published by \citet{Ripepi2022}: 

\begin{equation}
\begin{aligned}
W=(-5.988\pm0.018)-(3.176\pm0.044)(\log P-1.0)\\-(0.520\pm0.090){\rm [Fe/H]}
\label{eq:pwz}
\end{aligned}
\end{equation}

To calculate the value of $w$ for the DCEPs in our sample we used different {\it Gaia} $(G,G_{BP},G_{RP})$ magnitude data sets. For the DCEPs in the DR3 {\tt vari\_cepheid} table, the SOS Cep\&RRL pipeline provides intensity-averaged magnitudes in the three {\it Gaia} bands, that is, magnitudes which are calculated as to resemble as much as possible the magnitude that the DCEPs would have if they were  non-variable stars. Instead, for the 499 literature DCEPs 
that are not in the DR3 \linktotable{vari_cepheid} table, we only have mean magnitudes estimated in the {\it Gaia} photometric processing \citep[see][for details]{Riello2021} 
and available for all sources in the {\it Gaia} source catalogue.
However, using mean magnitudes in the {\it Gaia} source catalogue for the literature DCEPs sample does not bias our results because it was found that the difference between $w$ magnitudes calculated in the two different ways is of only $-0.01\pm$0.03 mag \citep{Ripepi2022}.
Obtaining reliable values of $w$ is possible only for sources with reliable values of the $G$, $G_{BP}$ and $G_{RP}$ magnitudes. 
Objects with magnitude close to or fainter than $G=20$ mag are expected to have very poor $G_{BP}$ photometry, thus  resulting in unreliable mean  $G_{BP}$ magnitudes. We discuss how to clean the sample for this effect at the end of this section. 

The other ingredient needed to calculate the distance to each DCEP  is $W$, for 
which we  need the period and iron abundance of each pulsator. The periods were taken from the {\tt vari\_cepheid} table  or the literature, while the ${\rm [Fe/H]}$ values are more difficult to obtain. One of the products of \gdrthree\ are iron abundances obtained with the Radial Velocity Spectrometer (RVS) on board {\it Gaia} \citep{DR3-DPACP-104},  which are available in the \linktotable{astrophysical_parameters} table published with DR3.
Here we found spectroscopic values of ${\rm [Fe/H]}$ and their uncertainties for 949 DCEPs of the \gdrthree\ sample and for 27 of the literature DCEPs. 
An estimate of the iron abundance for the remaining DCEPs was obtained adopting the metallicity gradient of the MW disc as measured by \citet{Ripepi2022}: ${\rm [Fe/H]}=(-0.0527\pm0.0022) R+(0.511\pm0.022)$ with rms=0.11 dex \citep[this estimate is in agreement with many other literature estimations, see][for details]{Ripepi2022}. Even if not particularly precise, the iron abundances obtained in this way allow us to use Eq.~\ref{eq:pwz} to derive reliable distances. According to the $PWZ$ relation in Eq.~\ref{eq:pwz}, the impact on the distance to a DCEP produced by an uncertainty of 0.11 dex in metallicity is $\sim$2.5\% and, even considering a conservative uncertainty of 0.2 dex in [Fe/H], the uncertainty on the distance would be a still tolerable 5\%.      

Having estimated both the values of $w$ and $W$ as explained above, it is  straightforward to calculate the distance to each DCEP in our sample and its uncertainty.  
This was calculated by error propagation: $\sigma_d=0.4605 \sigma_\mu d$, where $\sigma_\mu$ is the uncertainty on the distance modulus, calculated by adding in quadrature the uncertainties on $w$ and $W$, which, in turn, were estimated by propagating the uncertainties on the {\it Gaia} magnitudes 
for $w$, and for $W$ the uncertainties in the coefficients of Eq.~\ref{eq:pwz}, and the uncertainty on the iron abundance (the uncertainty on the periods are negligible). 

\begin{figure}
    \centering
    \includegraphics[width=0.49\textwidth]{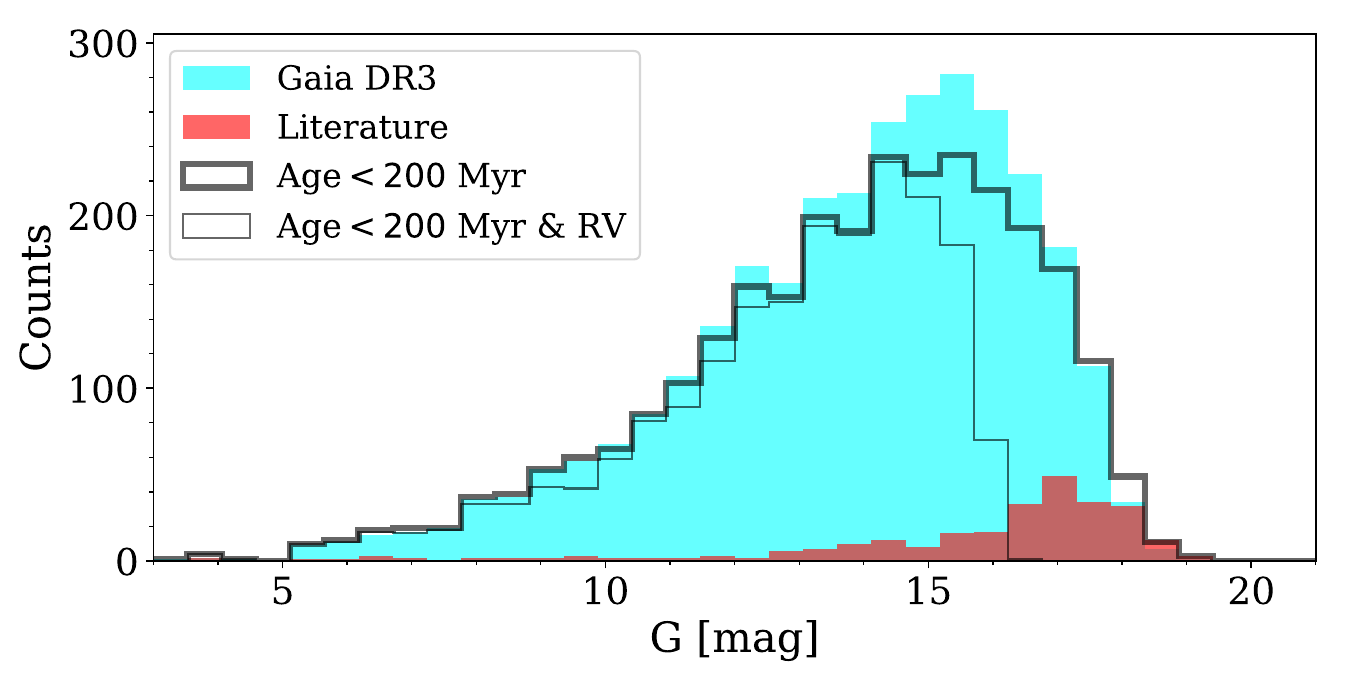}
    \caption{Histogram of the $G$ magnitude for the DCEP sample used in this work. We plot in cyan and red the histograms for the objects from the Gaia DR3 catalogue and those taken from the literature, respectively. The thick line shows the magnitude distribution of the DCEPs after the selection in age ($<$200 Gyr), while the thin line shows the histogram of the objects selected in age that also possess a line-of-sight measurement. In the last two cases the Gaia DR3 and literature samples were merged.
    }
    \label{fig:histoMagDcep}
\end{figure}

An analysis of the derived distances and relative uncertainties revealed that many faint objects had unreliable distances and/or very large uncertainties, due to the large photometric uncertainties, especially in the $G_{BP}$ band. To provide a cleaner sample for further analysis we experimented with the data, reaching the conclusion that retaining only DCEPs with distances smaller than 30 kpc and relative distance uncertainty better than 10\% is a good compromise between precision and completeness. This selection removed 240 DCEPs from the \gdrthree\ sample and 230 literature DCEPs, leaving us with a total final dataset of 3306 DCEPs useful for further analysis. The $G$-magnitude distribution of this selected sample is shown in Fig.~\ref{fig:histoMagDcep}. The number of faint objects, especially among the literature sample is now drastically reduced.

\subsubsection{Line-of-sight velocities.}

The spectra collected with the RVS spectrometer on board {\it Gaia} allows the measurement of time series line-of-sight velocity (RV) values for millions of stars with $G$ magnitude brighter than $\sim$15-16 mag \citep[for details see sect. 6.4.8 of][]{CU6-DR3-documentation}. In DR3 the RV time series are released for a selected sample of 774 DCEPs\footnote{Among these objects, 15 and 9 belong to the Large and Small Megallanic Clouds, respectively}. For this subsample, the SOS Cep\&RRL pipeline computed average RV values and relative uncertainties by fitting the RV curves folded according to the stars' periods and makes them available in the {\tt vari\_cepheid} table \citep[see][for details]{DR3-DPACP-168,DR3-DPACP-169}. 
Nevertheless, for a much larger number of DCEPs in our sample, mean RVs estimated from the spectroscopic pipeline are available in the main {\tt gaia\_source} table. 
We checked whether these arithmetically computed mean RV values are usable for variable objects such as DCEPs by comparing the RV values in the {\tt vari\_cepheid} and the {\tt gaia\_source} tables. We found a perfect agreement with a mean difference of  0.6$\pm$6 km/s and no visible trend \citep[see also][]{DR3-DPACP-168}. On this basis we decided to use the {\tt gaia\_source} catalogue RVs for our DCEP sample. In total we have RV estimates for 2059 DR3 and 67 literature DCEPs, respectively. The uncertainties on these values can be evaluated on the basis of fig. 6.13 by \citet{CU6-DR3-documentation}.

\subsubsection{Ages of DCEPs}

It has been known for a long time that the DCEPs follow a period-age (PA) relation \citep[see e.g.][and references therein]{Bono2005,Anderson2016}. More recently, \citet{Desomma2021}, based on an updated theoretical pulsation scenario, devised a more accurate period-age-metallicity (PAZ) relation. Since the DCEP PWZ relation allows us to obtain individual accurate distances, it follows that the DCEP PAZ allows us to date whatever region of the Galactic disk where a DCEP is present. To take advantage of this powerful tool, we adopted the following equation \citep[see Table 9 of][]{Desomma2021}:

\begin{equation}
\begin{aligned}
\log t = (8.423\pm0.006)-(0.642\pm0.004) \log P \\-(0.067\pm0.006) {\rm [Fe/H]} 
\label{eq:PAZ}
\end{aligned}
\end{equation}

\noindent
with rms=0.081; $t$, $P$ and [Fe/H] are the age (years), the period (days), and the iron abundance (dex). This relation is valid for F-mode pulsators and it was calculated using evolutionary tracks including overshooting. As the PAZ relation is not available for 1O-mode DCEPs \citep[see][for full details]{Desomma2021}, we decided to fundamentalise their periods according to the \citet{Feast1997} relation: 
$P_F = P_{1O}/(0.716-0.027 \log P_{1O})$, where $P_F$ and $P_{1O}$ are the F and 1O mode DCEP periods, respectively. 
In this way we were able to calculate the ages for every DCEP in our sample. 
As we want to use the DCEPs to trace the MW arms, we decided to use only DCEPs younger than 200 Myr. Therefore the sample used in the following is composed of 2808 pulsators, 1948 of which also have line-of-sight velocity measurements.   
Table~\ref{tab:meanDCEPsParameters} shows selected DCEPs properties which are not published in other \gaia\ catalogues or papers.

\begin{table*}
\begin{center}
	\caption{ \label{tab:meanDCEPsParameters} Selected parameters for the DCEPs sample. }
	\small\addtolength{\tabcolsep}{-2pt}
	\begin{tabular}{ c  c  c  c  c  c  c  c  c  c  c  c }
	\hline
	\hline
             source\_id                     & $\ell$ & $b$  &    $d$    &    $\mu$      &  $\sigma_\mu$  & [Fe/H] &$\sigma_\mathrm{[Fe/H]}$ &flag   &   source &     logAge &  $\sigma_\mathrm{logAge}$  \\
   & (deg)         &      (deg)        &   (kpc)  &  (mag)  & (mag)  & (dex) & (dex)   &          &              &      (dex)     &  (dex)     \\
   \hline
           4060910068247394432  &    0.38444   &      1.96190 &  8.02   &  14.522 &    0.129  & 0.50  &  0.20  & 1  &           P21      &   8.594  &   0.016 \\
           4049125051634137600  &   0.45968    &   $-$5.42326  &14.46   & 15.801  &   0.117  & 0.19   & 0.20   &1   &          P21       &  8.706   &  0.015  \\
           4048895253682114432  &    0.47058   &    $-$6.65562 & 19.14  &  16.410  &   0.114 & $-$0.05  & 0.20  & 1  &      Gaia\_DR3  &   8.623  &   0.015  \\
           &    &    &  &    &   &   &   &     &    &     &   \\
          4056461478623363968   & 359.98638  &   $-$1.45395  &15.81   & 15.995  &   0.157  & 0.11   & 0.20  & 1   &     Gaia\_DR3   &  7.715   &  0.015   \\
          \hline
\end{tabular}
\tablefoot{The full table is available as an electronic table via CDS. The source\_id is the \gaia\ source\_id, $\ell$ and $b$ are the galactic coordinates; $d$ is the heliocentric distance in kpc; $\mu$ and $\sigma_\mu$ are the distance modulus and its uncertainty; [Fe/H] and $\sigma$[Fe/H] are the iron abundance and its uncertainty; flag=0 or 1 means that the metallicity was taken from the astrophysical parameters or calculated from the metallicity gradient of the Galactic disc; source lists the provenance of the DCEP source: Gaia\_DR3 means that the star is included in the \gaia\ DR3 {\tt vari\_cepheids} catalogue, P21 or Inno indicate that the DCEPs were taken from the \citet{Piet2021} or \citet{Inno2021} catalogues, respectively; logAge and $\sigma_\mathrm{logAge}$ are the decimal logarithm of the age and its uncertainty.}
\end{center}
\end{table*}

\subsection{OB stars}

To select young stars on the upper main sequence we use the effective temperatures provided in DR3, selecting for stars with $\teff > 10\,000$K. For hot stars there are in general two sets of effective temperatures provided in DR3. One set is provided by a General Stellar Parameterizer from Photometry \citep[\gspphot, see][]{DR3-DPACP-156}, that estimates stellar parameters using the \gaia\ \bpmag/\rpmag\ spectrophotometry, astrometry and $G$ band photometry.  \gspphot makes different sets of parameter estimates using different stellar libraries, then for each source chooses one of these as the "best". Here we use this set of "best" parameters, as reported in the main \gaia\ source table. Another set of parameters is estimated from a software module (ESP-HS) that was optimised specifically for hot stars and which uses the BP/RP spectrophotometry, without the astrometry, together with the RVS spectra if also available \citep{DR3-DPACP-157, DR3-DPACP-160}. This second set of parameters is made available in the \linktotable{astrophysical_parameters} 
table.  Because of different quality filters for these different methods, \gaia\ sources may have one or both sets of effective temperatures, or remain without a temperature estimate. Indeed, only about half the sample of stars with $\teff > 10\,000$K from either method have temperatures from both. From a detailed comparison of those sources with stellar parameters from one or both methods, we settled on the following criteria:
\begin{itemize}
    \item For the stars with only \gspphot temperatures we use the spectral type determined by \esphs for all sources with \gaia\ BP/RP
    spectrophotometry as an additional assurance of quality. That is $\teff > 10\,000$K and the \esphs spectral type flag set to "O", "B" or "A". 
    
    \item For the stars with only \esphs temperatures we require that the effective temperature be in the range $10\,000 < \teff < 50\,000$K, as it was found that the small fraction of sources with $\teff > 50\,000$K are likely to be unreliable \citep{DR3-DPACP-160}. 

    \item For sources with both sets of stellar parameters we require that the effective temperature $\teff > 8000$K for \gspphot and $\teff > 10\,000$K for \esphs, letting the confirmation from \gspphot verify the sources with \esphs hotter than 50\,000K.
\end{itemize}
We note that we only use the effective temperature for selection. A comparison of our temperatures against those found in the literature \citep{Mathur2017,2018ApJS..235...42A,2021arXiv210802878X} for stars in the range 8000 to 10\,000K show an RMS difference  of less than 900K and offsets less than 400K. While the differences increase for higher \teff measured, they remain small enough to that we remain confident that they should nevertheless be in our sample. We also note that the use of \esphs products for all three of the cases above effectively poses an apparent magnitude limit on this sample of 17.65 in $G$.

\begin{figure}
    \centering
    \includegraphics[width=0.45\textwidth]{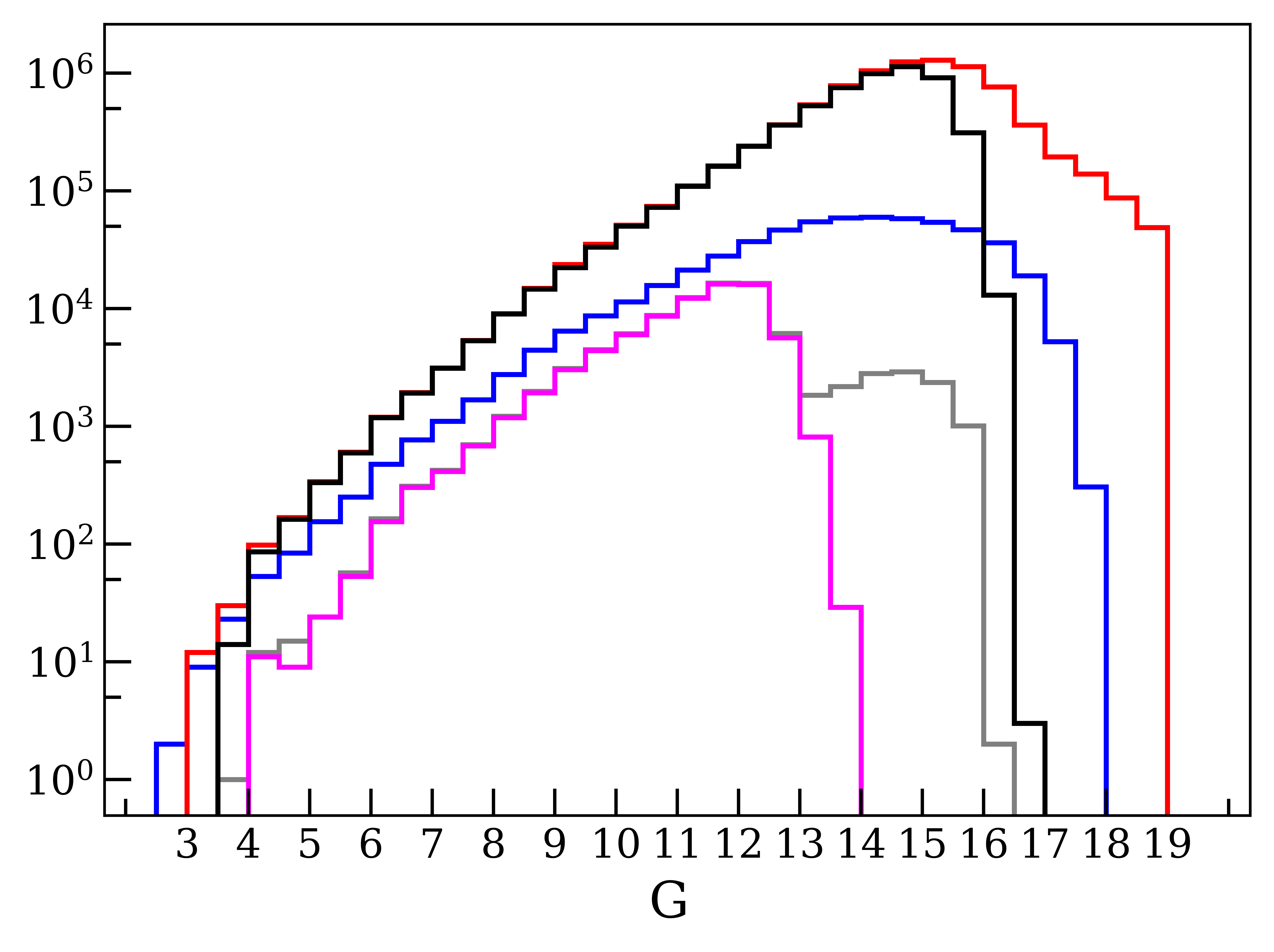}
    \caption{$G$ magnitude distribution of the selected giants (red) and OB (blue) samples. Black and grey histograms show the subsamples with line-of-sight velocities, for the giants and OB stars respectively. The magenta histogram are OB stars with line-of-sight velocities estimated with templates having a \teff > 7000K, and is the selected sample for mapping the velocities of the OB stars.  
    }
    \label{fig:Gmag_ob_hist}
\end{figure}

As a temperature selection will introduce undesired sub-dwarfs and white dwarfs in our sample, we impose the additional criterion $G + 5 \log(\varpi/100) < 2. + 1.8(\gbp - \grp)$ to remove sources fainter than our target upper main sequence stars. The colour term takes into account extinction, where 1.8 is approximately the slope of the reddening vector in $(\gmag, \bpminrp)$ space. To capture distant sources with negative parallaxes we rewrite this criteria in the form 
\begin{equation}
    (\varpi/100.)^5 < 10.^{(2.- \gmag +1.8*(\gbp - \grp))}\, .
\end{equation} 
(See appendix \ref{sec:queries} for an example query.) Together the above criteria give us 923\,700 stars, but we find that outside the plane of the Galaxy we have a significant number of stars in the direction of the Large (LMC) and Small (SMC) Megallanic Clouds, as well as a number of globular cluster members. We remove these contaminants by keeping only stars whose distance from the galactic plane is less than 300pc. This reduces the sample to 621\,609 stars. Finally we use the astrometric fidelity indicator $f_a$ (with values $0 < f_a < 1$) of \citet{Rybizki2022},   to remove sources with suspect astrometry, keeping stars with $f_a > 0.5$ as recommended in \citet{Zari2021}. This last criteria removes only 7\% of the sample, leaving us with a final sample of 579\,577 stars, of which 91\,836 (15.8\%) have line-of-sight velocities. However, we note that a fraction of these have line-of-sight velocities estimated using an RVS spectral template with temperatures very different from the effective temperatures finally estimated for them, and therefore likely to have errant line-of-sight velocities \citep{DR3-DPACP-151}. Removing stars whose  \linktoparam{gaia_source}{rv_template_temp}) is less than 7000K gives us 77\,659 stars with valid line-of-sight velocities.  Figure \ref{fig:Gmag_ob_hist} shows the $G$ magnitude distribution of our OB sample. 

\begin{figure}
    \centering
    \includegraphics[width=0.45\textwidth]{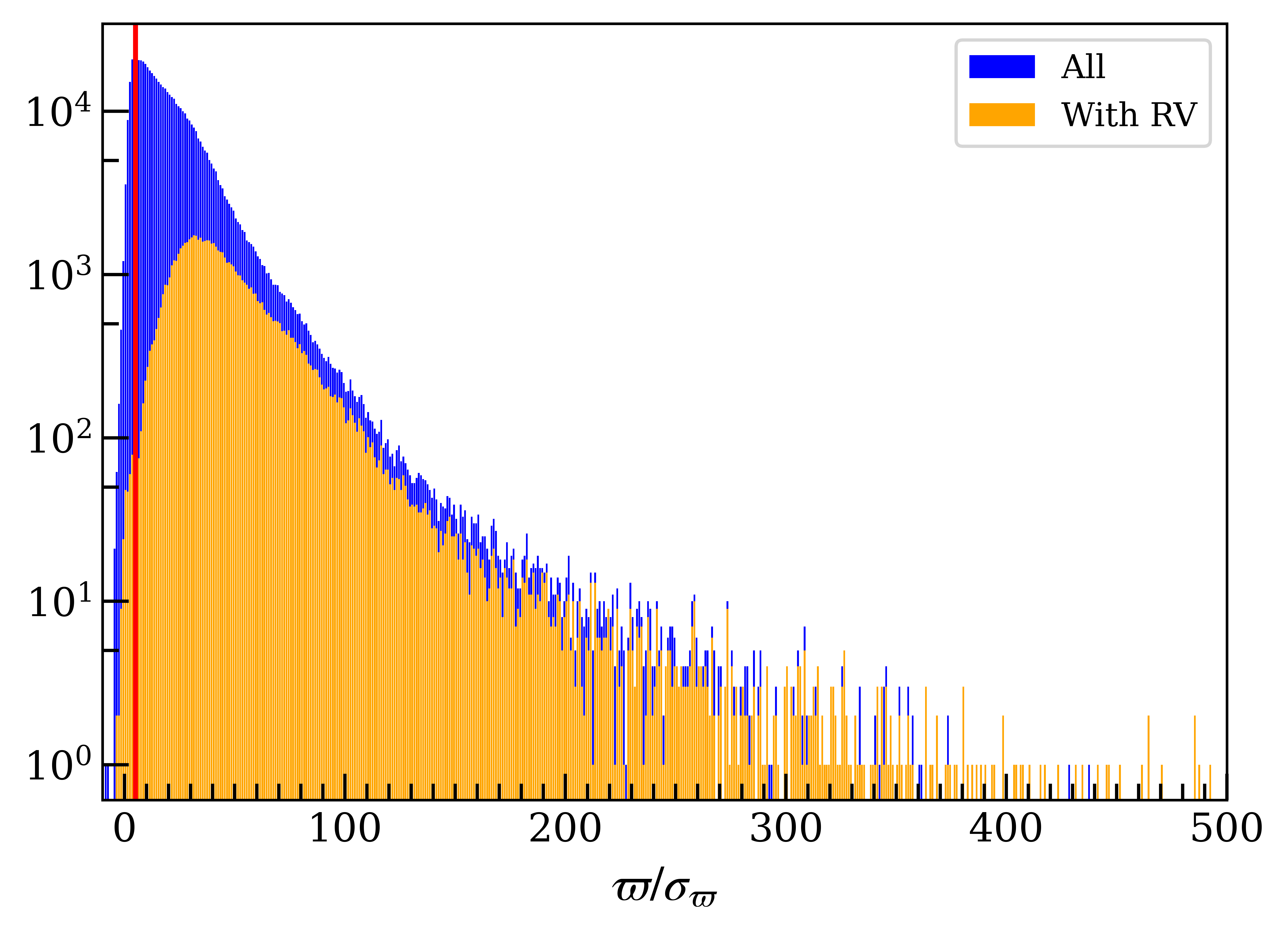}
    \caption{Distribution of $\varpi / \sigma_\varpi$ for the OB sample (blue) and the subsample with line-of-sight velocities (orange). The vertical red line is at $\varpi / \sigma_\varpi = 5$.
    }
    \label{fig:Gmag_plxerr_hist}
\end{figure}

For the purpose of mapping we need distance estimates for our sources. While \gaia\ provides parallaxes, about 40\% of our sample have significant ($\sigma_\varpi/\varpi > 0.20)$ parallax uncertainties, so that a simple inversion of the parallax cannot be considered reliable \citep{CBJ2015, Luri2018}.
Figure \ref{fig:Gmag_plxerr_hist} shows the distribution of $\varpi / \sigma_\varpi$ for our sample as well as the subsample with line-of-sight velocities.
For the subset of our stars with \gspphot parameters, \gdrthree\ also gives us distance estimates based on both astrometry and photometry.  However, 43\% of our sample are sources that have only temperature estimates from \esphs, and therefore no distance estimate from \gspphot. We therefore adopt the "photogeo" distances from \citet{CBJ2021} that are also based on both astrometric and photometric data, as recommended in \citet{DR3-DPACP-160}, and which are available for our entire sample\footnote{These distances can be found in the \gaia\ Archive as the external table, \linktotable{external.gaiaedr3_distance}.}.  

We mention that an alternative selection of high-fidelity OB stars is presented in \citet{DR3-DPACP-123}. The young B-stars from that sample are used for a basic modelling of the Milky Way rotation curve which results in parameters consistent with the mean OB star $V_\phi$ curve derived below in section \ref{sec:velomaps}. 


\subsection{Giants}

\begin{figure}
    \centering
    \includegraphics[width=0.49\textwidth]{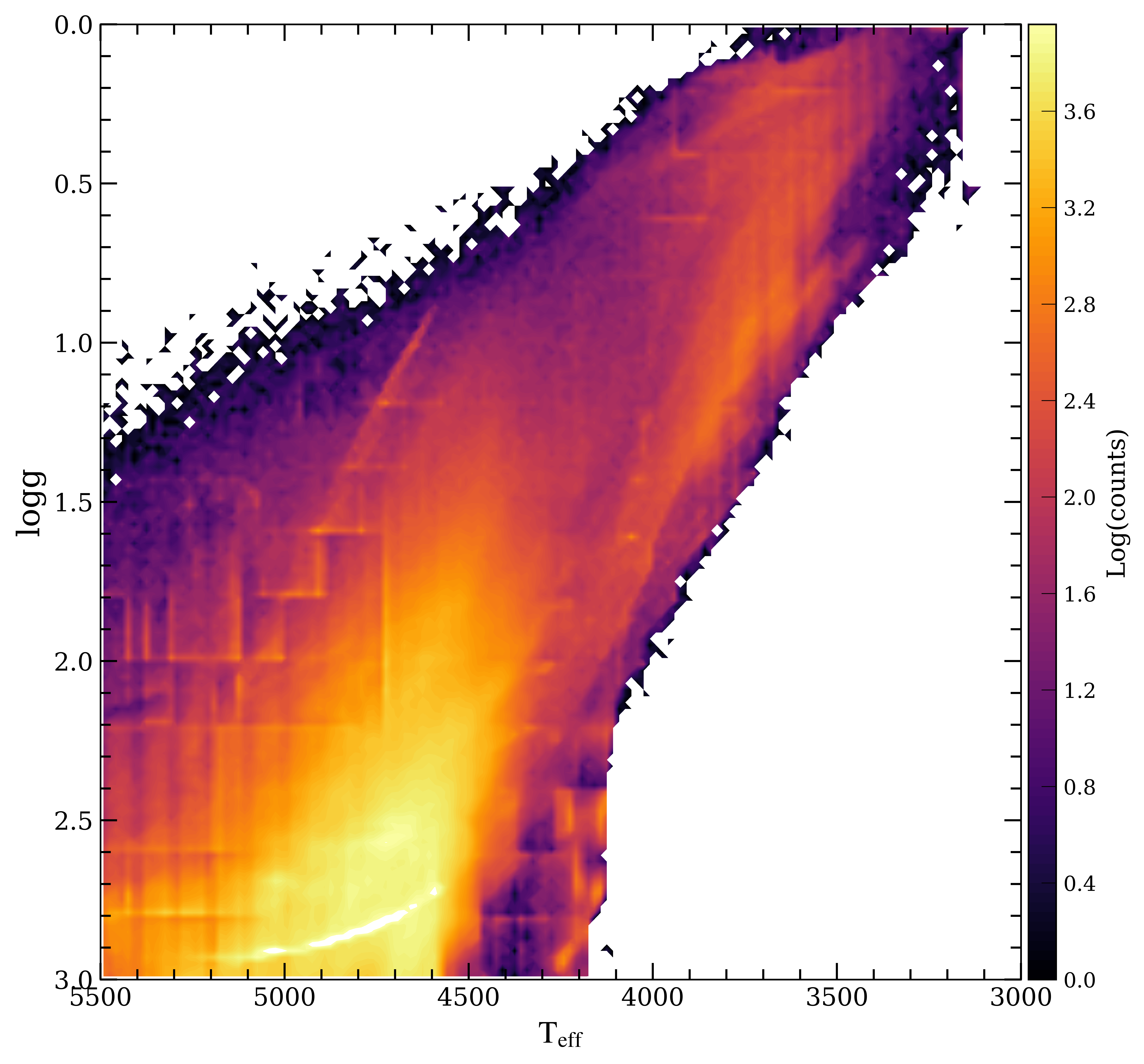}\\
    \caption{Kiel diagram for the selected RGB sample with astrometry. 
    }
    \label{fig:KielGiants}
\end{figure}

To select stars on the Red Giant Branch (RGB), we use the effective temperatures and surface gravities provided in DR3, selecting for stars with $3000< \teff < 5\,500$K and $\logg < 3.0$, as provided by \gspphot \citep{DR3-DPACP-156} in the main \gaia\ source table. (See appendix \ref{sec:queries} for an example of the query used.) These are given as the "best" set of parameters using a multi-spectral library approach, which for the RGB correspond to either the MARCS and PHOENIX libraries \citep{DR3-DPACP-160}. The Kiel diagram for these sources is shown in  Fig.~\ref{fig:KielGiants}. We refer to this set as the full RGB sample and it consists of $11\,576\,957$
sources. The magnitude $G$ distribution for the full RGB sample is shown in Fig.~\ref{fig:Gmag_ob_hist}, together with the RGB sample with RVS line-of-sight velocities.
 
\begin{figure*}
    \centering
    \includegraphics[width=\textwidth]{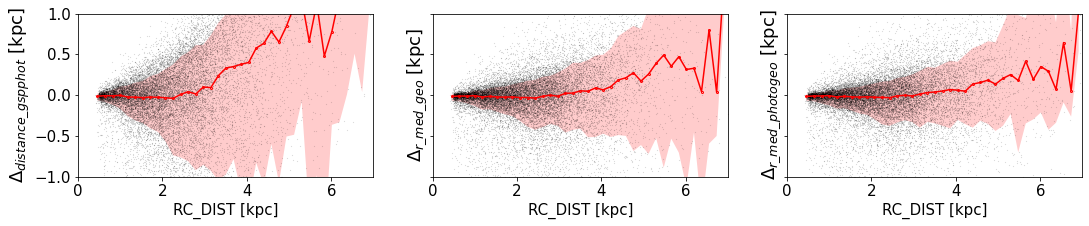}
    \caption{Absolute difference between the APOGEE Red Clump distance and three distance estimators considered in this work, \gspphot (left), "geo" (middle) and "photogeo" (right). }
    \label{fig:distanceComparisonRC}
\end{figure*}

As in the OB sample, and in order to perform density and velocity maps, we need to choose a distance estimator. The distribution of $\varpi/\sigma_{\varpi}$ for the RGB sample is very similar to that of the full OB sample shown in Fig.~\ref{fig:Gmag_plxerr_hist}, with about 39.5\% of the giant sample having a $\varpi/\sigma_{\varpi}<5$. 
All stars in the giant sample have \gspphot parameters, thus, one option can be to use the provided distance in \gdrthree\ \citet{DR3-DPACP-156}. Another option is the "geo" and "photogeo" distances from \citet[][hereafter CBJ2021]{CBJ2021}. In order to choose the appropriate distance estimator for the large extent of the RGB sample, we crossmatch the recent catalogue of red clump stars by APOGEE~DR17 \citep{Abdurrouf2021} with our RGB sample, using sky coordinates and a radius of $1$\, arcsec, resulting in a common sample of $18\,322$. By comparing the absolute difference between the reported photometric distance and the three different possibilities mentioned above of the stars in common, we observe that the distance estimator with less bias and dispersion is the "photogeo" distance by \citet{CBJ2021} (see Fig.~\ref{fig:distanceComparisonRC}). Also, and as shown in \citet{DR3-DPACP-127}, the large parallax uncertainties and the prior used makes the derived \gspphot distances to concentrate in density forming a ring around $2$\,kpc, making them inappropriate to study the inner disc. Therefore, and as in the OB sample, we adopted the "photogeo" as distance estimator for the RGB sample.

In addition, as in the selection of OB stars, we keep only sources with good astrometry, that is, with an astrometric fidelity $f_a> 0.5$. This criteria removes only $14\%$ of the sample, leaving us with a sample of $9\,959\,807$ stars, of which $6\,586\,329$
have line-of-sight velocities.

We found that the above selection also includes many sources from the Large and Small Magellanic Clouds, as well as a number of globular clusters. Since the goal of the paper is to study the Galactic disc, in the following we perform an extra cut on the altitude with respect to the Galactic plane, i.e. $|Z|<1$\,kpc, see Sect.~\ref{sec:confspace}. We have also explored whether this selection could include sources from the Sagittarius Dwarf galaxy, which could bias our kinematic study. We find no evidence of Sagittarius Dwarf sources in the proper motion map, and while red clump stars in the dwarf galaxy have a magnitude range of $17-18$ mag \citep[e.g.,][]{Antoja2020}, which falls in the faint end of our sample, so will be a negligible fraction in our full sample, and are not expected in the sample with RVS line-of-sight velocities, so we do not make any attempt for removing them.

The final RGB sample consists then of $8\,727\,344$ sources, with 
$5\,730\,578$ with RVS line-of-sight velocities.

In Table~\ref{table:Numbers} we summarise the number of stars in each of the selected tracers used throughout the paper. We provide both the number of sources with full 5D astrometry and the subsample including the 6D (astrometry and line-of-sight velocities).

\begin{table}
\caption{Number of stars according to the selected tracer.}             
\label{table:Numbers}      
\centering                          
\small
\begin{tabular}{l l l}        
\hline\hline                 
Sample & Astrometry & 6D \\    
\hline                        
Clusters (all ages) &  2681 & 2162\\
Young clusters (age < 100 Myr)  & 988  & 698\\
\hline
Classical Cepheids (DCEP)  & 3312 &  2127 \\
Classical Cepheids (age < 200 Myr)  &   2812 &  1949\\
\hline
Young field stars (OB) & 579577  & 77659\\
\hline
Giant field stars (RGB) & 8727344 & 5730578 \\
\hline                                   
\end{tabular}
\end{table}

\section{From observations to 6D phase space}
\label{sec:three}

\subsection{Mapping to configuration space}
\label{sec:confspace}

To map our tracers in a 3D Cartesian coordinate space we must transform astrometric angular measurements $(\varpi, \alpha, \delta)$ to associated lengths. While each of these measurements have associated uncertainties, we are here concerned with objects reaching large distances (i.e. small parallaxes).  In this case our positional uncertainties are completely dominated by the uncertainties in parallax (whose associated positional uncertainty is proportional to the square of the distance), and we can safely ignore the uncertainties (and correlations) with the angular positional measurements. Our problem is now just reduced to determining the heliocentric distance to each tracer and its uncertainty. In the previous section the distance estimate adopted for each tracer population is described: for the OB and RGB samples we adopt the "photogeo" distances of CBJ2021, for the clusters the inverse of their median parallax, and for the Cepheids a photometric based distance.



\begin{figure}
    \centering
    \includegraphics[width=0.49\textwidth]{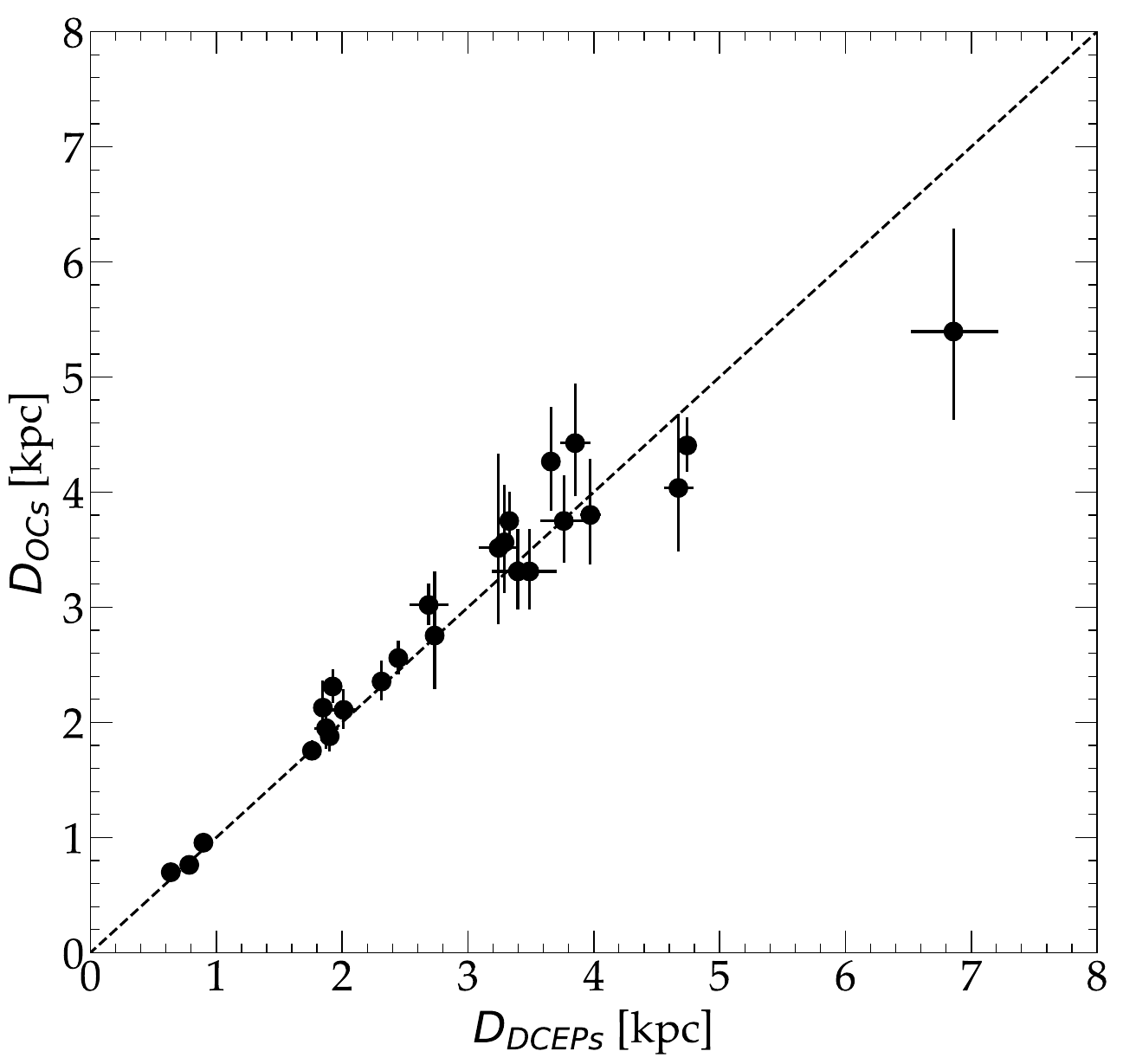}
    \caption{Comparison of the distances calculated for DCEPs and their host OCs.  
    }
    \label{fig:distanceComparisonDCEPs_OCs}
\end{figure}

It has been known for a long time that some MW DCEPs are members of OCs \citep[see e.g.][and references therein]{Anderson2013}. As in this paper we use both DCEPs and OCs to study the spiral arm properties of the MW, it is useful to compare the distances inferred for DCEPs as explained in Sect.~\ref{sect:dceps} with those of their host OCs, which are based on the median parallax of the OC members. To select possible DCEPs belonging to OCs we cross-matched the DCEPs list with that of all the known OC members. The cross-match was carried out using {\it Gaia} identifiers and returned 25 matches. The distance comparison is shown in Fig.~\ref{fig:distanceComparisonDCEPs_OCs}. The overall agreement is very good, approximately below 1$\sigma$ in all the cases, except for the farthest OC of the sample, namely UBC\,608 for which the discrepancy is however less than 1.5$\sigma$. 

Now that we have heliocentric distances $d$ we can easily derive heliocentric Cartesian coordinates $(x,y,z)$ under the assumption of being in a (non-relativistic) 3D Euclidean geometry. Since we are assuming the positions as known, we can take advantage of the provision of $(l,b)$ of each tracer in the \gaia\ Archive and use the usual transformations
\begin{equation}
 \begin{pmatrix}
x \\ 
y \\ 
z
\end{pmatrix}  = 
d \begin{pmatrix}
\cos b \cos l    \\ 
\cos b \sin l     \\ 
\sin b 
\end{pmatrix},
\label{Hcoord} 
\end{equation}
where positive $x$ is toward the Galactic centre. galactocentric Cartesian coordinates are then typically derived as a simple translation of the origin:
\begin{equation}
 \begin{pmatrix}
X \\ 
Y \\ 
Z
\end{pmatrix}  = 
\begin{pmatrix}
x - R_\odot   \\ 
y     \\ 
z + Z_\odot
\end{pmatrix} 
\label{HtoG}
\end{equation} where $R_\odot$ and $Z_\odot$ are the distance of the Sun from the Galactic centre and above the Galactic ($Z=0$) midplane. 

For our choice of $R_\odot$ we use the geometrical determination based on the line-of-sight velocity and relative astrometry of the resolved SagA*-S2 binary, as measured by the latest contribution of the GRAVITY Collaboration \citep{GRAVITY2021}, namely $R_\odot = 8277 \pm 9$(stat) $\pm 30$(sys) pc, though we note that this value does not agree within the uncertainties with the independent determination of $R_\odot$ by \citet{Do2019} (7.959 +/- 59 (stat) +/- 32 (sys)), nor with their previous determinations, indicating that our assumed $R_\odot$ {\em may} in fact be in error by as much as 200 to 300 parsecs. Our adopted value of $R_\odot$ assumes of course that the position of SagA* marks the Galactic centre, which is expected from dynamical considerations: A supermassive black hole not already at the centre of a large stellar system will eventually migrate to the centre due to dynamical friction \citep{GM2008}. Most recently \citet{Leung2022} have independently determined $R_\odot = 8.23 \pm 0.12$kpc based on observed stellar kinematics toward the Galactic centre, a value that is consistent with our assumed value, but with a more realistic estimate of its uncertainty.
For mapping in configuration space, any systematic error in $R_\odot$ only results in a trivial offset in the maps. However, as discussed further in section \ref{sec:sys_errors}, it can introduce rather undesirable effects when mapping the velocities in galactocentric cylindrical coordinates.

It should be noted that the transformation to galactocentric coordinates (Eq. \ref{HtoG}) is an approximation, as it assumes that the $b = 0$ and $Z=0$ plane are parallel to each other. While this was the original intent when the galactic coordinate system was defined, 
\citep{Gum1960, Blaauw1960}, there may well be a residual offset due to the Sun's height above the $Z=0$ plane. Indeed, it was already noted at the time that determinations of $Z_\odot$ from Hydrogen radio emission do not coincide with those based on nearby stellar samples.
In their pre-\gaia\ review of our knowledge of the Milky Way, \citet{JnO2016} found that estimates for the distance of the Sun above the midplane fall between 20 and 30 pc, while more recent estimates have generally been smaller \citep{Yao2017,Widmark2019,Anderson2019, Reid2019}. 
However, the evidence of vertical oscillations in the disc of the Milky Way \citep{Bennet2019} and evidence of its disequilibrium state \citep{Antoja2018} rather complicates this discussion, as the local stellar mid-plane of the Galaxy could very well not coincide with a $Z=0$ plane as defined by the average vertical density distribution of the inner disc. Indeed, different vertical modes may be present in the gas (and star formation tracers) and in the stars, or even between different stellar populations, explaining some of the observed variance between the different determinations of $Z_\odot$. As a case in point, using a very local sample of stars, \citet{GaiaEDR3_Smart_100pc} find that $Z_\odot$ varies from -4pc to 15pc for young to older stellar populations. The observed small negative offset of SagA* from $b=0$ also suggests that there is a residual tilt of $\delta \theta$ between the galactic ($b=0$) plane and the Galactic ($Z=0$) plane of the order of only $0.1 ^\circ$ \citep[see also the discussion in][]{JnO2016}.
However, since for mapping the large scale asymmetries in the disc we are primarily concerned with the positions and velocities of our tracers in the $(X,Y)$ plane, where the effect of this tilt is negligible, we conveniently assume $Z_\odot = 0$, so that $Z = z$.

\subsection{mapping to velocity space}

\begin{figure}
    \centering
    \includegraphics[width=0.49\textwidth]{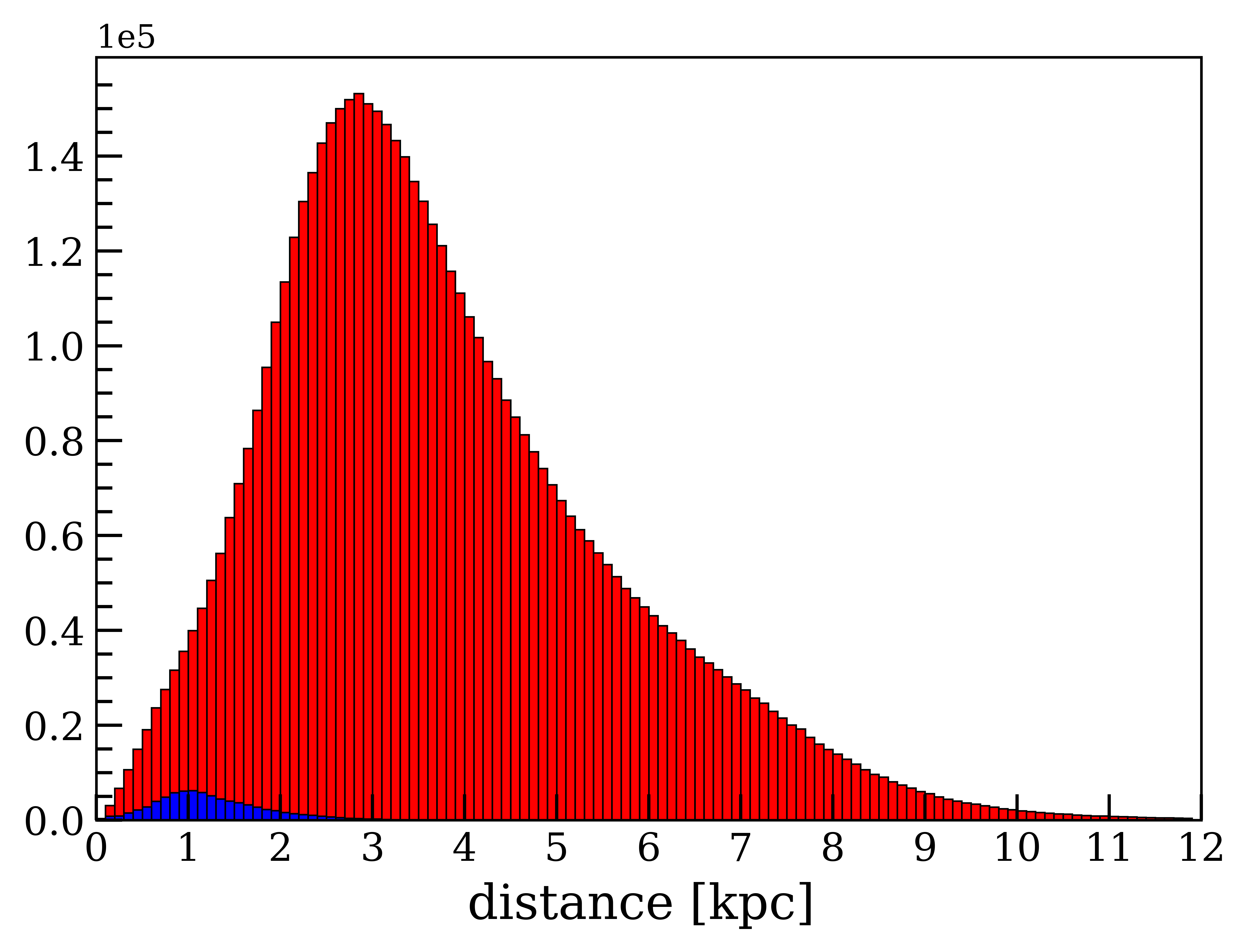}
    \caption{Distribution of the heliocentric distances from  \citet{CBJ2021}, of the RGB (red) and OB stars (blue) with line-of-sight velocities. 
    }
    \label{fig:distance_hist_RVsamples}
\end{figure}

To map the velocities we must now use the spectroscopically measured line-of-sight velocities $v_r$ together with the measured proper motions and the distance estimator described in the previous section
. As mentioned above, the OB and RGB samples with measured line-of-sight velocities are 77\,659 and 5\,730\,578 in number, respectively. In addition, the brighter magnitude limit of the OB sample with line-of-sight velocities also means that the area that can be mapped by the OB stars is much smaller than that of the RGB sample. (See figure \ref{fig:distance_hist_RVsamples}). 

The relative velocity components in the heliocentric Cartesian coordinates defined by Eq. \ref{Hcoord} are 
\begin{equation}
{\bf{v}}_{rel} =  
 \begin{pmatrix}
u\\ 
v\\ 
w
\end{pmatrix} = A^\prime_G\, A\, \begin{pmatrix}
4.74047~ {\mu_{\alpha^*}}~ d\\ 
4.74047~ {\mu_{\delta}}~ d\\
v_r 
\end{pmatrix},
\label{relVfromPar}
\end{equation}
where $(\mu_{\alpha^*},\mu_{\delta}$) are the proper motion components, $v_r$ the line-of-sight (i.e. radial) velocity, $A^\prime_G$ is the transformation matrix from equatorial to galactic coordinates, as given by equation 4.62 of the \gaia\ EDR3 \href{https://gea.esac.esa.int/archive/documentation/GEDR3/Data_processing/chap_cu3ast/sec_cu3ast_intro/ssec_cu3ast_intro_tansforms.html#SSS1}{online documentation},
and the matrix $A$ is the normal triad at the star\footnote{Alternatively one could first convert the astrometry to galactic coordinates, removing $A'$ from Eq.~\ref{relVfromPar} and substituting the proper motion components in ICRS with those in Galactic coordinates, and then computing $\bf{v}_{rel}$ from the Galactic coordinates. In this case the triad (Eq.~\ref{Arot}) would be in $(l,b)$ rather than $(\alpha, \delta)$.}:
\begin{equation}
A = \begin{pmatrix}
-\sin \alpha & -\sin \delta \cos \alpha  & \cos \delta \cos \alpha \\
 \cos \alpha & -\sin \delta \sin \alpha  & \cos \delta \sin \alpha\\
     0       &        \cos \delta        &          \sin \delta\\ 
\end{pmatrix}
\label{Arot}
\end{equation}

We note that for the line-of-sight velocities $v_r$
of the OB stars, we applied the following correction
\cite[as prescribed in][]{DR3-DPACP-151}:
    \begin{equation}
        v_{\mathrm los}  = \linktoparam{gaia_source}{radial\_velocity}
        - 7.98 + 1.135 \linktoparam{gaia_source}{grvs_mag}.
    \end{equation}
This is applied to stars where $8500 \le \linktoparam{gaia_source}{rv_template_teff} \le 14\,500$~K
and $6 \le \linktoparam{gaia_source}{grvs_mag} \le 12$.
 
The velocities ${\bf{v}}_{rel} = (u,v,w)$ derived above are relative to the Sun. To put them in a galactocentric reference frame we must add the Sun's velocity with respect to the galactic centre, $\bf{v}_\odot$: 
\begin{equation}\label{eq:galcentricVelCart}
{\bf{v_*}} = {\bf{v}}_{rel} + {\bf{v}}_\odot \, .
\end{equation}
Traditionally the Sun's velocity $\bf{v}_\odot$ has been estimated from the solar motion with respect to a local standard of rest (LSR) and an adopted value of the velocity of the LSR, usually assumed to be in circular motion about the galactic centre.  However, thanks to the recent precise measurement of the proper motion of the SagA*, together with $R_\odot$, the azimuthal components of the Sun's galactocentric velocity can be derived in a more direct and precise way. From \citet{Reid2020} we have $(\mu_l, \mu_b) = (-6.411 \pm 0.008, -0.219 \pm 0.007)$ \masyr for the proper motion of SagA*, which together with $R_\odot$ gives the Sun's Y and Z-velocity components. Meanwhile, the same reduction of the SagA*-S2 data by the GRAVITY Collaboration that yielded $R_\odot$ also yields the line-of-sight velocity toward SagA*, interpreted as the reflex motion of the Sun's velocity toward the Galactic centre. This gives us:
\begin{equation}
{\bf{v}}_\odot =
\begin{pmatrix}
9.3 \pm 1.3  \\
251.5 \pm 1.0  \\
8.59 \pm 0.28 
\end{pmatrix} \kms
\end{equation}
if we assume that SagA* is stationary with respect to the galactic centre. \citep[see][for further discussion on this approach to deriving $\bf{v}_\odot$]{Drimmel_2018}. The uncertainties in ${\bf{v}}_\odot$, as well as any error in our adopted $R_\odot$, gives us a systematic error common to all our galactocentric velocities ${\bf{v_*}}$. See section \ref{sec:sys_errors} below for further discussion.

One should note that the $(u,v,w)$ components of our galactocentric velocities ${\bf{v_*}}$ are rigorously in the same coordinate system defined by Eq. \ref{Hcoord}, so slightly tilted with respect to a $(U,V,W)$ coordinate system whose $U$-$V$ plane is parallel to the mean $Z=0$ plane of the Galaxy by the angle $\delta \theta$ mentioned above, if $Z_\odot \neq 0$. However, the systematic error introduced by ignoring this (unknown) tilt is much smaller than the systematic error introduced by the uncertainties in ${\bf{v}}_\odot$. 

Assuming $Z_\odot=0$, the galactocentric radial and azimuthal velocities can be found from 
\begin{eqnarray}\label{eq:cylindricalVel}
      v_R         & = & -u \cos \phi + v \sin \phi \nonumber  \\
      v_\phi   & = & u \sin \phi + v \cos \phi     \\ 
      v_z         & = & w \,  ,  \nonumber
\end{eqnarray}
where $\phi$ is the galactocentric azimuth, taken as positive in the direction of galactic rotation:
\begin{equation}\label{eq:phi}
\phi  = \arctan \left( \frac{Y}{-X} \right) = \arctan \left( \frac{d \cos b \sin l}{R_\odot - d \cos b \cos l} \right)\, ,
\end{equation}
\noindent making our $(R, \phi, z)$ galactocentric cylindrical coordinates a left-handed system.
 
\subsection{Propagation of uncertainties}\label{sec:uncertainties}


\begin{figure}
    \centering
    \includegraphics[width=0.49\textwidth]{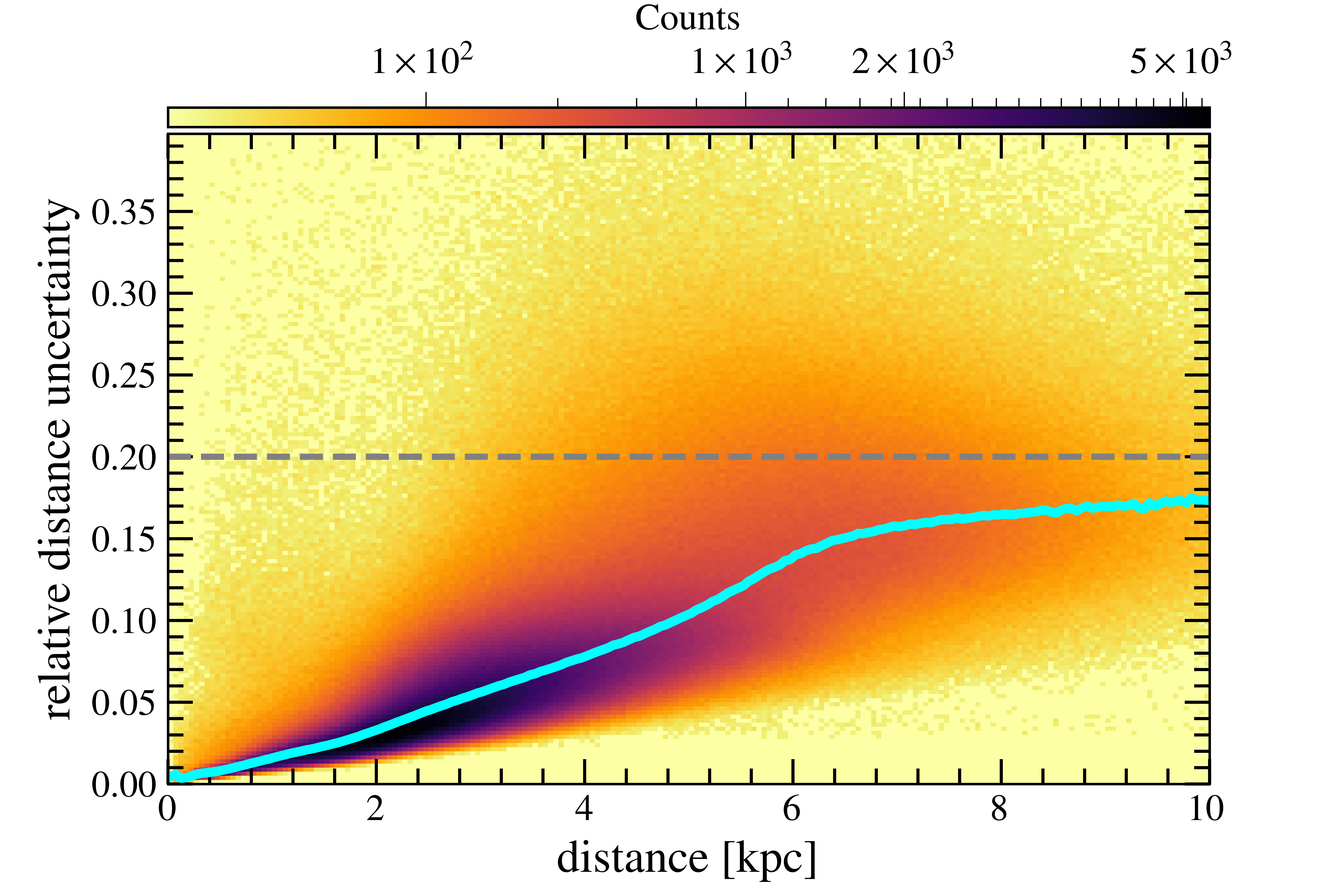}
    \caption{The relative distance uncertainties of the individual sources, as a function of distance, for the RGB sample with radial velocities. The blue curve shows the median in bins of 50~pc. 
    }
    \label{fig:distance_err_vs_d}
\end{figure}

To estimate the uncertainties in positions and velocities from the formal errors in astrometry, we choose to concatenate the Jacobian matrices of the consecutive transformations necessary to move from the initial reference frame to the desired one. In doing so, we are implicitly linearizing the functions that allow us to convert astrometry into positions and velocities. In other words, we are simplifying the sequence of non-linear transformations (e.g., see eqs.~\ref{relVfromPar},\ref{eq:cylindricalVel}, and \ref{eq:phi}) that convert the coordinates in one reference frame, $\vec{x}_1$, to the coordinates in another frame, $\vec{x}_2$, with a single matrix product of the form
\begin{equation}\label{eq:linear_transform}
    \vec{x}_2 = J\vec{x}_1,
\end{equation}
\noindent by taking only the linear term of the Taylor expansion of the transforming functions. In this case, $J$ corresponds to the Jacobian matrix of the functions used to transform from one frame to the other:
\begin{equation}\label{eq:jacob2}
    J_{i,j} = \frac{\partial f_{i}(\vec{x})}{\partial x_j},
\end{equation}
\noindent where $f_{i}$ is the function that calculates the $i^{\mathrm th}$ component of the vector $\vec{x}_2$ in the desired coordinate frame given the vector $\vec{x}_1$ in the original coordinate frame. 

In practice, we do the following: first, we construct the covariance matrix in the frame of the \gaia\ astrometry using the proper motion and line-of-sight velocity uncertainties, and their associated correlations. As said above, we neglect the uncertainties in sky position and consequently all the terms involved are set to zero. At the same time, we also set to zero the uncertainty in parallax and its correlations, replacing it with the uncertainty in our distance estimate, $\sigma_d$ (discussed below).  The resulting initial covariance matrix for any source is, thus,

\begin{equation}\label{eq:cov}
    \Sigma_{ICRS} = 
\begin{pmatrix}
0 & 0 & 0 & 0 & 0 & 0 \\
0 & 0 & 0 & 0 & 0 & 0 \\
0 & 0 & \sigma_d^2 & 0 & 0 & 0 \\
0 & 0 & 0 & \sigma_{\mu_\alpha^*}^2 & \rho_{\mu_\alpha^* \mu_\delta} \sigma_{\mu_\alpha^*} \sigma_{\mu_\delta}  & 0 \\
0 & 0 & 0 & \rho_{\mu_\alpha^* \mu_\delta} \sigma_{\mu_\alpha^*} \sigma_{\mu_\delta} & \sigma_{\mu_\delta}^2 & 0 \\
0 & 0 & 0 & 0 & 0 & \sigma_{v_{los}}^2 \\
\end{pmatrix},
\end{equation}

\noindent where $\rho_{\mu_\alpha^* \mu_\delta}$ is the correlation coefficient between the proper motion components, as given in the \gaia\ source table.

Then, with the Jacobian calculated as in Eq.~\ref{eq:jacob2}, we obtain the covariance matrix in heliocentric Cartesian coordinates (eq.~\ref{HtoG} and eq.~\ref{relVfromPar}):
\begin{equation}\label{eq:jacob1}
    \Sigma_{2} = J\Sigma_{1}J^T\,,
\end{equation}
After that, we use again equations \ref{eq:jacob1} and \ref{eq:jacob2} to move, successively, first into galactocentric Cartesian (eq.~\ref{HtoG} and eq.~\ref{eq:galcentricVelCart}) and, finally, to galactocentric cylindrical (eqs.~\ref{eq:cylindricalVel} and \ref{eq:phi}). At each step, we obtain a full 6x6 covariance matrix that encodes not only the estimated uncertainty in each quantity along its diagonal, but also their correlations. The transformation itself contributes significantly to some of these correlations like, for instance, the correlation between the $u$ and $v$ components of the velocity that arises naturally from using the distance in eq.~\ref{relVfromPar}. Nonetheless, the correlation between \gaia\ measurements are also a source of correlations regardless of the coordinate frames. 
However, it is the correlations introduced by the coordinate transformations, in concert with the distance uncertainties, that dominate the final correlations between the velocity components, and these are highly direction dependent. (See also the following section for further discussion.) 

Alternatively, we could have chosen to estimate the uncertainties by randomly sampling new "fake" observables based on the covariance matrix in the initial \gaia\ frame. However, there are a couple shortcomings in this approach: i) it requires a large amount of samples to obtain a robust estimation of the uncertainties, which can be computationally expensive, ii) we do not know the true error distribution function since the formal errors are estimated from the observables themselves, and iii) we do not have access to the full posterior distribution function for the distance estimators. While point (i) can be dealt with some patience, points (ii) and (iii) force us to make similar assumptions to those made for the strategy described above, thus limiting the usefulness of this alternative approach.

As a consequence of our decision to use the Jacobians to propagate the uncertainties, we are implicitly assuming that the uncertainties of the measured (input) quantities are symmetric. For the proper motions and line-of-sight velocities this is satisfied, their errors being well described by Gaussian distributions \citep{EDR3-DPACP-133}. However, as described in CBJ2021, the probability distribution functions of the distances are typically skewed, and this is also true of photometric distances, such as those used for the Cepheids. For the CBJ2021 distances, we render the  distance uncertainties symmetric by taking the mean of the distances of the provided 16th and 84th percentiles from the median distances, that is, we take as the individual distance uncertainty:
 \begin{equation}\label{eq:distance_error}
        \sigma_d  = (\mathtt{r\_hi\_photogeo} - \mathtt{r\_lo\_photogeo})/2
\end{equation}
where $\mathtt{r\_hi\_photogeo}$ and $\mathtt{r\_lo\_photogeo}$ 
 are respectively the 84th and 16th percentiles
of the ``"photo-geometric'' distances provided by CBJ2021. Figure \ref{fig:distance_err_vs_d} shows how the distance uncertainties in our sample varies with distance for the subset of RGB stars with radial velocities. We note 
that the median relative uncertainties are below 20\%. The distance uncertainties for the OB sample with line-of-sight velocities follows the same trend, though covering a much smaller range of distances. 

Using this distance estimator comes at the cost of loosing the correlations between the distances and the proper motions. Such correlations are not expected in purely photometric based distances, as is the case for the Cepheids, but will be present in as much as the distance is informed by the parallax for the other distance estimates. For the CBJ2021 distances, used for the OB and RGB samples, it will be the relatively nearby sources, which generally have small relative distance uncertainties, that will be constrained by the parallaxes, while the more distant sources with larger uncertainties will be primarily constrained by the photometry and the CMD prior, so will be only weakly correlated with the proper motions. Therefore, to propagate the uncertainties on the velocities, we only concern ourselves with the correlations between the proper motion components in the initial covariance matrix (eq. \ref{eq:cov}).





\begin{figure}
    \centering
    \includegraphics[width=0.49\textwidth]{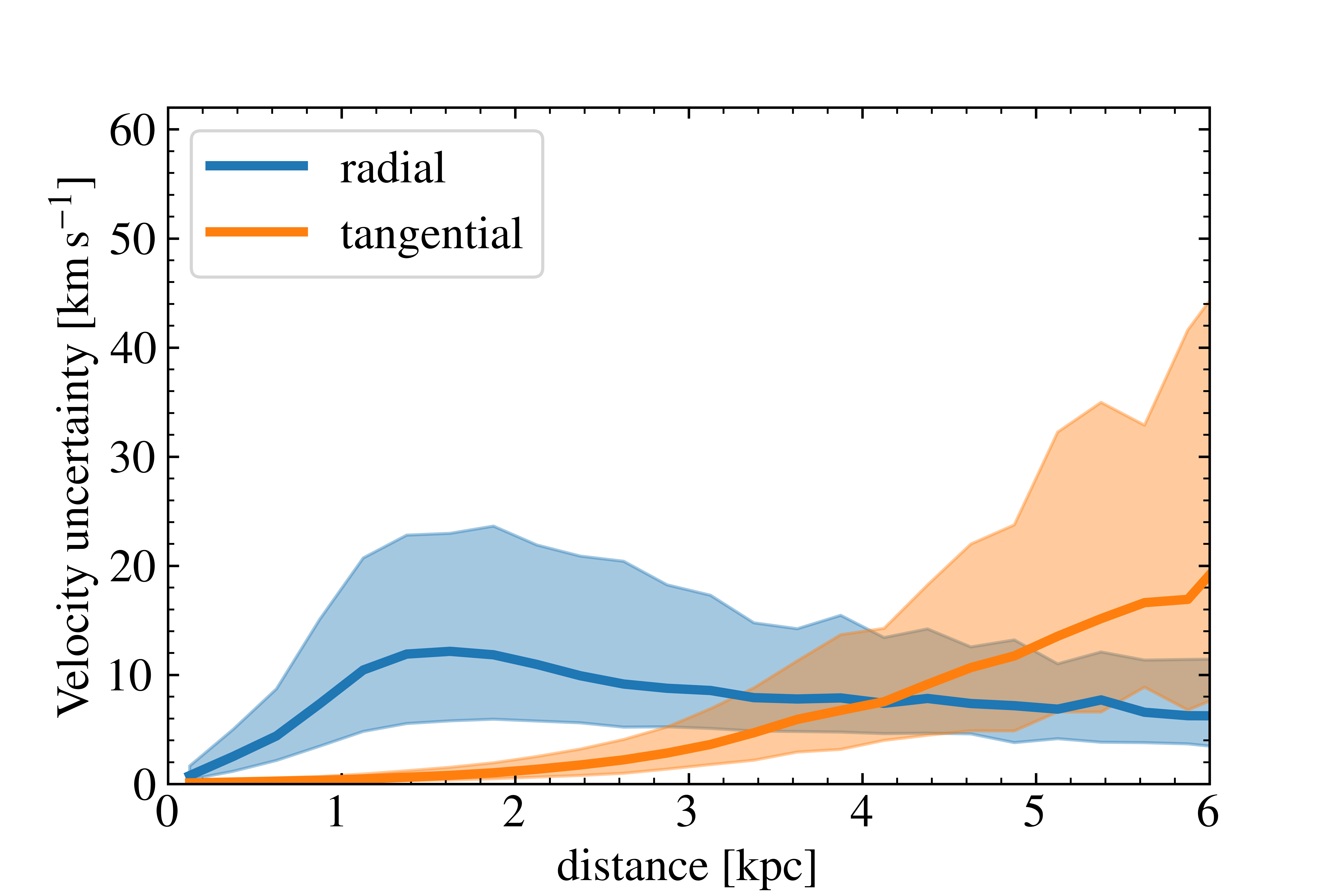}
    \includegraphics[width=0.49\textwidth]{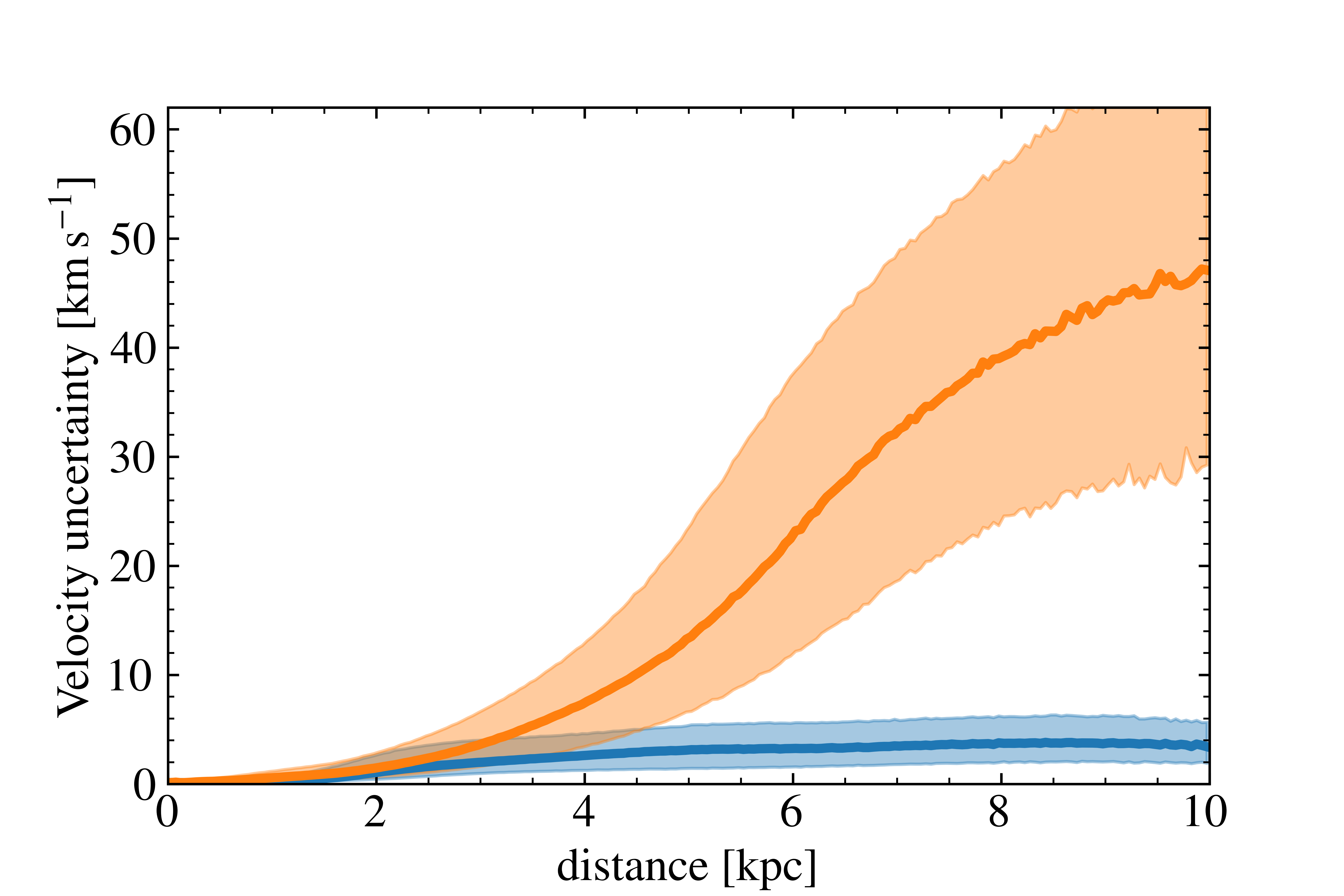}
    \caption{The median radial (blue) and tangential (orange) velocity uncertainties of the individual sources, as a function of distances. The shaded areas show the range between the 16th and 84th percentiles. The upper plot is for the OB stars, while the lower plot is for the RGB stars, for the samples with full velocity information. 
    }
    \label{fig:velocity_err_vs_d}
\end{figure}
 
We note that, since the tangential (perpendicular to the line-of-sight) velocity components are dependent on both distance and proper motion, their uncertainties will be perfectly correlated with the distance uncertainties. As the line-of-sight velocity does not suffer from correlations with the astrometry, and may be of a different magnitude with respect to the uncertainty in the tangential velocity, we should expect that the correlations and uncertainties of our velocity components will have a strong directional dependence that can potentially introduce false signals or patterns in our maps. 
Figure \ref{fig:velocity_err_vs_d} shows the median and range of the velocity uncertainties for the OB and RGB samples as a function of distance. We note that the velocity uncertainty of the OB stars with full velocity information, being limited to within a few kiloparsecs of the Sun, will be dominated by the uncertainty in the line-of-sight velocities. On the other hand, the uncertainties for both velocity components of the RGB stars are quite comparable to about 2~kpc, beyond which the tangential component then dominates the uncertainty.

Finally, we also note that our distance uncertainties will again introduce further direction dependencies in the errors and correlations in the final transformation to velocities in galactocentric cylindrical coordinate, via equations \ref{eq:cylindricalVel} and \ref{eq:phi}. Indeed, errors in $\phi$ can become quite large near the galactic centre. Also, by performing the coordinate transformation to ($v_R,v_\phi$) above, we have made ourselves vulnerable to systematic error in our assumed values of $R_\odot$ and ${\bf{v}}_\odot$. 
We explore the effect of these kinds of uncertainties in the following section.

\subsection{Effect of systematic errors}
\label{sec:sys_errors}


\begin{figure*}
\centering
\includegraphics[height=7.2cm]{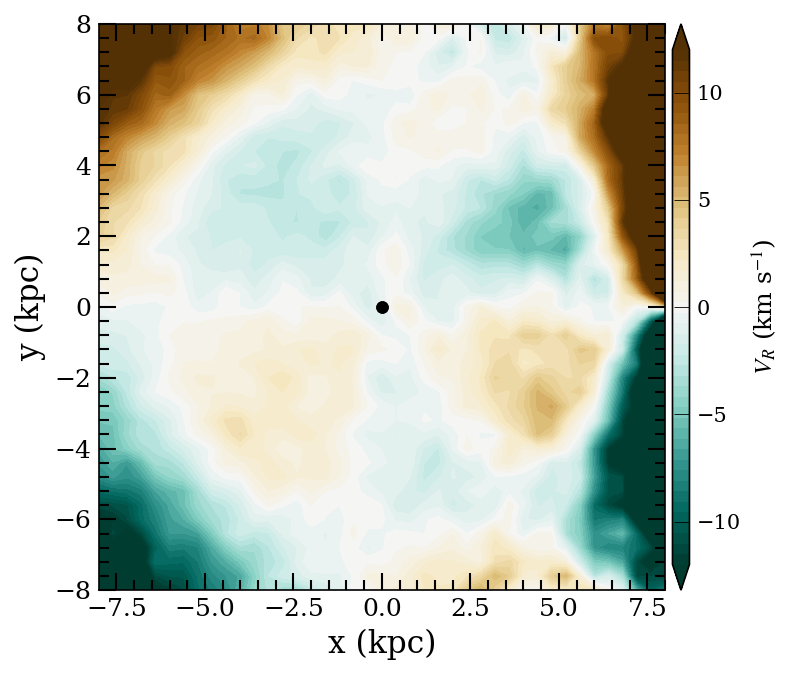}\includegraphics[height=7.2cm]{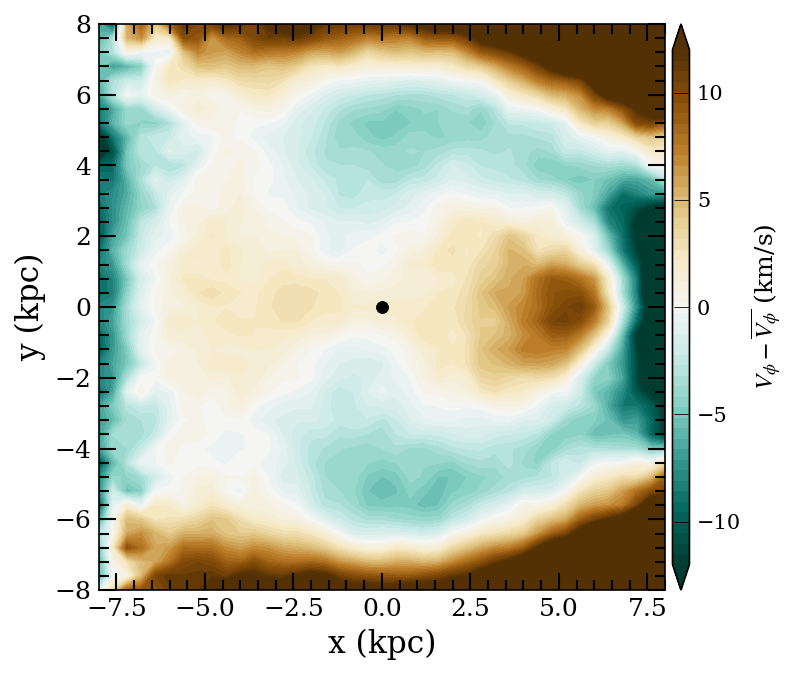}
\includegraphics[height=7.2cm]{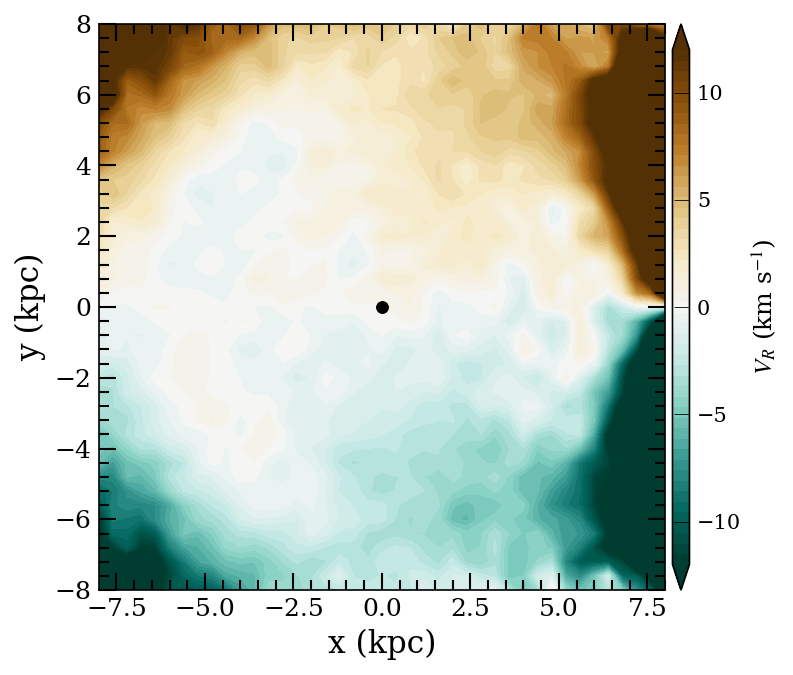}\includegraphics[height=7.2cm]{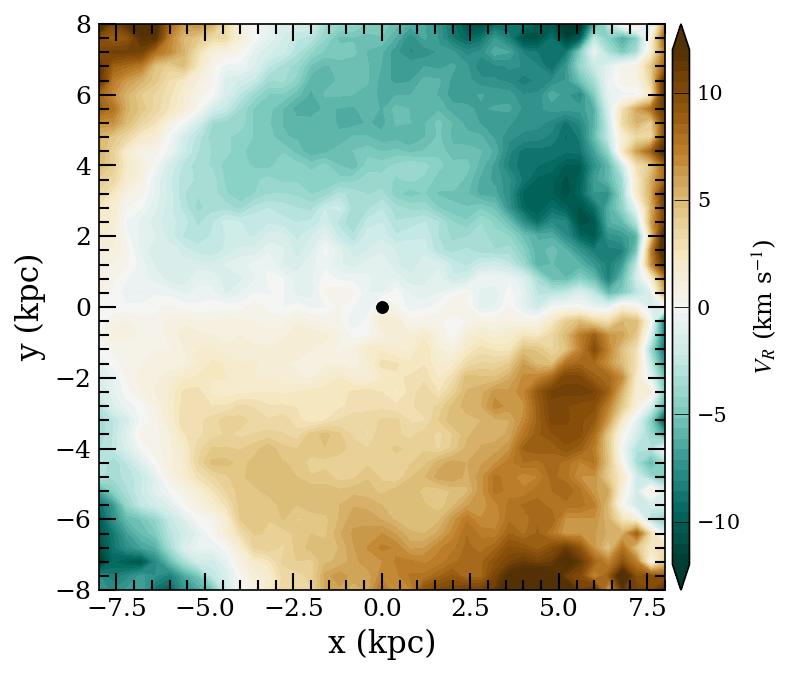}
\caption{Systematic velocity trends induced by the simulated uncertainties in our mock catalogue.
\emph{Upper left panel:} map of the observed $V_R$ assuming a relative parallax uncertainty = 0.2. \emph{Upper right panel:} same as left panel, but showing the observed $\Delta V_{\phi}$.
\emph{Lower left panel:} the observed $V_R$ assuming that the true value of $R_{\odot}$ is 200 pc larger than the adopted value, together with the distance uncertainty model explained in the text. \emph{Lower right panel:} the observed $V_R$ assuming that the true value of the solar velocity component $V_{\phi,\odot}$ is 10 km/s larger than the adopted value, together with the distance uncertainty model explained in the text.
\label{fig:impact_errors}
}
\end{figure*}
   
As mentioned above, while the formal errors are quite small there may be significant systematic errors in our assumed values for $R_\odot$ and ${\bf{v}}_\odot$. In addition, as we will demonstrate below, even random distance errors can introduce systematic errors in the mean galactocentric velocity components. To investigate the possible effects that such errors can introduce, we construct a mock catalogue from a rather artificial distribution of stars: We uniformly populate a disk of stars centred on the Sun's position with a radius of 8 kpc, add a Gaussian velocity dispersion and an azimuthal velocity with respect to the galactic centre at $R_{GC}$ that is consistent with what we observe for the RGB sample (see section 5), but assuming no mean radial motion (i.e. $v_R=0$). We also assume a motion of the observer at the Sun and from the relative motions derive the proper motion and line-of-sight velocities, to which we add fractional uncertainties of 0.01 in the proper motions and 0.1 in the radial velocities.  We then re-derive the ($v_R,v_\phi$) velocity components from the observed (noisy) proper motions, distances and line-of-sight velocities, and construct maps of the observed mean velocity field, as described in the section 5. 

To model the effect of distance uncertainties we add a 20\% gaussian uncertainty to the true parallax, then take the inverse of the observed parallax as our distant estimate. We note that this is much larger than the actual uncertainties of our dataset, which only reach a level of 20\% relative uncertainty in the distances at about 10 kpc from the Sun, but is used here simply for the purpose of illustration. However, like the actual distance errors, the probability distribution function of the distance from our inverse-parallax distance estimate is skewed toward larger distances. More importantly, for our rather artificial uniform disc of mock stars, the mean estimated distance will be slightly systematically underestimated for distances less than 8kpc, but strongly overestimated for distances beyond 8 kpc. These mean biases in the distances of our sample introduce systematic motions in the inferred mean velocity field ($V_R,V_\phi$), as shown in the upper two panels of figure \ref{fig:impact_errors}. In the following, we will refer to those biases as 'systematic errors', in the sense that the simulated uncertainties in our experiment systematically induce artefacts in the observed trends, as shown below.

Assuming a more realistic model for our parallax uncertainties of $\sigma_\varpi/\varpi = 0.02 d_{\mathrm kpc}$ (i.e. $\sigma_\varpi = 20\mu{\mathrm as}$), we investigate the effect of assuming erroneous values for $R_\odot$ and the velocity of the Sun with respect the values used to generate the mock observations. We first introduce a systematic error in our assumed value for $R_\odot$, taken to be 200 pc larger than the distance to the Galactic centre used to assign the mean rotational velocities to the mock stars, when we transform the observed proper motions, distances and line-of-sight velocities to ($v_R,v_\phi$) velocity components. The resulting inferred $V_R$ velocity field is shown in the lower left panel of figure \ref{fig:impact_errors}.  Similarly, assuming a correct value for $R_\odot$ consistent with the mock velocities, but assuming an incorrect velocity for the Sun's $v_\phi$ component, we see a similar but inverted pattern in the inferred $V_R$ velocities, with the Sun lying on an axis of symmetry of the inferred $V_R$ velocity field. (Lower right panel of figure \ref{fig:impact_errors}.)

Another important feature in the velocity maps to be noted are that the velocities beyond the actual distance limit of our mock sample (8\,kpc) are very strongly biased. This is an unavoidable feature of any magnitude limited sample, where the real density of the sources will sharply decrease beyond some limiting distance. Beyond this distance the sources used to estimate velocities will have systematically over-estimated distances, and consequently overestimated velocities (in modulus) perpendicular to the line-of-sight.  This limiting distance, beyond which the inferred velocities cannot be trusted, can vary significantly for different lines-of-sight, thanks to the effects of interstellar extinction.  In any case, {\em one should resist giving astrophysical significance to features at the edge of velocity maps.} 

While our assumed systematic errors in the above discussion may be larger than we expect, the purpose of this discussion is to warn anyone interpreting features or patterns seen in velocity maps in galactocentric cylindrical coordinates. As in coordinate space, {\em one should be most suspicious of any patterns in the kinematics that show any symmetry with respect to the Sun's position.}

\section{Coordinate maps}\label{sec:coord_maps}


\begin{figure*}[th]
    \centering
     \includegraphics[height=7.8cm,trim={0 0 2cm 0},clip]{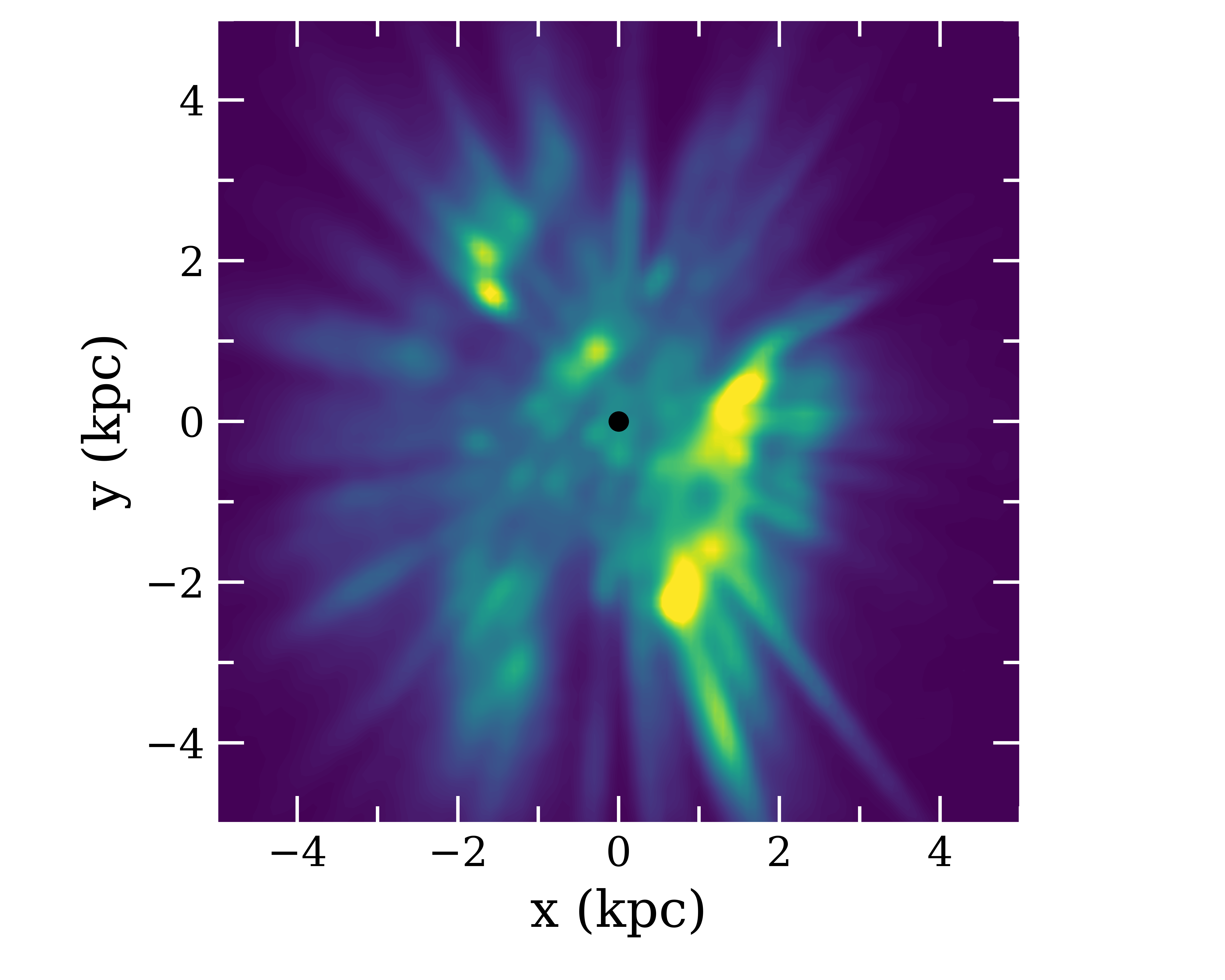}
    \includegraphics[height=7.8cm,trim={0 0 2cm 0.2cm},clip]{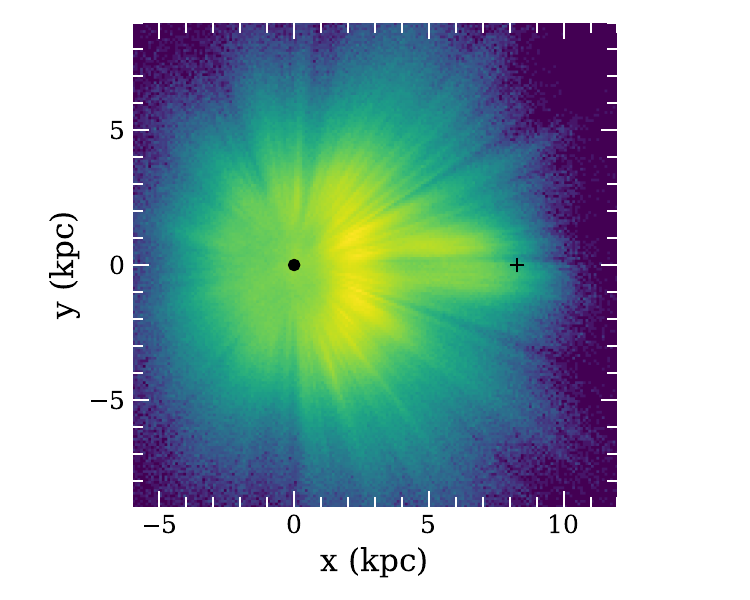}  
    \caption{Spatial distribution in the galactic plane, in heliocentric coordinates, of the selected OB stars (left panel) and RGB stars with line-of-sight velocities (right panel). For the OB stars, we perform a bivariate kernel density estimation using an Epanechnikov kernel with a smoothing length of 200 pc. For the RGB stars, we construct a 2-dimensional histogram, using cells of 100x100 pc. The galactic centre is toward the right, shown with a cross in the right hand plot, with galactic rotation being clockwise. Sun's position is shown with a black dot. A logarithmic stretch is used for the map of the RGB stars to enhance lower density regions.
    }
    \label{fig:XY_OB_and_giants}
\end{figure*}

In the present Section, we map the spatial distribution of the OB and RGB stars, open clusters, and Cepheids samples described in Section \ref{sec:samples}. Additionally, we present a comparison with some models available in the literature, and discuss the observed similarities and differences.  

The left panel of Figure \ref{fig:XY_OB_and_giants} shows the distribution of the OB stars in the XY-plane of the Galaxy. Far from being homogeneous, the distribution of the OB stars is highly structured, and has numerous regions where the stellar density is markedly higher than in other ones. Such high-density regions are not randomly distributed, but appear to be organised in spiral arm segments. Specifically, we can identify three quasi-diagonal segments crossing the left panel of Figure \ref{fig:XY_OB_and_giants}: from left to right, we discern the Perseus arm, the Local (Orion) arm, and an inner stripe corresponding to the Sagittarius-Carina and (possibly) the Scutum arms. Figure \ref{fig:XY_OB_and_giants} (left panel) maps with unprecedented detail the structural features of the OB stellar population, especially within about 3 kpc from the Sun. Beyond this distance the distribution becomes increasingly dominated by radial features produced by foreground extinction. 

The right panel of Fig.~\ref{fig:XY_OB_and_giants} shows the spatial distribution of the 5.7M RGB stars with line-of-sight velocities in the Galactic plane. Again, as for the OB sample, radial "shadow cones" from foreground extinction are clearly visible in the RGB sample, though with higher angular frequencies, as the density here is not derived using a smoothing kernel. Aside from this difference, and in contrast to the OB stars, the RGB sample exhibits a smooth spatial distribution, as expected from a dynamically old stellar population. No clear spiral structure is apparent from the stellar counts, possibly due to the fact that the giant sample contains typically old stars (compared to the other populations considered in this work). However, we note that a density enhancement is present toward the Galactic centre, presumably due to the Galactic bar. Additionally, we note that an overdense region is apparent at $x \simeq$ 2-3 kpc, running across a range of y. This is due to the combination of two main factors: (i) we are using a magnitude limited sample, implying that the density decreases with heliocentric distance; (ii) the intrinsic distribution of RGB stars in the Galactic disc is increasing toward the inner parts of the Galaxy (as expected from an exponential disk).


\begin{figure}
    \centering
    \includegraphics[width=0.49\textwidth]{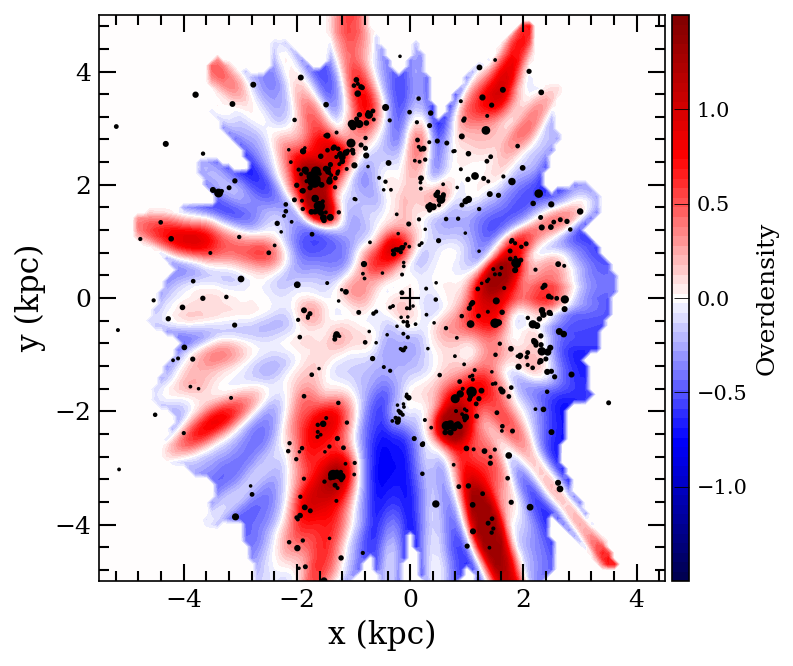}
    \caption{Overdensity map of the OB stars over plotted with the positions of the open clusters younger than 63 Myr and with $n_0 > 5$, plotted with filled circles whose size is proportional to $\sqrt{n_0}$. The cross indicates the position of the Sun.
    }
    \label{fig:XY_OB_and_clusters}
\end{figure}

To better explore the Galactic spiral structure, we apply to our OB sample the same approach adopted by \citet{Poggio:2021} to map the stellar over-density, defined as
$$ \Delta_{\Sigma} (X,Y) = \frac{ \Sigma (X,Y) \, -  \langle \, \Sigma (X,Y) \, \rangle }{\langle \, \Sigma (X,Y) \, \rangle } \, $$ 
where the local surface density $\Sigma (X,Y) $ and the mean surface density $\langle \, \Sigma (X,Y) \, \rangle$ are constructed using kernel density estimators with bandwidths of 0.3 and 2 kpc, respectively, adopting an Epanechnikov kernel. The resulting map is shown in Figure \ref{fig:XY_OB_and_clusters}. As we can see, the red diagonal stripes in Figure \ref{fig:XY_OB_and_clusters} correspond to the segments of the nearest spiral arms, consistently with the features identified in the left panel of Figure \ref{fig:XY_OB_and_giants}. Based on Figures \ref{fig:XY_OB_and_giants} and \ref{fig:XY_OB_and_clusters}, the emerging picture of the Galactic spiral structure is in good agreement with the one found in \citet[][]{Poggio:2021} (see their Fig. 1B and 1C).

The same Figure \ref{fig:XY_OB_and_clusters} shows a comparison between the OB over-density map and the young ($<63$ Myr) and bright (i.e. with at least 5 members brighter than $M_G =0 $) open clusters (OCs). Distances to the OCs were obtained by inverting the median parallax of each cluster, as described above in section \ref{sec:OCdata}.  As we can see in Figure \ref{fig:XY_OB_and_clusters}, there is a good agreement between the open clusters distribution and the spiral structure mapped by the OB sample. 

\begin{figure}
    \centering
    \includegraphics[width=0.49\textwidth]{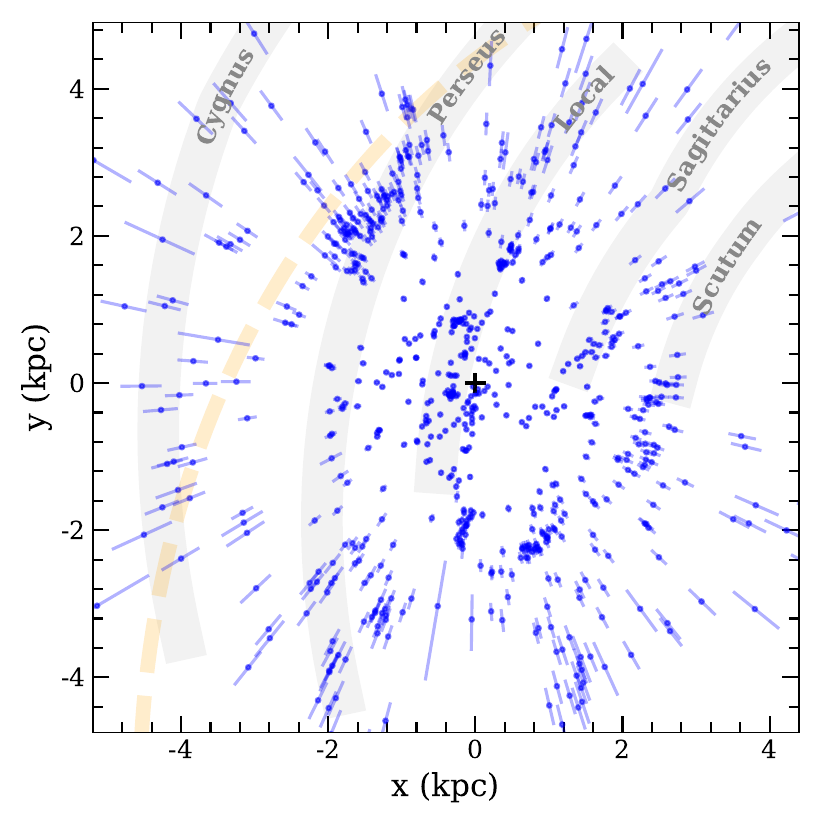}
    \caption{Heliocentric coordinates of the clusters younger than $\log t$=7.6 (63\,Myr). Thick grey lines are the spiral arm model of \citet{Reid2019}, and the dashed line is the trace of the Perseus arm modelled by \citet{Levine06}. The bars represent the 1-$\sigma$ uncertainty on the distance, taking into account statistical and systematic parallax errors.
    }
    \label{fig:xy_clusters}
\end{figure}

Figure~\ref{fig:xy_clusters} shows the spatial distribution of the clusters younger than $\log t$=7.8 ($\sim$63\,Myr). Although the young clusters clearly trace multiple elongated structures, evocative of arms and inter-arm regions, their distribution is not continuous. The young OCs alone do not constitute a sufficient sample to clearly define the main spiral arms.
The most striking difference when comparing this distribution to the spiral arm model of \citet{Reid2019} (shaded grey arms in Fig.~\ref{fig:xy_clusters}) is that the Perseus arm appears interrupted for two kiloparsecs. This discontinuity has been observed before in cluster distributions \citep[e.g.][]{CantatGaudin2020} or CO clouds \citep[][]{Peek21perseus}. On the other hand, the model of \citet{Levine06} traces very nicely the orientation of the Perseus arm in the upper main sequence stars \citep[see also][]{Poggio:2021}, and appears to be in reasonable agreement with the distribution of the OCs as well (see the orange dashed line in Fig.~\ref{fig:xy_clusters}). According to this model, the two groups that \citet{Reid2019} consider to define a low-pitch-angle Perseus arm (one in the second and one in the third Galactic quadrant) would in fact belong to two different arms.

\begin{figure*}
    \centering
    \includegraphics[width=0.49\textwidth]{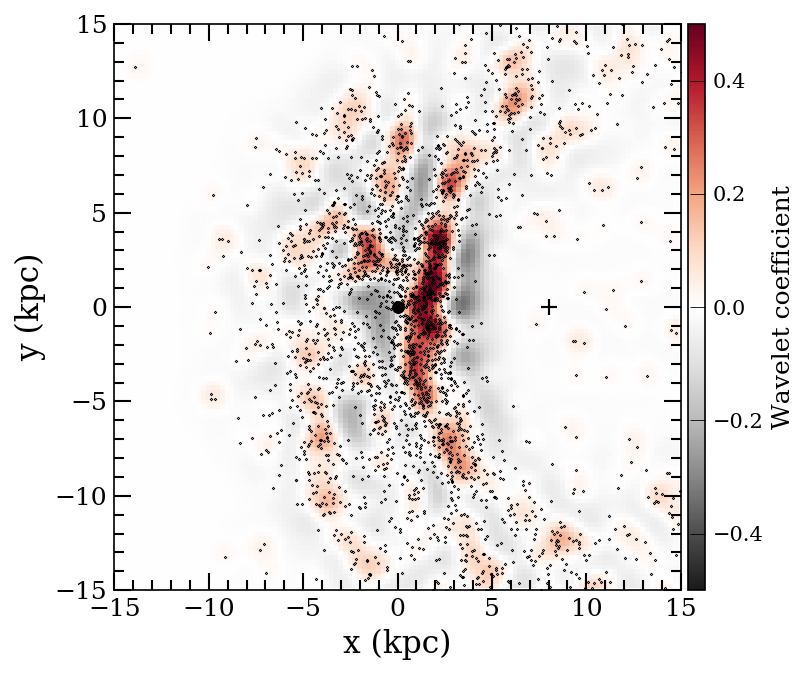}
      \includegraphics[width=0.49\textwidth]{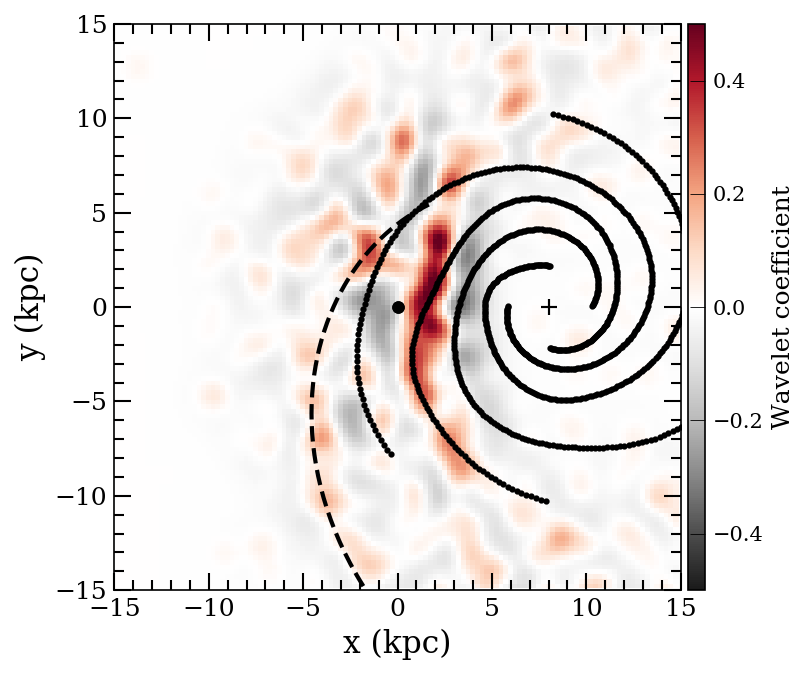}
    \caption{Wavelet transformation of the Cepheids with age < 200 Myr. Black dots in the left panel show the positions of the single sources, while the right panel shows, on a larger scale, an overlay of the model from \citet{Taylor:1993} (solid lines) and the one from \citet{Levine06} (dashed line).
    }
    \label{fig:WT_Ceph}
\end{figure*}


Finally, we map the spatial distribution of Cepheids younger than 200 Myr using a wavelet transformation (WT). The left panel of Figure \ref{fig:WT_Ceph} shows a comparison between the single sources (shown as black dots) and the WT coefficients \citep[colored map, see technical details in][]{Ramos:2018,Poggio:2021}. The right panel of Figure \ref{fig:WT_Ceph} shows a comparison between the Cepheids WT and some models available in the literature. Solid lines show the 4-armed model of Taylor and Cordes \citep{Taylor:1993}; in the region where the Cepheids are more abundant ($x < 5$ kpc), the Sagittarius Carina arm appears to be in good agreement below approximately $y \simeq$2.5 kpc. At approximately $x \simeq 2.5$ kpc and $y \simeq$ 5 kpc, the overdensity of Cepheids seem to diverge with respect to the Taylor and Cordes model of the Sag-Car arm in this direction. Meanwhile, in the outer regions of the Galaxy, the orientation of the Perseus arm in the Cepheids seems more consistent with the Levine model \citep{Levine06} than either the Taylor and Cordes or Reid models. This outer arm, based on HI data, is remarkably traced by the Cepheids out to a galactocentric radius of at least 16 kpc. Unfortunately the Cepheids are too sparse to trace the weaker Local (Orion) arm.

\section{Velocity maps}
\label{sec:velomaps}
This section studies the kinematics of RGB and OB stars in the disc of the Milky Way to highlight the effects of the  disc asymmetries on velocities. We use the selections of OB and RGB stars within $|z| \le 0.3$ (77\,659 stars) and $|z| \le 1$\,kpc (5\,730\,578 stars) of the Galactic plane, respectively. We first describe the construction of the maps of the mean velocity and velocity dispersion from the individual velocities derived in section \ref{sec:three}, and analyse the resulting velocity fields of the RGB sample, which cover a much larger extent of the disc than the OB sample, discussed at the end of this section. Table \ref{table:Not} summarising our notation is provided for convenience. 

\begin{table}
\caption{Notation and Nomenclature}             
\label{table:Not}      
\centering                          
\begin{tabular}{c l}        
\hline\hline                 
variables &  \\    
\hline                        
   $\varpi,\vec{\mu}$    & astrometry  \\ 
   $\sigma_\varpi,\sigma_{\mu_\alpha},\sigma_{\mu_\delta},$  & their uncertainties  \\
   $v_\mathrm{los}$ & line-of-sight (los) velocities \\
   $d, \sigma_d$        & distance and related uncertainty \\
   $x,y,z$              & heliocentric Cartesian coordinates \\  
   $X,Y,Z$              & galactocentric Cartesian coordinates \\
   $u,v,w$              & $x,y,z$ velocity components \\
   $R, \phi, z$         & LH galactocentric cylindrical coordinates \\
   $v_R, v_\phi, v_z$   & $R, \phi, z$ velocity components \\ 
   $\sigma_{v_R}, \sigma_{v_\phi}, \sigma_{v_z}$         & their uncertainties \\
   $V_R, V_\phi, V_z$   & mean $v_R, v_\phi, v_z$ velocities \\
   $\sigma_{V_R}, \sigma_{V_\phi}, \sigma_{V_z}$         & their uncertainties \\
   $V_R$       & radial velocity \\
   $V_\phi$    & azimuthal velocity \\
   $\overline{V}_\phi(R)$    & mean azimuthal velocity at $R$ \\
  $\sigma^*_R, \sigma^*_\phi, \sigma^*_z$ & velocity dispersions in $R,\phi, z$ directions \\
\hline                                   
\end{tabular}
\end{table}



\subsection{Construction and analysis of velocity maps}
\label{sec:velocitymaps}

We built maps of the ordered and random motions for the radial, azimuthal, and vertical velocity components. We designed these maps with a constant  $100$\,pc resolution, with $341 \times 341$ pixels. These characteristics result from empirical choices  to have sufficient numbers of stars per bin for the current analysis. To perform robust derivations of the velocity per pixel, we considered only the cells with a minimum of $20$ stars, and masked all others. This gives grids with a median and a maximum numbers of $58$ and $267$ stars per pixels for OB stars, respectively, 
and $152$ and $2\,541$ stars per pixel for RGB stars, respectively.

At a given heliocentric position $(x, y)$ corresponds a cell $j$ of $100\times100$ pc in our three velocity components $k = R,\phi,$ or $z$ maps, which contains $N_*$ stars. We estimate the stellar mean velocity $V_k$ and its associated dispersion\footnote{Most inconveniently, $\sigma$ is the traditional notation for velocity dispersion, but also that for Gaussian uncertainties. To avoid potential confusion, we add a superscript $*$ to the stellar velocity dispersion.} $\sigma^*_k$, 
by optimising the log-likelihood of the distribution of ``observed'' velocities $v_{k,j}$ (with uncertainties ${\sigma_v}_{k,j}$) of the stars located within the $j$-cell. We assumed Gaussian uncertainties on the individual independent $v_{k,j}$, so that the \textit{negative} log-likelihood to minimise is:
\begin{equation}\label{eq:negative-loglik}
\mathcal{L} (V_k,\sigma^*_k) =  \frac{1}{2}\sum^{N_*}_j\left(\ln({\sigma^*_k}^2 + {\sigma_v}_{k,j}^2) + \frac{(v_{k,j}-V_k)^2}{{\sigma^*_k}^2 + {\sigma_v}_ {k,j}^2} \right).
\end{equation}
Appendix~\ref{appendix:derive-Vk-uncertainty} details the derivations of the uncertainty ${\sigma_V}_{k,j}$ of our mean velocities ${V}_{k,j}$, which also demonstrates that under certain conditions we can approximate them as $\sigma_{V_k} = \sigma^*_k/\sqrt{N_*}$.

\begin{figure*}
\centering
\includegraphics[height=7.2cm,trim={0.75cm 1cm 0.1cm 0.4cm},clip]{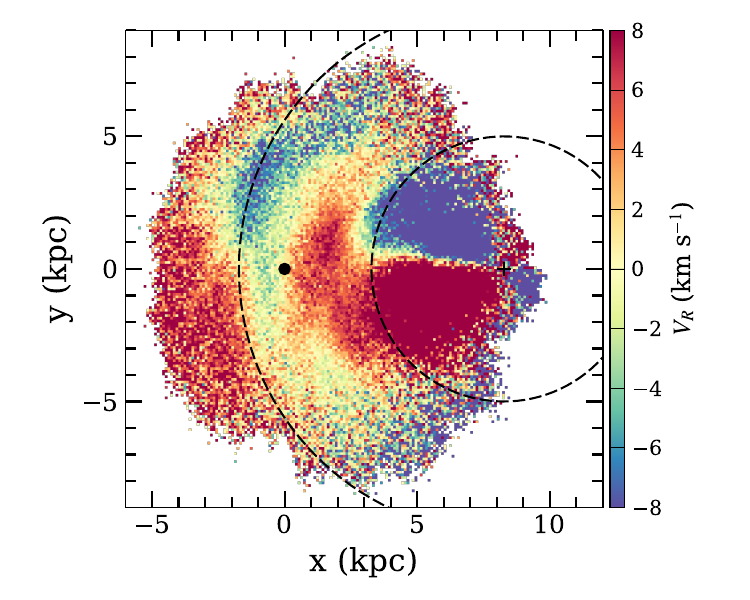}
\includegraphics[height=7.2cm,trim={2.05cm 1.02cm 0.22cm 0.25cm},clip]{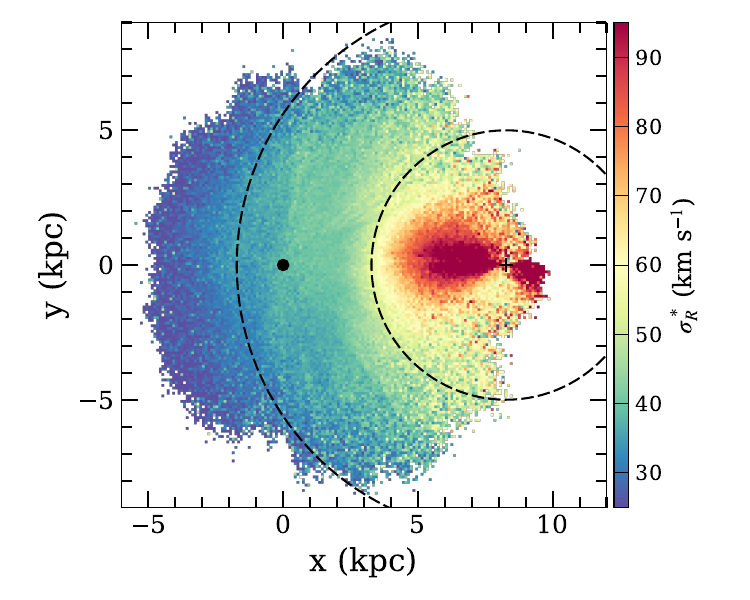}\\
\includegraphics[height=7.2cm,trim={0.75cm 1cm 0.2cm 0.4cm},clip]{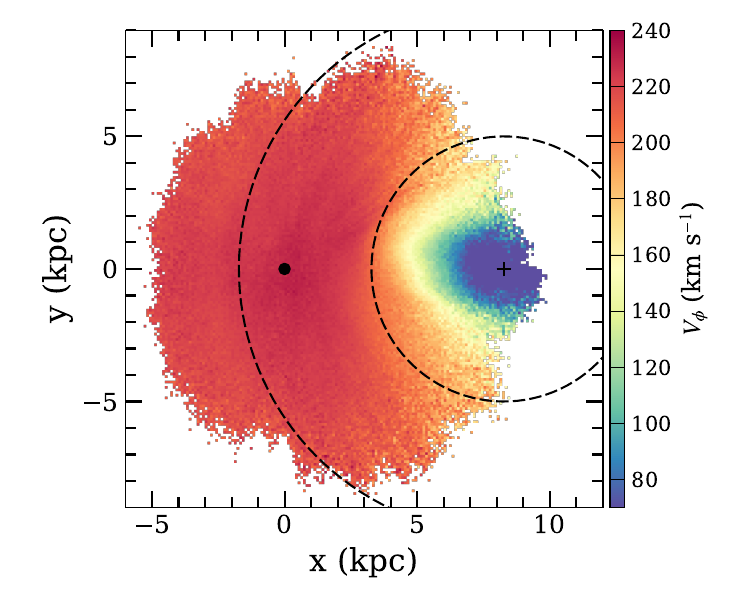}
\includegraphics[height=7.2cm,trim={2.05cm 1cm 0.3cm 0.4cm},clip]{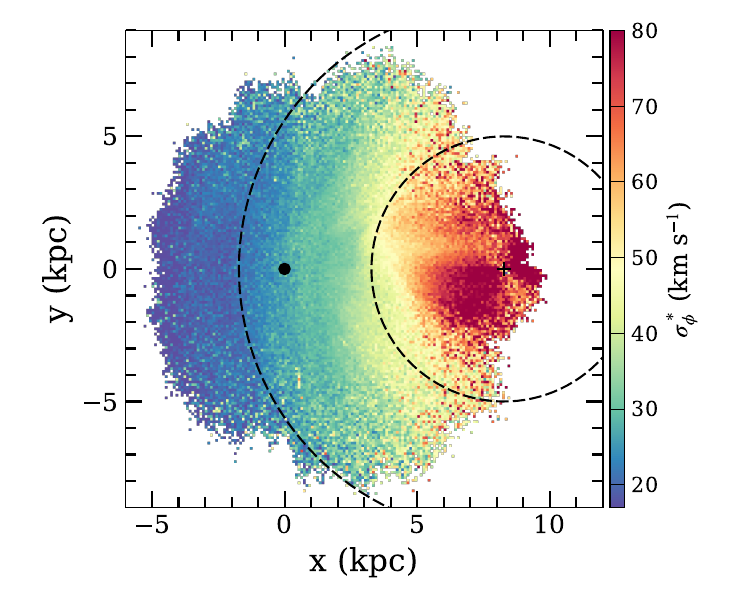}\\
\includegraphics[height=7.8cm,trim={0.7cm 0.3cm 0.2cm 0.3cm},clip]{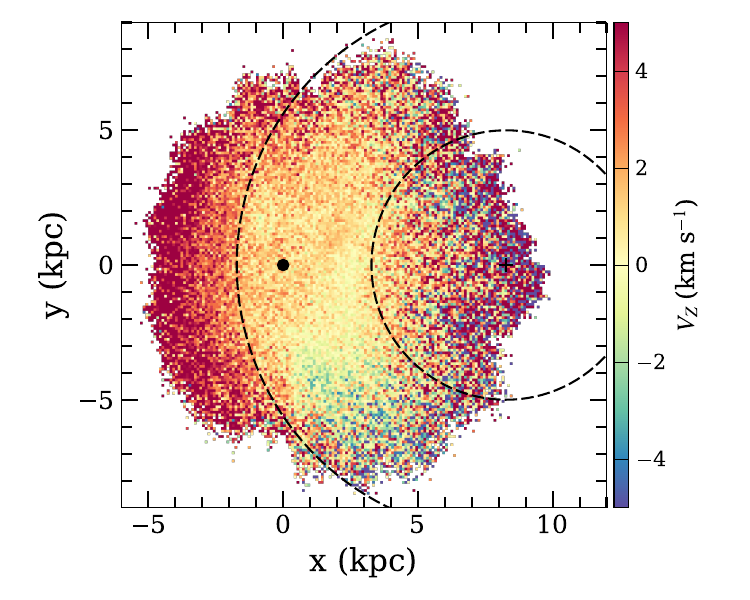}
\includegraphics[height=7.8cm,trim={2.05cm 0.3cm 0.3cm 0.4cm},clip]{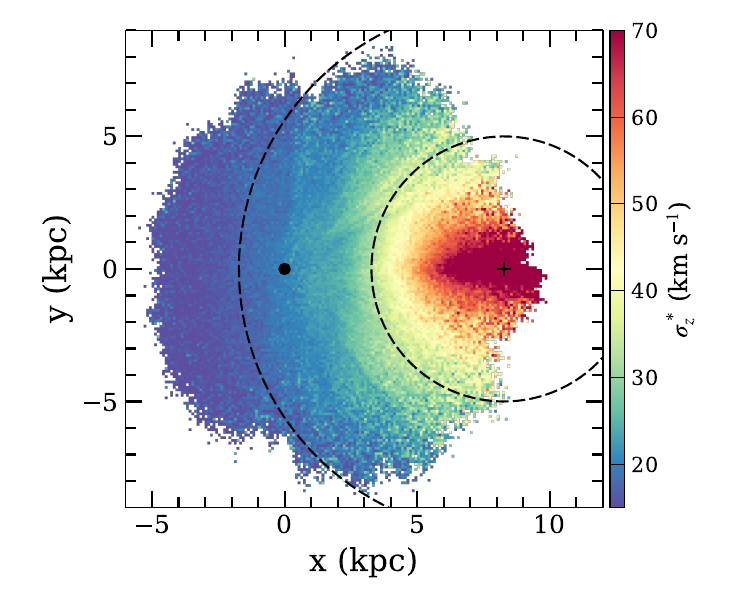}
\caption{Velocity maps of RGB stars. Left and right panels are for the ordered and random motions, respectively. From top to bottom, panels show the radial, azimuthal, and vertical velocity components. The origin, emphasised by the black dot, indicates the position of the Sun, and a plus symbol indicates the Galactic Centre. Dashed circles represent constant distance to the Galactic centre at $R=5$ and $R=10$ kpc.  The velocity ranges have been chosen to enhance contrasts to help identify by eye regions where streaming motions occur. Left panels of Fig.~\ref{fig:uncertaintyvelomaps} show the associated uncertainty maps. We detail the construction of these maps in Sect.~\ref{sec:velocitymaps}. These velocity and velocity dispersion maps can be found in CDS in FITS format. 
\label{fig:velomapsgiants}
}
\end{figure*}

Figure~\ref{fig:velomapsgiants} shows the resulting three component velocity fields for the sample of RGB stars (left panels), and the velocity dispersions (right panels). We optimised the displayed velocity ranges to help identify by eye regions where streaming motions occur. Left panels of Fig.~\ref{fig:uncertaintyvelomaps} shows the associated uncertainty maps for the velocities.

The \vrad\ map shows a remarkable bisymmetric feature on both sides of the GC, with negative and positive values on each side of the apparent bar's major axis. This quadrupole feature is a characteristic of the mean inward motion down to $\sim -40$ \kms\ ($y >0$) and the mean outward motion up to $\sim 45$ \kms generated by the Galactic bar. \gdrthree\ allows us to confirm the bar quadrupole \vrad\ pattern identified in \citet{Bovy2019} and \citet{Queiroz2021}, who used thousands of stars, but using \gdrtwo{} and EDR3 astrometry \citep{GaiaDR2_Brown,GaiaEDR3_Brown} together with APOGEE line-of-sight velocities \citep{Majewski:2017, 2018ApJS..235...42A}. Section~\ref{sec:bisymbar} and Appendix~\ref{sec:appsimu} describe in detail this quadrupole pattern. We also note that the RGB sample contains enough stars to apparently provide us with mean velocity estimates beyond the Galactic Centre (GC), though these should be interpreted with caution.

The \svrad\ map shows a bisymmetric pattern as well, but different from \vrad, with larger amplitudes which are aligned with the direction where \vrad\ changes its sign along the bar major axis, and lower along a perpendicular direction. Here again, the GC is the node of the quadrupole feature.
In addition to the central quadrupole, streaming motions in \vrad\ also occur at larger radii, e.g. $R = 6.5$ kpc  ($x \sim 1.5$ kpc), which shows  \vrad\ being larger for  $|y| < 2$ kpc than at smaller and larger azimuths. At $R \sim 10$ kpc ($x,y \sim -1.5,+1$ kpc), \vrad\ shows a clear change of sign with respect to azimuth. 

The distribution of the azimuthal  velocity \vtan\ is elongated in the bar: within the central 5 kpc, the rotation at a given radius is slower along the apparent bar axis than perpendicular to the bar axis. 
The \svtan\ map also seems to exhibit a bisymmetry in the bar region, but rotated by about $45\degr$ with respect to that seen in \svrad. The azimuthal random motion appears smaller when the radial random component is larger. In other words, the planar velocity dispersion is highly anisotropic in the bar region. 

The vertical velocity \vz\ is mostly positive. Unlike the other two velocity components, it does not show any pattern linked to the presence of the Galactic bar. A streaming is observed around $R= 7$ kpc, as \vz\ shows among the lowest values on one side  ($1 \le x \le 4,y<-2$ kpc), while being positive on the opposite side with respect to $y = 0$. However, this pattern, being symmetric about the x-axis, may be an artefact of systematic errors.  Vertical motions are also larger with galactocentric radius toward the galactic anticentre, clearly showing the kinematic signature of the warp of the Galactic disc beyond, in agreement with previous results \citep{DR2-DPACP-33, RomeroGomez:2019, Poggio:2018, Poggio:2020,LopezCorredoira:2019}. We do not observe any bisymmetric feature in the vertical dispersion, which is larger along $l \sim 0$. It should be noted that the vertical velocity component is more sensitive to systematic errors as, our sources being near the galactic ($b=0$) plane, this component is predominantly manifested in the tangential velocities that are sensitive to distance errors (see Fig.~\ref{fig:velocity_err_vs_d}).


In summary, these maps detect the significant signature of the Galactic bar of the MW in the inner disc.  We also find some streaming signatures of \vrad\ at larger galactocentric distances, which might be associated with corotation or the outer Lindblad resonance.
Section~\ref{sec:outerdisc} discusses the case of the young stellar populations and the similar maps for the OB star sample. 


\subsection{Analysis of radial profiles}
\label{sec:velocityprofiles}

\begin{figure}[t!]
\includegraphics[width=9cm,trim={0 0.88cm 0 0},clip]{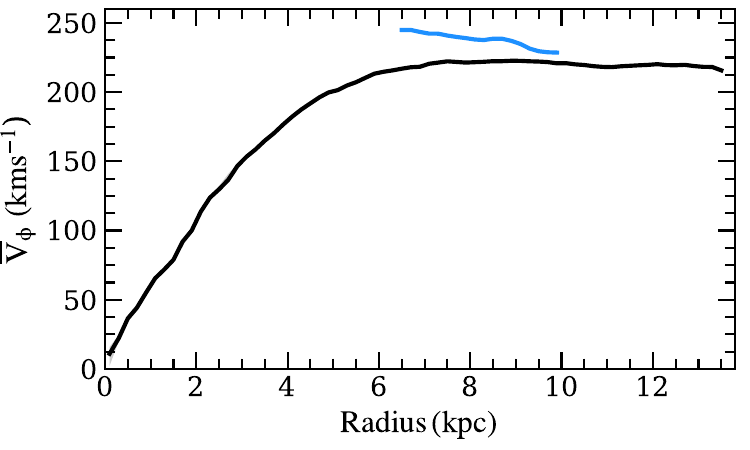}
\includegraphics[width=9cm,trim={0 0 0 0},clip]{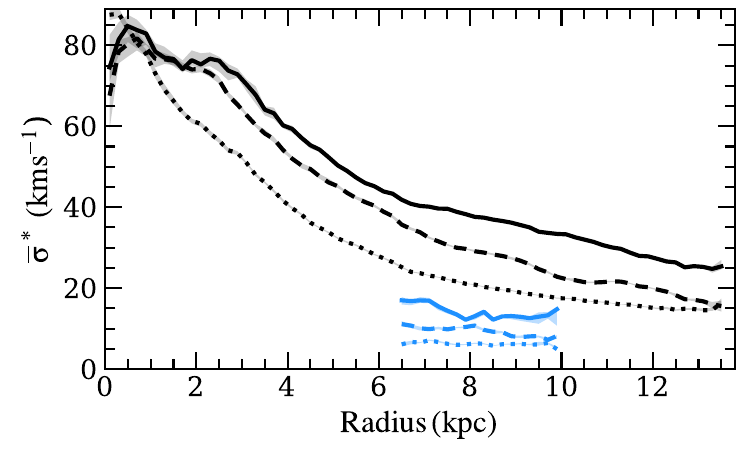}
\caption{Velocity profiles of the Milky Way RGB  and OB stars (black and blue lines, respectively).
\textit{Top panel:} Rotation curves. 
Shaded area represent the uncertainties. 
\textit{Bottom panel:} Radial, azimuthal and vertical velocity dispersions (solid, dashed and dotted lines, respectively). The data shown in these figures can be found in CDS.
\label{fig:vphiprof}
}
\end{figure}

From these maps, we inferred the average axisymmetric variation of velocities following the same procedure as in \citet{GaiaEDR3_Luri_LMC}, where for each bin in radius of 200 pc size, the median value of the cells in the radial bin define the azimuthally-averaged velocity $\overline{V}_\phi(R)$ at that radius. We discarded the radial bins where the number of pixels from the maps is less than 5 pixels. Bootstrap resamplings were performed at each radius to define the velocity uncertainties, measured at the 16th and 84th percentiles of the velocity distributions. In Fig.~\ref{fig:vphiprof}, we present the Galactic rotation curves, as well as the velocity dispersion profiles for the RGB and OB samples. Despite the asymmetries observed in the velocity field, especially at radii $R < 5$ kpc, the rotation curve of the RGB stars is regular. It smoothly increases like a solid body up to $\overline{V}_\phi \sim 220$ \kms\ to $R \sim 6$ kpc, then remains constant out to the last measured radius. The amplitude of the curve for the younger OB stars is larger than for RGB giants, because of their smaller asymmetric drift, and is decreasing with galactocentric radius, rotating on average 17 \kms\ faster than the RGB stars over the radial range $R=6.5-10$ kpc. These average rotation curves have been subtracted from the \vtan\ maps to produce residual velocity fields $V_\phi(x,y) - \overline{V}_\phi(R)$,  useful for evidencing  velocity streaming in the outer disc (Sect.~\ref{sec:outerdisc}). 

\begin{figure}[t!]
\includegraphics[width=9cm]{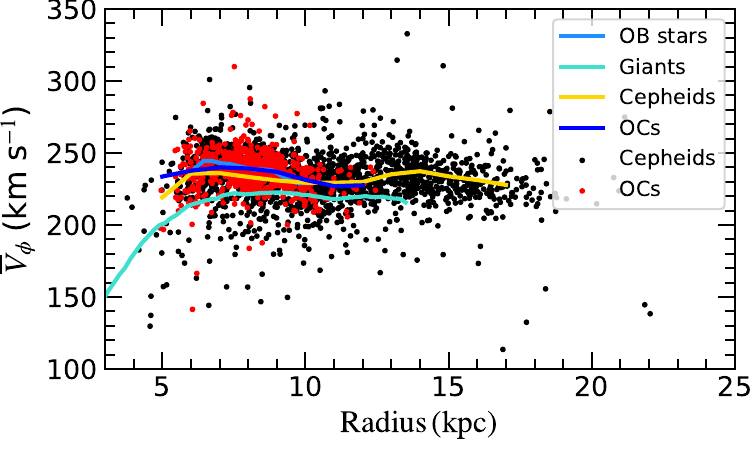}
\caption{Comparison of rotation curves of the different samples. Mean azimuthal velocity for the open clusters (dark blue) and Cepheids (yellow) as a function of galactocentric radius (calculated as explained in the text), compared to the profiles of the OB (light blue) and RGB stars (turquoise). Coloured points show the single sources for the Cepheids (black dots) and OCs (red points).
\label{fig:compvrotyoung}
}
\end{figure}

Figure ~\ref{fig:compvrotyoung} shows a comparison between the azimuthal velocity of the OB stars, the open clusters, the Cepheids, and the RGB giants with respect to galactocentric radii. The median velocities for Cepheids and open clusters are calculated using overlapping radial bins of 1 kpc step and 2 kpc width. As expected, there is a good agreement between the OB stars and the OCs. The Cepheids exhibit a mean azimuthal velocity similar, but slightly slower, to the OB and the open clusters, but reach significantly larger radii. 
Finally, the mean azimuthal velocity for the RGB giants is systematically lower than the other tracers, as expected for asymmetric drift for an older population with a larger velocity dispersion.
 
The radial and azimuthal velocity dispersion profiles of RGB stars show two distinct regions. Within the inner part of the bar ($R  \sim 2.5$ kpc), the profiles are shallow and the dispersions comparable ($\sim 75-80$ \kms), though we note that the $(x,y)$ map of the dispersion (Fig.~\ref{fig:velomapsgiants}) in this region shows very strong asymmetries.  Beyond 2.5 kpc, $\overline{\sigma}^*_z < \overline{\sigma}^*_\phi < \overline{\sigma}^*_R$ and the   profiles continuously decrease for all three components. 
 The radial profiles of the velocity dispersion $\overline{\sigma}^*_k$ of the OB stars do not vary strongly with radius, showing average radial, azimuthal and vertical dispersions of $14.2, 9.4$ and $6.2$ \kms, respectively, in very good agreement with the observed random motions of gas within $R=8$ kpc \citep[$4-9$ \kms, ][]{2017Marasco}. 
 The axis ratios of the velocity ellipsoid as averaged from these profiles for $R> 3$ kpc are ($\overline{\sigma}^*_\phi/\overline{\sigma}^*_R, \overline{\sigma}^*_z/\overline{\sigma}^*_R, \overline{\sigma}^*_z/\overline{\sigma}^*_\phi) = (0.81, 0.60, 0.75)$ for the RGB stars, and $(0.66, 0.44, 0.66)$ for OB stars.


\subsection{The kinematics of the bar}
\label{sec:bisymbar} 

In this Section, we estimate some fundamental parameters of the Galactic bar, guided by results obtained from Section~\ref{sec:velocitymaps} and with the help of a mock galaxy from a numerical test particle simulation of a barred galaxy.
We want to make clear that this is not a made-to-measure, customised simulation that intends to quantitatively reproduce the observed dataset of RGB stars. The simulation has to be seen as a simple diagnostic tool for qualitative comparison to the observations, and the results presented below as tentative possibilities. We refer the reader to App.~\ref{sec:appsimu}) for the description and analysis of the simulation. We thus applied the same recipes as those described in App.~\ref{sec:appsimu}) to find, by analogy, the orientation and pattern speed of the Galactic bar, as well as the location of the outer Lindblad resonance. 

We start to fit the non-uniformity of the kinematics in the bar region of the RGB sample, assuming that the Galactic bar perturbs velocities by adding a bisymmetric component to the axisymmetric motions. This should apply to most of the ordered and random motions, because all of them but $v_z$ were shown to be similarly structured in the bar region (Sect.~\ref{sec:velocitymaps}).  
We therefore performed a simple Fourier decomposition  up to second order to characterise the axisymmetric and 
the bisymmetric components seen in the maps. 
This approximation of \vrad\ and \vtan\ is referred to as $V_{R,\rm mod}$ and $V_{\phi,\rm mod}$, and is given by:
\begin{equation}
V_{R,\rm mod}(R,\phi) = \overline{V}_R(R) +A_R(R)\cos\left(2(\phi - \phi_R(R))\right),
\label{eq:bisym}
\end{equation}
where $\overline{V}_R(R)$ is the axisymmetric mean value, and $A_{R}$ and $\phi_R$ are the amplitude and phase angle of the bisymmetric Fourier harmonics of \vrad\, respectively. 
Another parameter of interest is the scatter in the modelling, $V_{R,s}$, which absorbs all other asymmetric departures from the bisymmetry in the velocity fields. We use an analogous expression for $V_{\phi,\rm mod}$.

\begin{figure}[th]
\includegraphics[width=9cm,trim={0 0.88cm 0 0},clip]{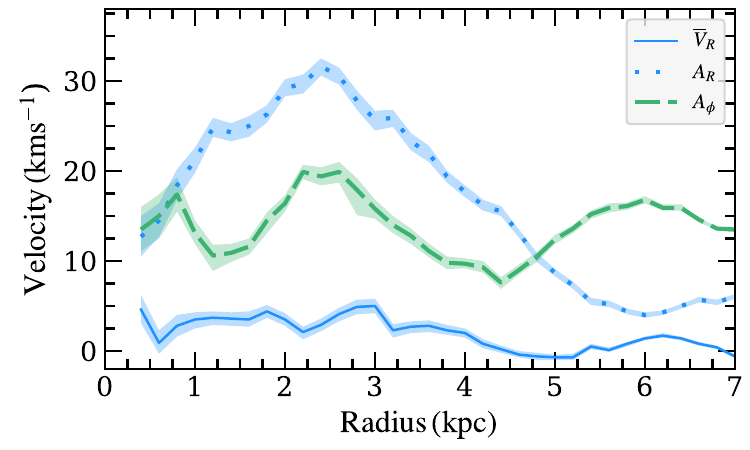}
\includegraphics[width=9cm,trim={0 0 0 0},clip]{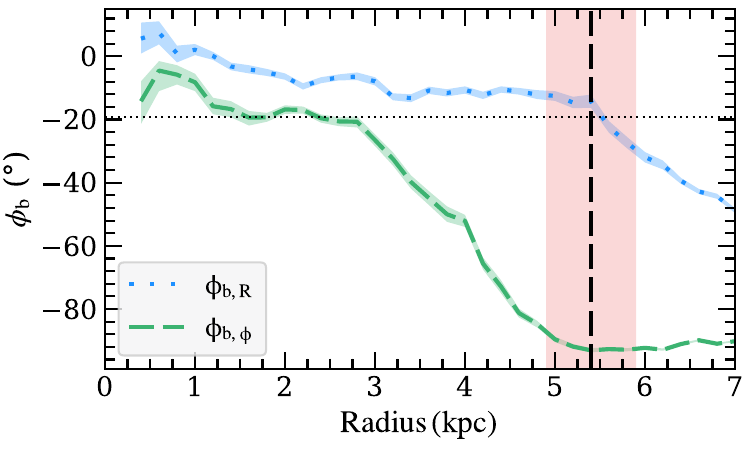} 
\caption{Results of the bisymmetric model of the \vrad\  and \vtan\ velocity maps. \textit{Top:} Amplitudes $A_{R}$ and $A_{\phi}$ of the bisymmetry. The axisymmetric component $\overline{V}_R$ is also shown. \textit{Bottom:} Phase angle of the perturbation with respect to the Sun-GC direction. Shaded area represent the uncertainties. The horizontal dotted line marks the estimated bar orientation angle. The vertical dashed line and purple   area show the estimated location of the bar corotation.}
\label{fig:vradasym}
\end{figure}

We performed Bayesian inferences of the model through Markov Chain Monte Carlo (MCMC) fits, using the Python library  \textsc{emcee} \citep{2013Foreman}. As this model best applies to regions where the kinematics is perturbed by a second order perturbation, we restricted the analysis to $0 \le R \le 7$ kpc, and considered a radial bin width of $200$ pc.
Defining the residual velocity as $V_{R,\rm res} = V_R - V_{R,\rm mod}$, the conditional likelihood function for each radial bin is expressed by: 
\begin{equation}
    \mathcal{L} (\overline{V}_R,A_R,\phi_R,V_{R,s})= - \frac{1}{2} \left(n_{\rm pix} \ln(2\pi) + \sum^{n_{\rm pix}} \left(V_{R,\rm res}^2 / \xi^2 + \ln(\xi^2) \right) \right), 
    \label{eq:likfun}
\end{equation}
and similarly for $\phi$, where $\xi^2 = \sigma^2_{V_R} + V^2_{R,s}$, and $n_{\rm pix}$ is the number of pixels inside the corresponding radial bin, imposing  $n_{\rm pix} > 25$ as a condition. Radial bins not satisfying this condition are not considered. 
We set a number of $32$ walkers and $2000$ steps in the MCMC fits, which is enough to converge towards  robust and stable solutions. The prior distributions are uniform and span $[-50,50]$ \kms\ for $\overline{V}_R$,  $[0,300]$ \kms\ for $\overline{V}_\phi$, $[0,50]$ \kms\ for $A_{R}$, $[0,30]$ \kms for $V_{R,s}$, and $[-80,110]\degr$ for $\phi_{R}$. Following prescriptions from  App.~\ref{sec:appsimu}, $\phi_{R}$ and $\phi_{\phi}$ are linked to the  direction of the bisymmetric perturbation of density with respect to the Sun-GC direction in the bar region, and that we call $\phi_{\textrm{b}}$, by $\phi_{\textrm{b}, R} = \phi_{R} - \pi/4$ in the case of the \vrad\ model, and $\phi_{\textrm{b}, \phi} = \phi_{\phi} - \pi/2$ in the case of \vtan.  We quote the uncertainties on the parameters at the 16th and 84th percentiles of the posterior distributions.

Figure~\ref{fig:vradasym} shows the resulting fits, where it can be seen that the bisymmetry is strongest at $R=2.5$ kpc in both the azimuthal and radial velocity fields. While the amplitude of the perturbation decreases beyond 2.5 kpc for $A_{R}$, it admits other maxima at $R \sim 0.8$ and $6$ kpc for $A_{\phi}$. Unsurprisingly, this simple model shows that \vrad\ is anything but axisymmetric, as $A_{R}$ exceeds $\overline{V}_R$. The axisymmetric radial velocity is mostly positive, showing a bulk outward motion of 3.6 \kms\ for $R < 3$ kpc, and null beyond 3 kpc, on average. As the radial velocity should be null on average, for a disc nearly relaxed, we attribute this non-negligible $\overline{V}_R$ to the incomplete coverage of azimuthal angles. In other words, this bulk motion  is only representative of the observed portion of the Galactic disc.  Another feature of interest is the low velocity scatter parameter $V_{R,s}$ and $V_{\phi,s}$, which is $\sim 6$ \kms at small radius and decreases to $<1$ \kms out to $R=7$ kpc (not shown in Fig.~\ref{fig:vradasym}). As for $\overline{V}_\phi$, we find it is very similar to the median rotation curve derived in the previous section.

The orientation of the bisymmetry with respect to the Sun-GC direction  is found to be different in the \vrad\ and \vtan\ models, particularly beyond $R=3$ kpc as seen in the bottom panel of Fig.~\ref{fig:vradasym}. At the peak of the strength at $R=2.5$ kpc, $\phi_{\textrm{b}}$ differ by $\sim 12\degr$. Interestingly, $\phi_{\textrm{b}}$ in the  \vtan\ model remarkably shows the same trend as the one seen in the numerical simulation (Fig.~\ref{fig:modelVrasymsimu} of App. ~\ref{sec:appsimu}), where it is seen that it faithfully traces the true bar orientation, even in the presence of uncertainties. As shown in Fig.~\ref{fig:modelVrVphiSimuThreeErrors}, the $\phi_{\textrm{b}}$ in the  \vrad\ model is more affected by the uncertainties. We can thus infer that the bar angle with respect to the Sun-GC direction is very close to the value for which $\phi_{\textrm{b},\phi}$ remains flat before the abrupt drop, that is $- 19.2\degr \pm 1.5\degr$ for $1.5 < R \le 2.8$ kpc. 
Then, as the location of the minimum of $\phi_{\rm bisym}$  of the \vtan\ model beyond the drop of phase and before it rises at larger radii corresponds to the corotation of the galactic bar in the simulation (see right panel of Fig.~\ref{fig:modelVrasymsimu}), we find by analogy that the range $R=5.2-6$ kpc is hosting the corotation radius of the Galactic bar, $R_{\rm CR}$. We adopt a conservative value $R_{\rm CR} = 5.4\pm 0.5$ kpc. This is surprisingly also the location where $\phi_{\rm bisym}$   starts to decrease for \vrad. This coincidence was not seen in  the numerical simulation, maybe because of the lack of spiral structure beyond the bar in the mock data.  

\begin{figure}
\includegraphics[width=9cm]{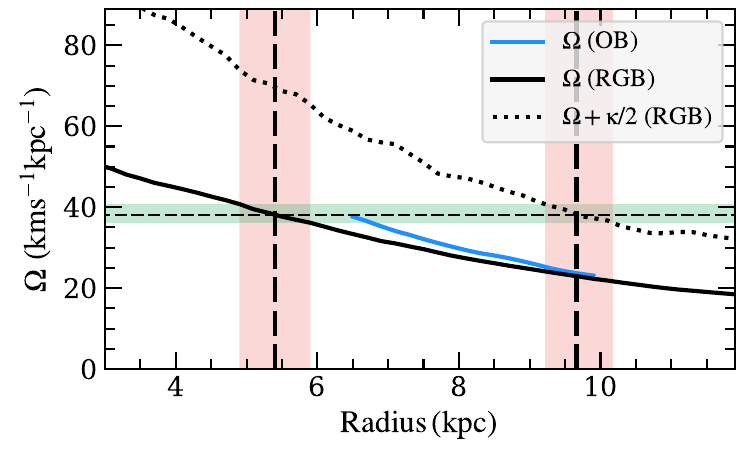}
\caption{Lindblad resonances of the Galactic bar. The black solid (dotted) line shows the angular frequency (angular frequency plus half of the epicyclic frequency) for the RGB stars, while the blue solid line shows the angular frequency for the OB stars. The horizontal dashed line represents the bar pattern speed, and the two vertical dashed lines at $R=5.4$ and $9.7$ kpc, with their uncertainty range in purple, are the location ranges of the corotation and outer Lindblad resonance, respectively.}
\label{fig:corotation}
\end{figure}

In Fig.~\ref{fig:corotation}, we show the angular velocity curve,  $\Omega=\overline{V}_\phi/R$, and its combination with the epicyclic frequency for the RGB stars (black) and OB stars (blue, only the angular velocity). We emphasise here that the epicyclic frequency is derived by assuming the epicycle approximation, which might not be perfectly correct in a radial range far from the Solar neighbourhood. Assuming the corotation radius range obtained above from the phase of the bisymmetry in \vtan, the intersection between $R_{\rm CR}$  and the $\Omega$ curve provides the bar pattern speed, which is $\Omega_{\rm bar}=38.1^{+2.6}_{-2.}$\,\kms kpc$^{-1}$. We can also provide an estimation of the Outer Lindblad resonance (OLR), which is the galactocentric radius where the fixed pattern speed intersects the curve $\Omega+\kappa/2$, that is, the position in the disc where stars rotate slower than the bar pattern and perform two radial oscillations for one revolution around the Galactic centre ($\kappa$ is the epicyclic frequency). In this case, the RGB sample provides an OLR of $9.7 \pm 0.5$\,kpc, outside the Solar radius. A word of caution has to be written here, because we estimate $\Omega_{\rm bar}$ and the OLR from the azimuthal angular velocity of the RGB sample, which is not the same as that from the circular velocity due to asymmetric drift.  
We note that the OB stars, being younger and supposedly less affected by asymmetric drift, have an angular velocity slightly larger than that of RGB stars. By extrapolation to the inner disc, our $\Omega_{\rm bar}$ value could thus be biased towards lower values and should be taken as a lower limit. 


\subsection{Kinematics of the outer disk}
\label{sec:outerdisc}

\begin{figure*}
\centering
\includegraphics[height=6.5cm,trim={0.65cm 1cm 0.1cm 0.25cm},clip]{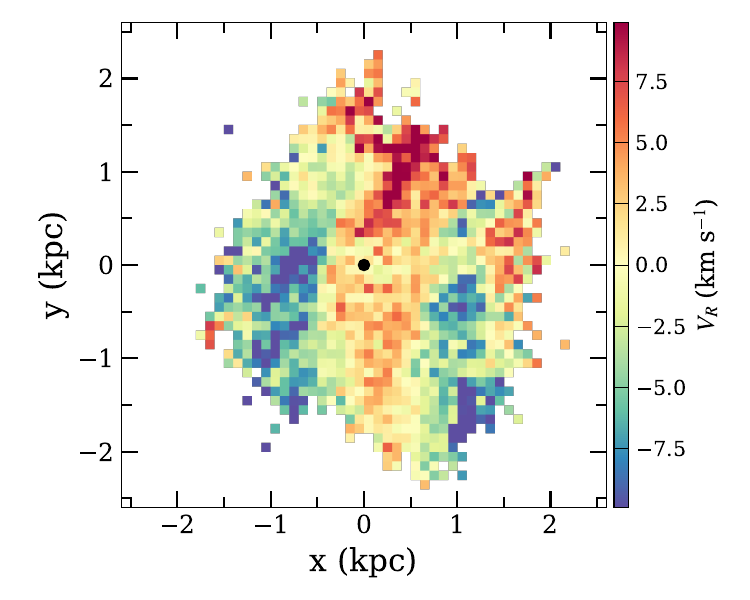}
\includegraphics[height=6.5cm,trim={0.65cm 1cm 0.22cm 0.25cm},clip]{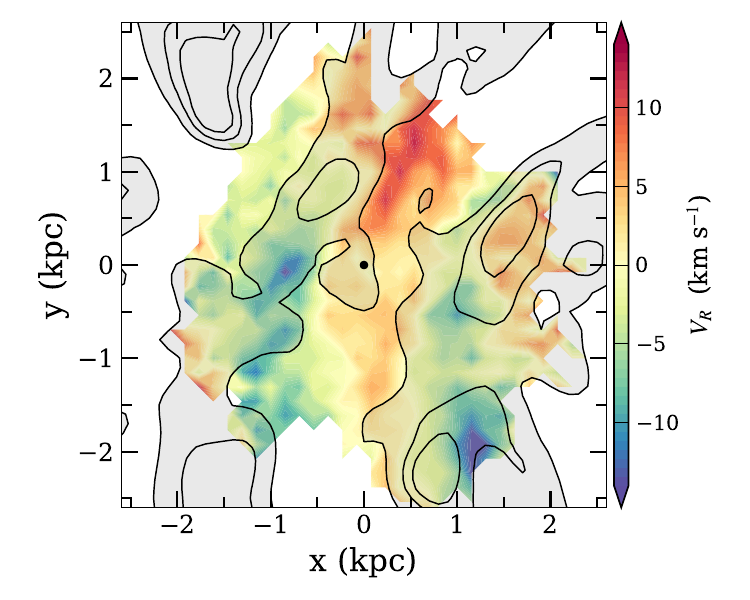}\\
\includegraphics[height=6.5cm,trim={0.65cm 1cm 0.1cm 0.25cm},clip]{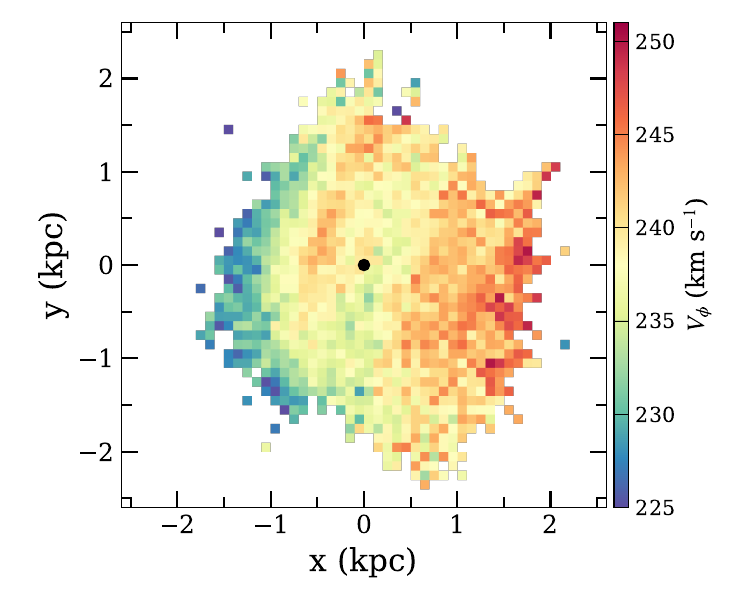}
\includegraphics[height=6.5cm,trim={0.65cm 1cm 0.22cm 0.25cm},clip]{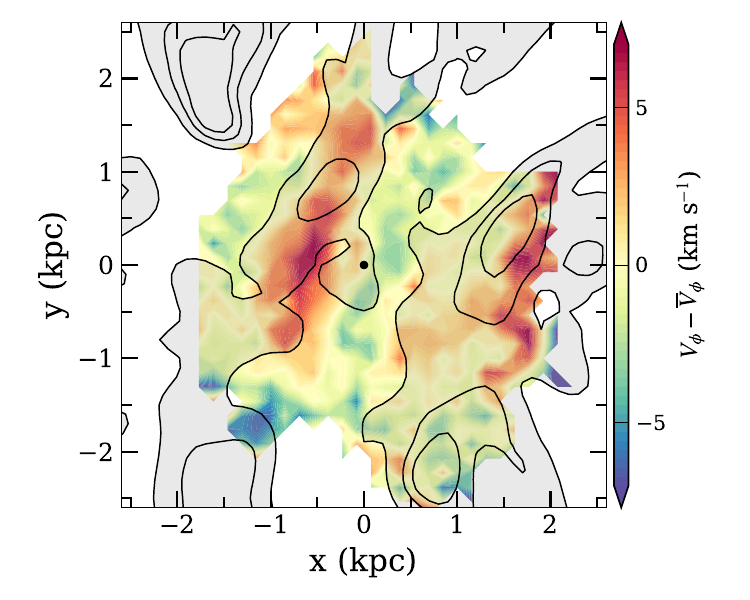}\\
\includegraphics[height=7.1cm,trim={0.22cm 0.3cm 0.5cm 0.3cm},clip]{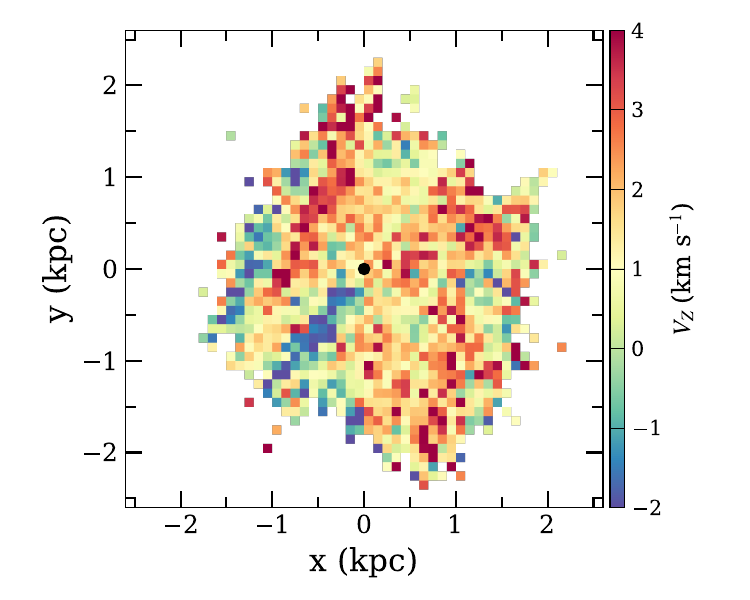}
\includegraphics[height=7.0cm,trim={0.38cm 0.3cm -0.3cm 0.25cm},clip]{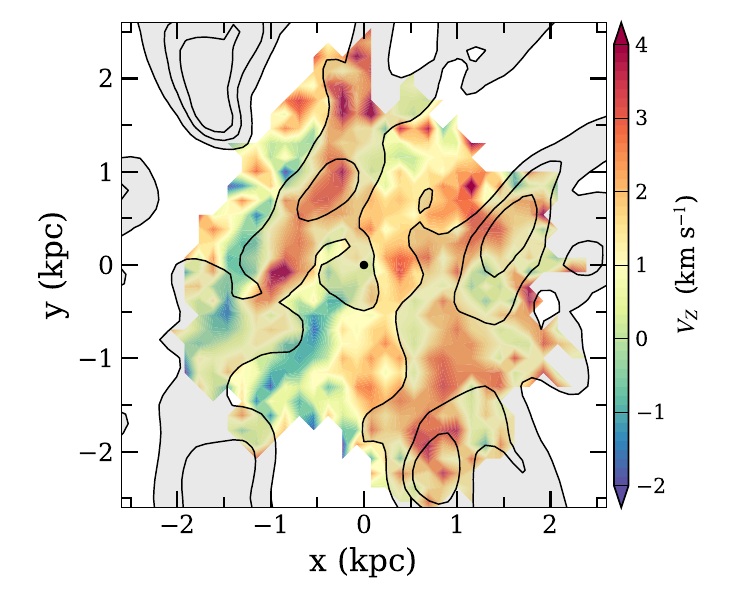}
\caption{Velocity maps of OB stars. Left panels show the inferred radial, azimuthal, and vertical velocity, from top to bottom. 
The velocity ranges have been chosen to enhance contrasts. The right panels show the maps of the radial velocity (top panel), the azimuthal residual velocity (middle panel) and the vertical velocity (bottom panel), compared to the overdensity contours of the OB stars (grey shaded areas), calculated as explained in Section \ref{sec:confspace}.
In all panels, the position of the Sun is marked by a filled circle. The maps were obtained considering only stars within $|z| \le 0.3$ kpc. The velocity and maps (left hand column) can be found in CDS in FITS format.
\label{fig:velomaps_OBstars}
}
\end{figure*}

In this Section, we present and discuss the velocity maps for the OB and RGB sample in the outer disk (i.e. with galactocentric radius $R>5$ kpc), and compare them to the spatial location of the spiral arms in the Galaxy.

Figure \ref{fig:velomaps_OBstars} shows the velocity maps of the OB stars, with the $V_R, V_{\phi}, V_Z$ maps on the left panels. On the right panels we show the same maps (with the exception of the middle panel, where $V_{\phi}$ has been replaced by the residual map of $V_{\phi}$, calculated as explained in Sec.~\ref{sec:velocitymaps}), but compared to the spiral arms found in the overdensity (see Sec.~\ref{sec:coord_maps}), overlayed as gray-shaded contours. We note that here, for the OB stars, the overdensity in the spiral arms and the streaming motions can be studied using \emph{the same stellar population} (although, of course, the two samples are not exactly the same, given that only the OB stars with line-of-sight velocities have been used to derive the velocity maps). This gives us confidence that the maps are self-consistent, and that the comparison between the spatial spiral arms and the corresponding streaming motions is appropriate.

The top-right panel of Figure \ref{fig:velomaps_OBstars} shows an alternating positive-negative pattern in $V_R$, the orientation of which is not aligned with the spiral arms in density. A prominent feature of stars with positive $V_R$ is apparent at approximately $x \approx 0$ kpc, crossing the map almost vertically and connecting the upper edge of the Local Arm with the lower side of the Sag-Car arm. The middle-right panel of Figure \ref{fig:velomaps_OBstars} shows the residual of $V_{\phi}$ with respect to the mean $\overline{V}_{\phi}(R)$. Based on this map,
the stars located just outside and inside the Local Arm move systematically slower than the mean $\overline{V_{\phi}}$ (yellow-green regions), while stars lying on the Local Arm appear to move systematically faster than $\overline{V_{\phi}}$ (red region). It should be noted, however, that the alignment of these azimuthal streaming motions with the density contours is not perfect in the lower part of the map. On the other hand, a striking alignment between the local arm and a systematically positive vertical velocities $V_z$ is shown in the bottom-right panel of Figure \ref{fig:velomaps_OBstars}, indicating that the OB stars exhibit both in-plane streaming motions as well as vertical bending waves, all with a relatively short radial wavelength. 

\begin{figure*}
\label{fig:vphiresmap}
\includegraphics[width=9cm,trim={0.65cm 0.3cm 0 0},clip]{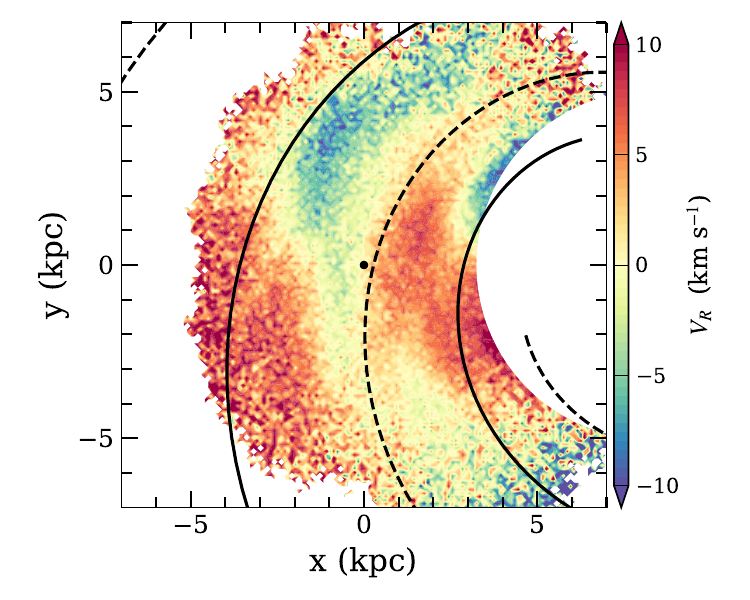}
\includegraphics[width=9cm,trim={0.65cm 0.3cm 0 0},clip]{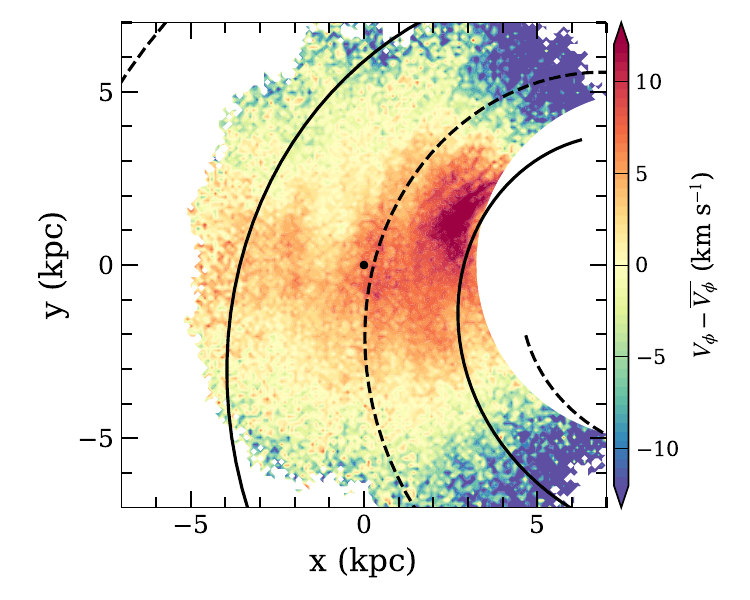}
\caption{ Velocity maps for the RGB sample. The left panel shows the inferred radial velocity, $V_R$. The right panel shows the map for $V_{\phi} - \overline{V_{\phi}}$, i.e., the residual azimuthal velocity. Overlaid on both maps is the 2-arm $NIR$ spiral model from \cite{Drimmel2000}, shown as solid lines. The dashed lines correspond to the minimum interarm density (see text). The position of the Sun is marked by a black dot.  \label{fig:giants_plus_NIR_model} } 
\end{figure*}

In contrast with the OB sample, the RGB sample is an older, dynamically relaxed stellar population extending much further from the Sun, allowing us to trace the outer disc kinematics over a much larger extent than the OB stars. In order to better highlight the observed features in the outer disc, we mask out the region $R<5$ kpc and
show the $V_R$ map and the $V_{\phi}$ residuals in \autoref{fig:giants_plus_NIR_model}. To compare the velocity maps of the RGB stars to the spatial position of the spiral arms in the NIR, we make a comparison with the 2-arm model from \cite{Drimmel2000}, shown as solid lines, while the $\pi/2$ phase shifted geometry, corresponding to the minimum interarm density, is shown as a dashed curve. While no clear spiral structure is evident in the RGB spatial distribution (see Section \ref{sec:confspace}), the large scale streaming motions in the radial velocities suggest some possible correspondence, especially along the inter-arm dashed curve. 

Also noteworthy, the $V_R$ map (left panel) shows a large positive $V_R$ feature in the third quadrant (between $180^\circ<l<270^\circ$), which becomes positive at about $l=170\degr$ between about $R = 9$ and 11 kpc. This feature is approximately aligned with an analogous change in the radial velocities in the inner bar region (see also the top-left panel of figure \ref{fig:velomapsgiants}), and is similar to a feature seen in simulations with a bar near the Outer Lindblad resonance. In contrast to the $V_R$ map, the $V_{\phi}$ residuals in the right panel of \autoref{fig:giants_plus_NIR_model} do not show any clear non-axisymmetric features in the outer disk. There is, however, a noticeable gradient in the $V_{\phi}$ residuals, about $y=0$, where the residuals are systematically lower for larger values of $|y|$. Being symmetric about the $x-$axis, this is likely a systematic in the velocity due to distance uncertainties, as noted in sec \ref{sec:sys_errors} (\autoref{fig:impact_errors}). In any case, we remind the reader that features near the edge of these maps may well be artefacts from oversampling sources with over-estimated distances, as also discussed in sec \ref{sec:sys_errors}.

\begin{figure*}
\includegraphics[width=0.7\columnwidth]{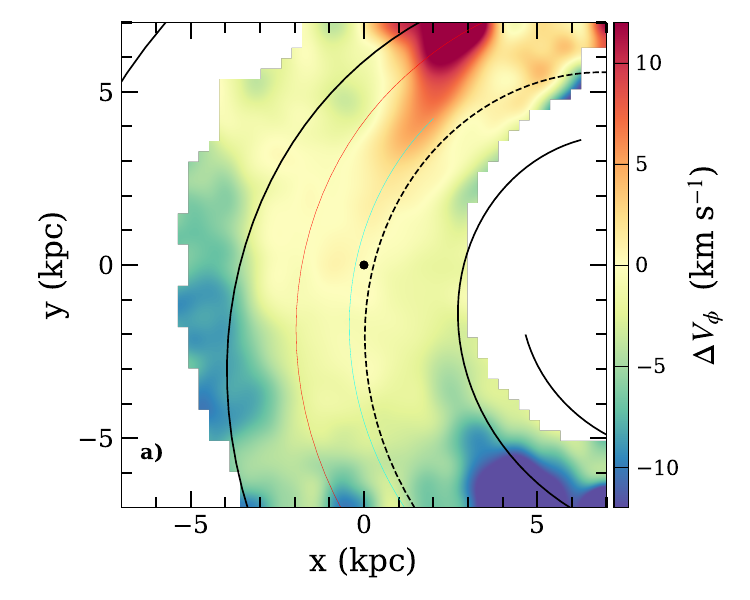}
\includegraphics[width=0.7\columnwidth]{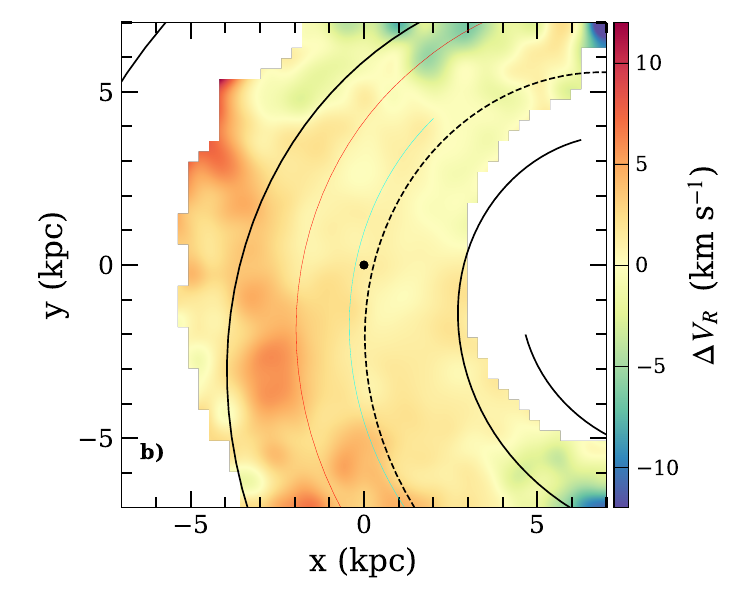}
\includegraphics[width=0.7\columnwidth]{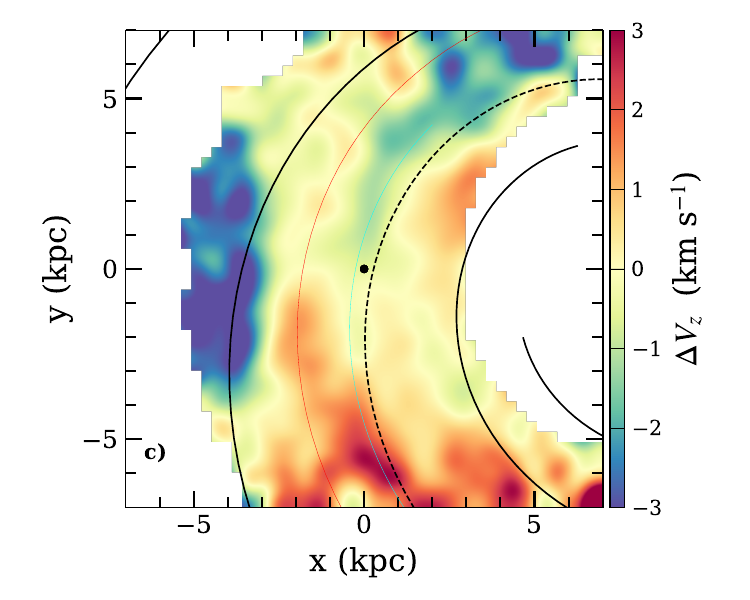}
\includegraphics[width=0.7\columnwidth]{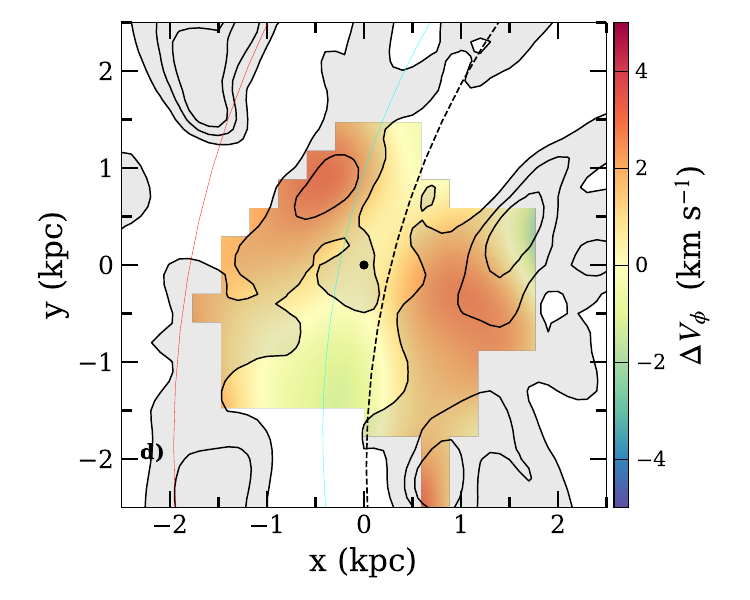}
\includegraphics[width=0.7\columnwidth]{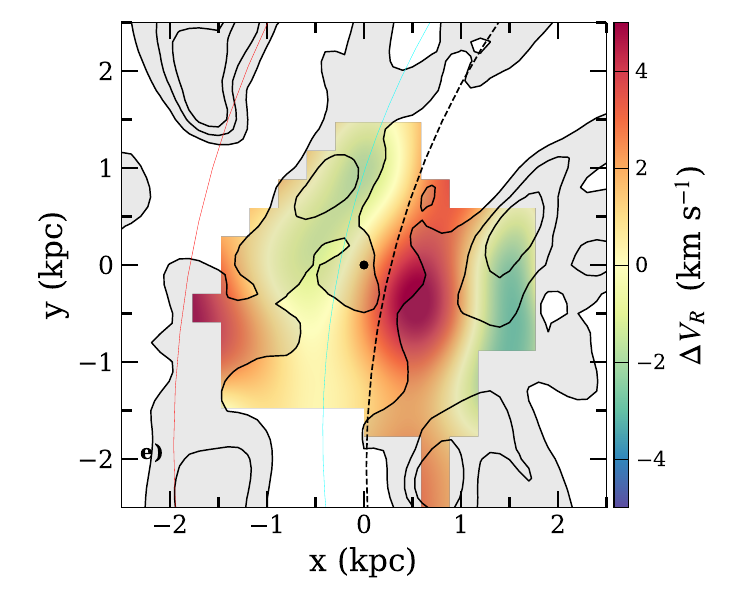}
\includegraphics[width=0.7\columnwidth]{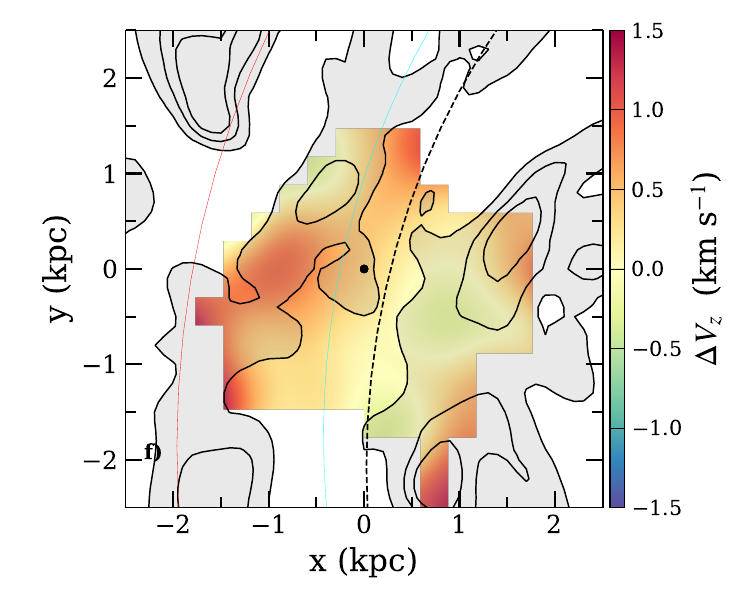}
%
\caption{Heliocentric maps of $\Delta V_{\rm i}$, i.e., ($V_{Z>0} - V_{Z<0}$), for the components s, $i ={\phi,R,Z}$, in the RGB sample (top), and the OB star sample (bottom). Also overplotted is the near infrared 2-arm spiral model (black) from \cite{Drimmel2000}, as well as the $Perseus$ (red), and the $local$ (cyan) arms model from \cite{Reid2019}. The position of the Sun is indicated by the black dot at $(x,y)=$(0,0). \label{fig:deltavi_giants}}
\end{figure*}

In addition to exploring non-axisymmetry in the disc, with the increase in spatio-kinematic coverage, we can also explore (a)symmetry about the plane of the Galaxy. In \autoref{fig:deltavi_giants} we present velocity difference maps between the $z>0$ \& $z<0$ hemispheres, for the RGB sample (top panels) and OB stars (bottom panels). In each of the components ($k ={R,\phi,Z}$) we make velocity maps for both the upper and the lower hemispheres, then take their difference (i.e. $\Delta V_k = V_{k,Z>0} - V_{k,Z<0}$).  As with the other velocity maps, we have also overplotted the 4-arm spiral model (black) based on near infrared (NIR) data from \cite{Drimmel2000}. Additionally, we plot the $Perseus$ (red), \textit{Sag-Car} (purple), and the Local (cyan) arms from \cite{Reid2019}. For the RGB stars, \autoref{fig:deltavi_giants}(a) shows that in $\Delta {V_{\phi}}$, within a heliocentric radius of about 3 kpc, there is no significant difference between the upper and the lower disc, while beyond $R>11$ kpc the disc is rotating faster below the disc plane ($z<0$) by up to 10\kms{}, in the third quadrant and in the anticentre direction. A similar feature is seen in the $\Delta {V_{R}}$ map, though here in the opposite sense, with more positive radial motion above the disk plane.  By comparing with \autoref{fig:giants_plus_NIR_model} (left panel), we see that this feature corresponds with the positive $V_R$ feature in the third quadrant already noted above. It becomes clear that the positive \vrad{} in this part of the outer disc is almost entirely located above the disc plane. It is interesting to note that, about the plane of the Galaxy, $\Delta {V_{R}} \sim0$\kms{} in the rest of the map. Finally, in \autoref{fig:deltavi_giants}(c), we show the $\Delta {V_{z}}$ map. This is equivalent to mapping breathing modes (contraction and expansion with respect to $z=0$) as in \cite{Widrow:2014}. The absolute amplitude of the $\Delta {V_{z}}$ is about 2-4 \kms{}, i.e., much lower than in the $\phi$ and $R$ components. Nevertheless, again the outer disc shows a marked feature, though more toward the anticentre. Also, we note that at least one of the features in the $\Delta {V_{z}}$ map roughly aligns with the location of the Perseus arm from \cite{Reid2019}. 
 
 The $\Delta V$ map for the OB stars cover a much smaller extent of the disk. Indeed we can only construct difference maps not much beyond 1 kpc from the disc, shown in the lower panels of \autoref{fig:deltavi_giants}, with the spiral overdensity contours overlaid as in \ref{fig:velomaps_OBstars}. In \autoref{fig:deltavi_giants}(d), we note that that residuals in $\Delta {V_{\phi}}$, are positive in between the arms. This would suggest that inside the arms, OB stars rotate faster in the upper disc. In \autoref{fig:deltavi_giants}(e), we note that $\Delta {V_{R}}$ is generally negative inside the arms, implying that the OB stars in the upper disc are moving towards the Galactic centre inside these arms. In \autoref{fig:velomaps_OBstars}, we already noted that for the OB population, $\Delta {V_{R}}$ is generally negative between the arms; the $\Delta {V_{R}}$ map for the OB stars shows that this is stronger in the upper disc. 
 Similarly, \autoref{fig:velomaps_OBstars} also showed that the $\Delta {V_{\phi}}$ residuals were also higher in the inter-arm region, and the $\Delta {V_{\phi}}$ map shows this is more prominent in the upper disc. 
 This suggests there may be some shearing motion in the $R$ and $\phi$ directions associated with these arms, or that the spiral perturbation that aligns with the OB population is stronger in the upper disc. 
 Finally, for completion, we also present the $\Delta {V_{z}}$ map for the OB stars, in \autoref{fig:deltavi_giants}(f), showing that there may be an associated breathing mode, though here we see a net expansion that may be aligned with the overdensity of the Local arm.  In general, the amplitude of the $\Delta V_{\rm k}$ for the OB stars is much lower compared to the RGB stars.


\section{Discussion}
\label{sec:discussion}
In section 4 we have summarised the distribution of our four samples in the $(x,y)$ plane. As expected, the younger samples indeed trace well-known spiral arm segments, confirming earlier results using \gedrthree\ astrometry and young stellar samples selected using NIR photometry \citep{Zari2021, Poggio:2021}.  Segments of the nearest spiral arms are evident, though we have made no attempt to derive the local stellar density. Doing so would require an accurate 3D map of extinction, as well as the selection function of the sample being used, which depends both on the survey selection function as well as the criteria used to construct the sample of tracers \citep{Rix2021}. For instance, for compiling our OB and RGB samples we have used the new astrophysical parameters in \gdrthree, which are only available for a fraction of the stars. As reported in \citep{DR3-DPACP-157, DR3-DPACP-160, DR3-DPACP-156}, about half of the brighter stars are missing astrophysical parameters, which leads to an artificial depression in the apparent stellar density within about 1\,kpc of the Sun. 

In contrast, as already discussed in section \ref{sec:coord_maps}, extinction is responsible for sampling bias with respect to angle, leading to the evident radial features centred on the Sun's position. 
The radial features so obviously visible in the distribution of our sources are only weakly visible in the velocity and dispersion maps in section \ref{sec:velomaps}, and are a consequence of biased distance estimates. The fact that they are not very evident gives us confidence that we are justified in assuming that our samples are not kinematically biased, and that the adopted distances are reliable. Nevertheless, as discussed in section \ref{sec:sys_errors}, beyond some limiting distance the majority of stars in a given volume element will have over-estimated distances, which will in turn lead to the mean velocity component tangential to the line-of-sight being over-estimated. This will cause a direction dependent distance limit to the velocity maps, beyond which systematic errors will dominate. Such systematic errors will show symmetries with respect to the Sun-galactic centre line. For example, such a signature is evident in the map of $V_\phi - \overline{V_\phi}$ (figure \ref{fig:giants_plus_NIR_model}), showing large negative azimuthal velocities for distances beyond about 6 kpc in the directions of $40\deg < |l| < 60\deg$. Keeping in mind the principles noted at the end of section \ref{sec:sys_errors}, we discuss further below the kinematic maps presented in section \ref{sec:velomaps}.

 
\subsection{Kinematics}


While the optical passband of \gaia\ limits its mapping capabilities in configuration space, it nevertheless samples the kinematics of the disc of the Milky Way to large distances. Here, as is the usual practice, we assume that our samples are not kinematically biased. This is a reasonable assumption as long as each sample either covers a limited range of ages, or can be considered as kinematically relaxed and of common origin. In particular, we have not made any effort to make a distinction between the young (thin) disc stars and the old $\alpha$-rich (thick) disc in our RGB sample \citep[][]{DR3-DPACP-104}. With these assumptions and caveats in mind, we here discuss further the astrophysical significance of some of the features that we have noted in the velocity maps in the previous section. 

The maps shown in Fig.~\ref{fig:giants_plus_NIR_model} can be directly compared to the corresponding velocity maps from \gdrtwo{} \cite[][their figures 19 and 20]{DR2-DPACP-33}. While many of the features out to about $3$ kpc were already mapped with \gdrtwo{}, thanks to the larger sample of stars with line-of-sight velocities, we can now map a significantly larger portion of the disc. The new data reveals a complex and rich velocity field in the Galactic disc all the way to the Galactic centre and the in-plane motions of the stars in the inner Galaxy spectacularly reveal the clear signature of the Galactic bar. In particular, we confirm the existence of a quadrupole pattern in the radial velocity field, as previously found by \citet{Bovy2019} and \citet{Queiroz2021}. Additionally, we now clearly see the imprints of the Galactic bar in the azimuthal velocity field, and in the stellar velocity dispersions. With the additional guidance provided from of a numerical simulation, we are able to estimate the aspect angle of the bar and its corotation radius through a simple bisymmetric model of the kinematic signature of the bar in the azimuthal velocity field. We find that the bar has an angle of $19.5 \pm 2.5\degr$ with respect to the Sun-GC direction, as measured within $1.5 < R \le 2.8$ kpc, and a corotation radius of $5.4\pm 0.5$ kpc. Then, using the observed angular frequency, we deduce a lower limit of the bar pattern frequency of $\Omega_{\rm bar}=38.1^{+2.6}_{-2}$\,\kms kpc$^{-1}$, and an upper limit for the location of the outer Lindblad resonance of $9.7 \pm 0.5$\,kpc  within the framework of the epicycle approximation. \citet{JnO2016} compiled previous estimates of the corotation radius and gave $\Omega_{\rm bar}=43 \pm 9$\,\kms kpc$^{-1}$,  and a  corotation radius range of $4.5-7$ kpc. Applying the Tremaine-Weinberg method to their kinematic sample, \citet{Bovy2019} estimated a bar pattern speed of $41\pm3$\,\kms kpc$^{-1}$ and a corotation radius of $5.5\pm0.4$ kpc. Other works found $37.5$\,\kms  kpc$^{-1}$ \citep{Clarke:2019}, $39\pm 3.5$\,\kms kpc$^{-1}$ \citep{dynPortail:2017} and $37.5-40$\,\kms kpc$^{-1}$ \citep{Li2022}, which is based on hydrodynamical simulations and made-to-measure models. Our findings are thus in good agreement with other estimates of the bar's fundamental parameters. As for the bar orientation angle,  \citet{JnO2016} compiled a range of $28\degr-33\degr$, while other recent estimates range from $20\degr$ to $28\degr$ \citep[e.g.][]{Wegg:2015,Portail:2017,Bovy2019,Queiroz2021}. Our value is thus at the low end of the proposed range. 



The outer disk has a large variety of features and streaming motions, which likely contain fundamental clues on the dynamical nature of the non-axisymmetric structures in the Milky Way. \citet{Monari:2016} modelled the impact of the bar and a two-armed quasi-static spiral pattern in the Galactic disc, both simultaneously and separately. The comparison of the observed kinematic maps of the RGB stars confirms that the influence of the Galactic bar likely dominates even the outer portions of the disc out to at least $R=11$kpc.  \citet{Faure:2014} carried out test particle simulations to predict the global stellar response to spiral perturbations in the Galactic disc, in the absence of an external excitation (such as due to an accreting satellite). They integrate stellar orbits in a 2-arm \textit{Lin-Shu} type spiral potential (without a bar) and produce maps of mean galactocentric radial velocity (\vrad). They show (their figure 6) that inside the spiral arms' corotation, in the region traced by the arm, the mean \vrad\ is negative (of the order of -7 \kms{}), i.e., stars have a bulk motion towards the Galactic centre. Meanwhile, in the region between the arms, the stellar radial motion is positive, i.e., stars have a bulk motion towards the anticentre. Outside corotation, the pattern is reversed. Trends in $V_R$ induced by the spiral arms are also shown in \cite{Antoja:2016} and other works. In our sample of RGB stars, a spiral feature might be visible in the $V_R$ maps, apparent as an elongated systematic positive feature, located internally to the dashed line in Fig. \ref{fig:giants_plus_NIR_model} (left panel), marking the minimum interarm density.


Other features are also seen in the $\Delta V$ maps that are worthy of note. Some of them might be explained as extinction artefacts (i.e. those that are elongated along the line-of-sight), while others are likely real. In particular, we see an asymmetry in all three components in the outer disk beyond $R=10-11$ kpc in the RGB sample that does not as yet have a clear explanation. This feature is likely related to an asymmetry already noted in \citet{DR2-DPACP-33}, where the authors follow the asymmetry in the azimuthal and vertical components,  (\vtan\ ,\vz)  plane, showing a clear bimodality of stars. Stars concentrate mainly in two clumps, one with negative \vz\ at lower \vtan\ which is more prominent in the north, and one with positive \vz\ at higher \vtan\ more visible in the south. The different proportions of the clumps of the bimodality at different Z seems to be the cause of the asymmetry.

For the OB sample, one of the main results is that the velocity field of the OB stars shows streaming motions that have a characteristic length similar to the spiral arm density. This has never been shown with such detail in 2-dimensional maps for OB stars. One should nevertheless bear in mind that our sample of OB stars is expected to trace the local (out to $\approx 2 - 2.5$ kpc in heliocentric distance) motions of the gas, rather than the large-scale features of the Galaxy. Indeed, OB stars are typically young, and therefore are expected to inherit the motion of the gas from which they were recently born. However, due to the relatively small region sampled by our OB sample, the streaming motions in the OB stars really map only the Local Arm. 

A comparison between the OB and RGB stars' velocity fields reveals numerous differences. 
The observed differences are not unexpected, since these two samples trace, respectively, dynamically cold and hot stellar populations of the Milky Way. In the OB stars we clearly see streaming motions that are associated with the spiral structure of this population.  In contrast, the signature of the spiral arms in the RGB sample is not clearly evident and, if present, is seen in the radial motions. In any case, if there is a signature of streaming motions related to the spiral arms in the RGB sample it is consistent with a two-armed structure, possibly driven by the bar.


\subsection{Caveats and shortcomings}

No method or procedure is perfect, and here we have adopted one that is not without imperfections. In this section we confess all of our sins. 

In section 5 we construct maps of the mean velocity field in galactocentric cylindrical coordinates based on the derived velocities in the same coordinates of the individual sources. These in turn were derived, in part, from the individual measures of the proper motions and estimated distances. While our methodology is relatively straight-forward it has several shortcomings which should be addressed in the future. Beyond a distance of about three kiloparsecs, the largest source of uncertainty in our velocity maps comes from the distance uncertainties, yet our treatment of these is not completely satisfactory.

Our distances are taken from CBJ2021, which uses a Bayesian approach to estimate the distances from both astrometry and photometry, incorporating a prior that includes both current astrophysical knowledge of stellar structure, as well as some informed assumptions about Galactic structure. 
The probability distribution of the individual distances is in general asymmetric, but we have rendered them symmetric in order to apply a traditional approach to propagating the uncertainties to galactocentric coordinates that implicitly assume that the uncertainties are symmetric. As discussed in sec 3.3, this is in part done out of necessity as we are not able to sample the probability distribution function distance for each star. 

We have assumed that there is no correlation between the distance and proper motion uncertainties, which cannot be true in as much as these distances are informed by the parallax. On the other hand, as pointed out in section 3, these correlations are likely to be unimportant at larger distances where the distances are more constrained by the photometry than the astrometry.

When propagating our uncertainties we did take into account the correlation between the proper motion components, which would result in a correlation between the two velocity components in galactocentric coordinates. However, we did not take these correlations into account when estimating the mean velocity field in section 5. 

Most grievously, when estimating the mean velocity and velocity dispersion for a given volume element, we only consider stars found in that small volume element, according to their estimated distances.  That is, though we've taken into account the uncertainties of the distances when deriving the uncertainties in the individual velocities, we've implicitly assumed that that these distances are perfect when binning the stars into cells for estimating the mean velocity field and velocity dispersions. 

Finally, it should be remembered that the CBJ2021 distance estimates are based on a prior that includes a bar with an assumed geometry. As this prior becomes increasingly important with increasing distance, there is reason for concern that our velocity field may be influenced by this prior. Apart from the CBJ2021 photogeometric distances, we have attempted using different distance estimates: CBJ2021 geometric distances show very similar results, Starhorse3 \citep{Anders2022} distances also show a similar bisymmetric behaviour, while \gspphot\ distances have a similar trend but at a wrong distance. As mentioned above, all these distance estimators have a prior that includes a bar.
In any case, we stress that this potentially introduces a bias in the derived velocity field that is not considered in our qualitative comparison with simulated data.


\section{Conclusions}
\label{sec:concl}

Using \gdrthree\ we have mapped the kinematics of the stars over an extensive area of the Milky Way's disc. This has been made possible thanks to the new line-of-sight (radial) velocities, as well as new and reliable astrophysical parameters that have allowed us to differentiate between young (OB) and old (RGB) stars. While our sample of OB stars with 6D space and velocity data is much more limited, their kinematics are seen to be distinctly different than the RGB stars, likely reflecting the complex motions of the gas from which they were recently born. Our RGB sample has allowed us to map the kinematics of the Galaxy over nearly a quarter of the disc, providing us a first clear picture of the large-scale kinematic signature of the bar. In order to interpret the features seen in the kinematic maps we have relied on simple comparisons with a simulation of a barred galaxy.  

In contrast to the bar signature, clear evidence of streaming motions associated with spiral arms is much less evident and, if present, is consistent with the two-armed structure that we see in the near-infrared. 
With regards to the non-axisymmetric structure in 3D configuration space, we have made no attempt to reconstruct the stellar density. Nevertheless, we are able to map the young OB sample and young clusters out to 4 to 5 kiloparsecs from the Sun, confirming previous findings that the Local arm has a length of at least 8kpc. While weaker than the inner Sag-Car arm or the outer Perseus arm, it seems its nomenclature is perhaps not so appropriate, and that we should return to its original designation as the Orion arm \citep{vandHulst1954}. 

This study should only be considered as a preliminary exploration of what \gdrthree\ has to offer with regards to galactic structure. Our kinematic maps are derived from velocities of individual stars based on individual distance estimates. A more ideal approach would be a derivation of the velocity field recognising that the velocities, like the distances, should be inferred quantities rather than derived quantities as we have treated them here. Indeed, with regards to galactic dynamics, the individual velocities are not of direct interest to us at all. Instead, an appropriate model of the mean velocity field should be adopted and incorporated in the prior, then the relevant parameters of the model (like the bar orientation) adjusted to arrive at the most likely set of parameter values given the data. Alternatively, one could generate a suite of simulations of the kinematics of the Milky Way, assuming different parameters, model the uncertainties to transform these to a suite of mock catalogues, then find the one that best matches the \gaia\ data. In either case, any "fitting" should ideally be done against the measurements,  in data-space rather than in model-space, taking into account the selection effects on the data.  

Though we have only taken a first look at the treasure of data that \gdrthree\ has to offer, it is clear that the \gaia\ mission continues to fulfil its promise to provide the information needed to eventually arrive at a complete understanding of the dynamical state and processes at work in shaping our Galaxy. Certainly we can expect further discoveries and insights from the community in the future as it digests this latest census of the stars in our Milky Way.

\begin{acknowledgements}
RD would like to thank Robert Benjamin for helpful discussions on the treacherous history of mapping the spiral structure of our Galaxy.  
This work presents results from the European Space Agency (ESA) space mission \gaia. \gaia\ data are being processed by the \gaia\ Data Processing and Analysis Consortium (DPAC). Funding for the DPAC is provided by national institutions, in particular the institutions participating in the \gaia\ MultiLateral Agreement (MLA). The \gaia\ mission website is \url{https://www.cosmos.esa.int/gaia}. The \gaia\ archive website is \url{https://archives.esac.esa.int/gaia}.
Full acknowledgements are given in Appendix \ref{appendixAck}.

This work has used the following software products:
\href{http://www.starlink.ac.uk/topcat/}{TOPCAT}, \href{http://www.starlink.ac.uk/stil}{STIL}, and \href{http://www.starlink.ac.uk/stilts}{STILTS} \citep{2005ASPC..347...29T,2006ASPC..351..666T};
Matplotlib \citep{Hunter:2007};
IPython \citep{PER-GRA:2007};  
Astropy, a community-developed core Python package for Astronomy \citep{2018AJ....156..123A}.

\end{acknowledgements}

   \bibliographystyle{aa} 
   \bibliography{mybib,dpac} 
%


\begin{appendix}

\section{Queries used to select samples}
\label{sec:queries}
For the OB sample, the following query was used: 
\begin{lstlisting}[language=sql]
SELECT g.*, ap.*
FROM gaiadr3.gaia_source AS g
INNER JOIN gaiadr3.astrophysical_parameters AS ap
ON g.source_id = ap.source_id
WHERE ((ap.teff_gspphot > 10000 AND (ap.spectraltype_esphs = 'O' OR ap.spectraltype_esphs = 'B' OR ap.spectraltype_esphs = 'A') AND ap.teff_esphs IS NULL) OR (ap.teff_esphs > 10000 AND ap.teff_gspphot > 8000) OR (ap.teff_esphs > 10000 AND ap.teff_esphs < 50000 AND ap.teff_gspphot IS NULL))
AND power(g.parallax/100.,5) < power(10.,2.-g.phot_g_mean_mag+1.8*g.bp_rp)
\end{lstlisting}

For the RGB sample, the following query was used:
\begin{lstlisting}[language=sql]
SELECT g.*
FROM gaiadr3.gaia_source AS g
WHERE (g.teff_gspphot<5500 and g.teff_gspphot>3000) and (g.logg_gspphot<3.)
\end{lstlisting}
Subsequent cross-matches to the tables containing the CBJ2021 distances and the astrometric fidelity flag were done after saving the results of the above query as a user table. The final selection on $|Z|<300$\,pc and $|Z|<1$\,kpc for the OB sample and the RGB samples, respectively, was done during the creation of the maps in Section~\ref{sec:coord_maps} and \ref{sec:velomaps}. 

As written, these queries will pull all columns from the tables \linktotable{gaia_source} and \linktotable{astrophysical_parameters}. To save space the user is recommended to pull only the columns of interest. 

\section{Maps of velocity uncertainties}
\label{sec:velo_uncertainties}
\subsection{Velocity uncertainties with Fisher formalism}
\label{appendix:derive-Vk-uncertainty}

We briefly explain how we estimate the uncertainties of the velocity $V_k$ from the negative log-likelihood given in Eq.~(\ref{eq:negative-loglik}): First, we minimise the negative log-likelihood in order to obtain estimates of the mean velocity $V_k$ and the dispersion $\sigma_k^*$. Second, we employ the Fisher formalism by computing the Hessian matrix of second derivatives of Eq.~(\ref{eq:negative-loglik}) w.r.t.~the two fit parameters $V_k$ and $\sigma_k^*$ and the covariance matrix of the two fit parameters is then given by the inverse of the Hessian.

We compute the first derivatives of Eq.~(\ref{eq:negative-loglik}) w.r.t.~$V_k$ and ${\sigma_k^*}$, respectively, equate to zero and try to solve:
\begin{equation}\label{eq:solution-V_k}
\frac{\partial\mathcal{L}}{\partial V_k} =  -\sum^{N_*}_j\frac{v_{k,j}-V_k}{{\sigma_k^*}^2 + \sigma_{v_{k,j}}^2}=0
\quad\Leftrightarrow\quad
V_k = \frac{\sum^{N_*}_j\frac{v_{k,j}}{{\sigma_k^*}^2 + \sigma_{v_{k,j}}^2}}{\sum^{N_*}_j\frac{1}{{\sigma_k^*}^2 + \sigma_{v_{k,j}}^2}}
\end{equation}
\begin{equation}
\frac{\partial\mathcal{L}}{\partial{\sigma_k^*}} =  \frac{1}{2}\sum^{N_*}_j\left[\frac{2{\sigma_k^*}}{{\sigma_k^*}^2 + \sigma_{v_{k,j}}^2} - 2{\sigma_k^*}\frac{(v_{k,j}-V_k)^2}{({\sigma_k^*}^2 + {\sigma_k^*}_{v_{k,j}}^2)^2} \right]=0
\end{equation}
Since ${\sigma_k^*}=0$ is not an allowed solution because the dispersion has to be strictly positive, we can cancel out from this and obtain:
\begin{equation}\label{eq-help-1}
\sum^{N_*}_j\frac{1}{{\sigma_k^*}^2 + \sigma_{v_{k,j}}^2} = \sum^{N_*}_j\frac{(v_{k,j}-V_k)^2}{({\sigma_k^*}^2 + {\sigma_k^*}_{v_{k,j}}^2)^2}
\end{equation}
This cannot be solved analytically for ${\sigma_k^*}$ and we employ a numerical solution instead.

Next, we need to compute the Hessian matrix of second derivatives of $\mathcal L$:
\begin{equation}
\frac{\partial^2\mathcal{L}}{\partial V_k^2} =  \sum^{N_*}_j\frac{1}{{\sigma_k^*}^2 + \sigma_{v_{k,j}}^2}
\end{equation}
\begin{equation}
\frac{\partial^2\mathcal{L}}{\partial V_k\partial{\sigma_k^*}} =  2{\sigma_k^*}\sum^{N_*}_j\frac{v_{k,j}-V_k}{({\sigma_k^*}^2 + {\sigma_k^*}_{v_{k,j}}^2)^2}
\end{equation}
\begin{equation}
\frac{\partial^2\mathcal{L}}{\partial{\sigma_k^*}^2} = \sum^{N_*}_j\left[
4{\sigma_k^*}^2\frac{(v_{k,j}-V_k)^2}{({\sigma_k^*}^2 + {\sigma_k^*}_{v_{k,j}}^2)^3}
- \frac{2{\sigma_k^*}^2}{({\sigma_k^*}^2 + {\sigma_k^*}_{v_{k,j}}^2)^2} 
\right]
\end{equation}
The last equation has already been simplified by using Eq.~(\ref{eq-help-1}). We now write all the second derivatives into the Hessian matrix:
\begin{equation}\label{eq:Hessian}
H = \left(\begin{array}{cc}
\frac{\partial^2\mathcal{L}}{\partial V_k^2} & \frac{\partial^2\mathcal{L}}{\partial V_k\partial{\sigma_k^*}} \\
\frac{\partial^2\mathcal{L}}{\partial V_k\partial{\sigma_k^*}} & \frac{\partial^2\mathcal{L}}{\partial{\sigma_k^*}^2} 
\end{array}\right)
\end{equation}
Given the numerical best-fit solutions for $V_k$ and ${\sigma_k^*}$, the Hessian $H$ is just a 2$\times$2 matrix of numerical values that we can invert. According to the Fisher information criterion, the covariance matrix of the fit parameters is equal to the negative of the inverse Hessian. Since Eq.~(\ref{eq:negative-loglik}) is the \textit{negative} log-likelihood, the covariance matrix is the inverse Hessian \textit{without} the minus sign:
\begin{equation}
C = H^{-1}
\end{equation}
The standard deviation on $V_k$ is then simply given by $\sqrt{C_{00}}$. Note that we cannot give an analytic expression for it because we cannot solve Eq.~(\ref{eq-help-1}) analytically for ${\sigma_k^*}$. Yet, we can compute $\sqrt{C_{00}}$ from the numerical solution for ${\sigma_k^*}$.

\subsection{Solution for ${\sigma_k^*}\gg\sigma_{v_{k,j}}$}
\label{appendix:approximate-Vk-uncertainty}

The dispersion ${\sigma_k^*}$ may be much larger than all individual measurement errors $\sigma_{v_{k,j}}$. In that case, we can approximate ${\sigma_k^*}^2+\sigma_{v_{k,j}}^2\approx {\sigma_k^*}^2$ such that Eq.~(\ref{eq:solution-V_k}) and Eq.~(\ref{eq-help-1}) simplify to:
\begin{equation}
V_k = \frac{1}{N_*}\sum^{N_*}_j v_{k,j}
\quad\textrm{and}\quad
{\sigma_k^*}^2 = \frac{1}{N_*}\sum^{N_*}_j(v_{k,j}-V_k)^2
\end{equation}
In this specific case, there now is an analytic solution for $\sigma_k^*$.
The Hessian from Eq.~(\ref{eq:Hessian}) reduces to:
\begin{equation}
H 
 = \left(\begin{array}{cc}
\frac{N_*}{{\sigma_k^*}^2} & \frac{2}{{\sigma_k^*}^3}\sum^{N_*}_j(v_{k,j}-V_k) \\
\frac{2}{{\sigma_k^*}^3}\sum^{N_*}_j(v_{k,j}-V_k) & 
4\frac{\sum^{N_*}_j(v_{k,j}-V_k)^2}{{\sigma_k^*}^4}
- \frac{2N_*}{{\sigma_k^*}^2} 
\end{array}\right)
\end{equation}
Given $V_k = \frac{1}{N_*}\sum^{N_*}_j v_{k,j}$ we have $\sum^{N_*}_j(v_{k,j}-V_k)=0$ such that the off-diagonal elements of $H$ disappear and we are left with:
\begin{equation}
H 
 = \left(\begin{array}{cc}
\frac{N_*}{{\sigma_k^*}^2} & 0 \\
0 & 
4\frac{\sum^{N_*}_j(v_{k,j}-V_k)^2}{{\sigma_k^*}^4}
- \frac{2N_*}{{\sigma_k^*}^2} 
\end{array}\right)
 = \left(\begin{array}{cc}
\frac{N_*}{{\sigma_k^*}^2} & 0 \\
0 & 
\frac{2N_*}{{\sigma_k^*}^2}
\end{array}\right)
\end{equation}
Since this Hessian has format 2$\times$2, we can easily invert it by hand to obtain the covariance matrix $C$. In fact, we are only interested in the first element of the diagonal of $C$:
\begin{equation}
C_{00}  = \frac{H_{11}}{H_{00}H_{11} - H_{01}H_{10}}
= \frac{\frac{2N_*}{{\sigma_k^*}^2}}{\frac{N_*}{{\sigma_k^*}^2}\frac{2N_*}{{\sigma_k^*}^2} - 0}
=\frac{{\sigma_k^*}^2}{N_*}
\end{equation}
Consequently, the uncertainty on the mean velocity $V_k$ is approximately $\frac{{\sigma_k^*}}{\sqrt{N_*}}$ if the dispersion satisfies ${\sigma_k^*}\gg\sigma_{v_{k,j}}$.

\begin{figure*}
\centering
\includegraphics[height=7cm,trim={0.75cm 1cm 0.1cm 0.3cm},clip]{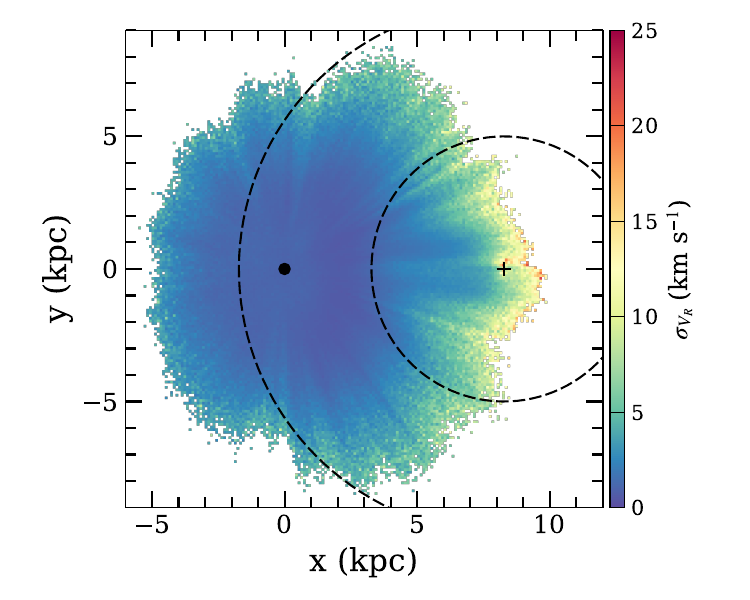}\includegraphics[height=7cm,trim={0.75cm 1cm 0 0.4cm},clip]{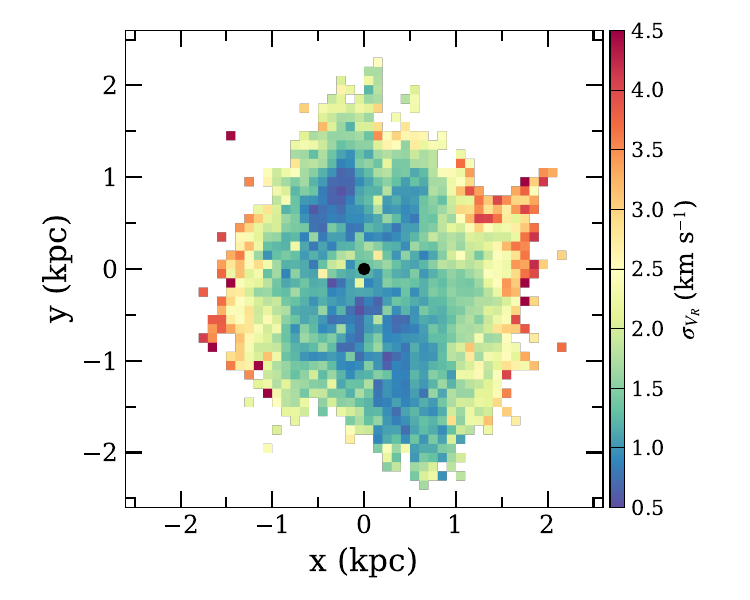}
\includegraphics[height=7cm,trim={0.75cm 1cm 0.2cm 0.3cm},clip]{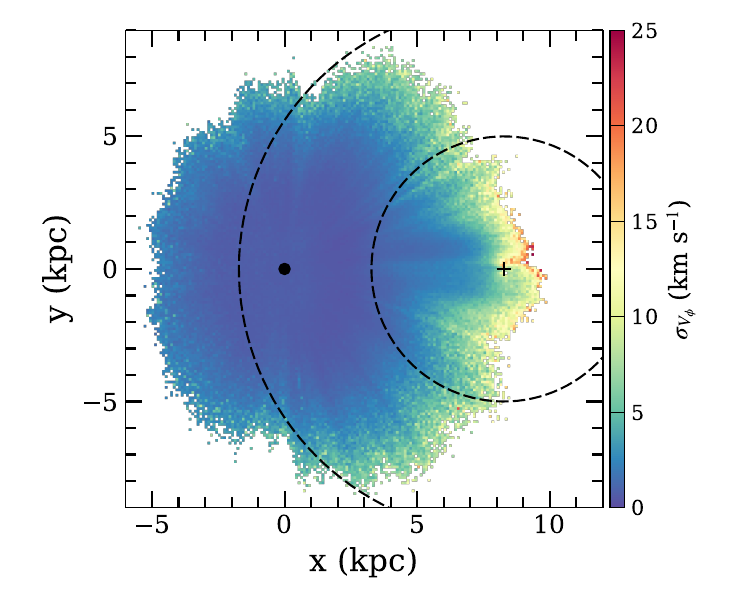}\includegraphics[height=7cm,trim={0.65cm 1cm 0 0.4cm},clip]{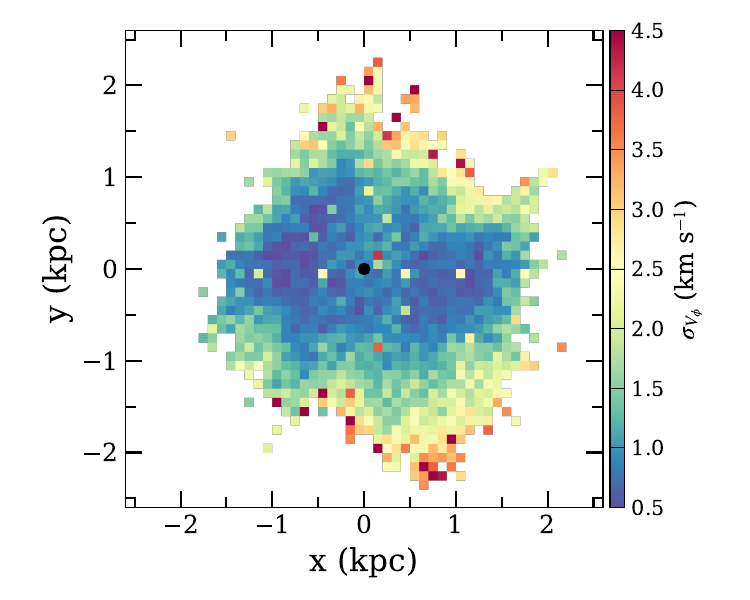}
\includegraphics[height=7.6cm,trim={0.75cm 0.3cm 0.25cm 0.3cm},clip]{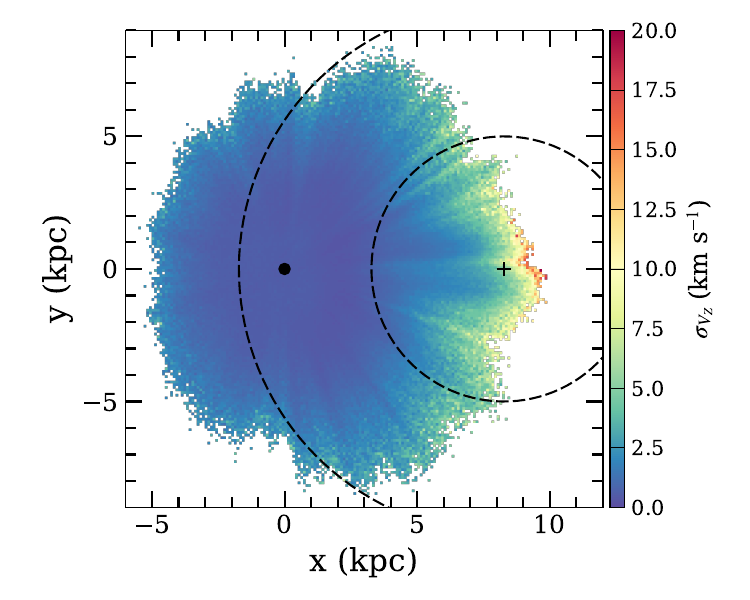}\includegraphics[height=7.6cm,trim={0.65cm 0.3cm 0.05cm 0.4cm},clip]{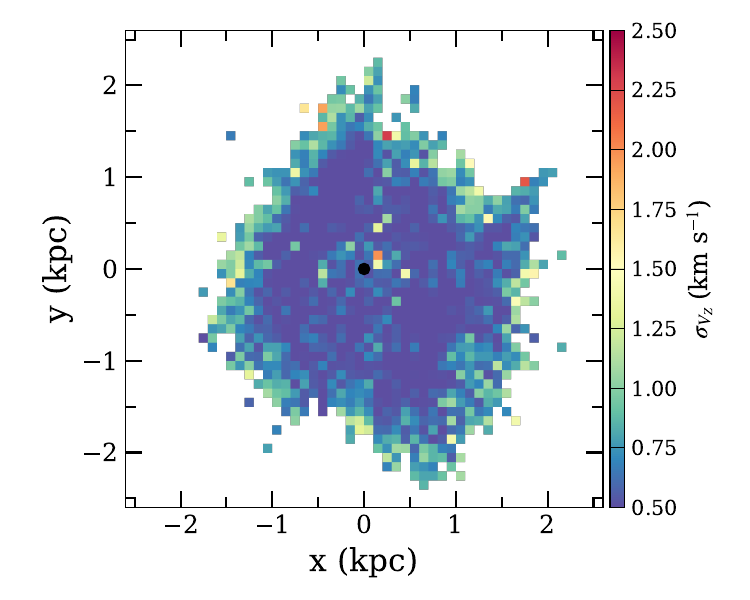}
\caption{Velocity uncertainty maps for Milky Way RGB giants (left column) and OB stars (right column). From top to bottom, the radial, azimuthal, and vertical uncertainties are shown. The position of the Sun is marked by a filled circle, and that of the Galactic Centre by a plus sign.   The velocity ranges have been chosen to enhance contrasts. These uncertainty maps can be found in CDS in FITS format.
\label{fig:uncertaintyvelomaps}
}
\end{figure*}
\FloatBarrier

\section{Test particle simulations and \gaia\, mock catalogues}
\label{sec:appsimu}
We use test particle simulations to guide the reader interpret the velocity maps shown by the RGB sample in the inner regions and no dynamical modelling has been undertaken. The initial conditions, the Galactic potential and the steps performed in the integration process are described in \citet{RomeroGomez2015}. Here we make a short summary. 

Initial conditions for positions and velocities are drawn for a disc density distribution following a Miyamoto-Nagai disc potential \citep{Miyamoto1975} with a typical scale-height ($h_z=300$\,pc) and radial velocity dispersion ($\sigma_{U}=30.3\,$\kms) of a red clump star. We first integrate the initial conditions in the axisymmetric potential of \citet{Allen1991} for $10\,$Gyr, we introduce the Galactic bar potential adiabatically during $4$ bar rotations and we integrate another $4$ bar rotations so that the particles get in statistical equilibrium with the final bar potential. The Galactic bar consists of the superposition of two aligned Ferrers ellipsoids \citep{Ferrers1877}, one modelling the triaxial bulge with semi-major axis of $3.13\,$kpc and the second modelling the long thin bar with semi-major axis of $4.5\,$kpc. The bar rotates as a rigid body with a constant pattern speed of $45\,$\kms kpc$^{-1}$, placing corotation and the Outer Lindblad Resonances at $4.7\,$kpc and $8.4\,$kpc, respectively.


In order to mimic the RGB sample, we generate two sets. One assuming for each particle an absolute magnitude of $M_K=-1.61$\,mag \citep{Alves2000} and no dispersion, and another one with an absolute magnitude of $M_K=-3.0$\,mag and a dispersion of $1$\,mag, corresponding to the Red Clump and the bright Red Giant Branch, respectively. We assign an observed colour typical of a red clump star \citep{Alves2000}. We balance the particles in each set so that it mimics the RGB proportion of Red Clump stars and brighter sources and the apparent magnitude and spatial distribution. Using absorption \citep{Drimmel2003} and extinction models \citep{Cardelli1989}, we obtain the \gaia\, G magnitude. We follow the prescriptions described in the \gaia\ Science performance webpage to obtain the astrometric and spectroscopic uncertainties. In this case, since we use ``photogeo'' distances, we adopt a constant relative uncertainty in distance of $15\%$, but we also made tests with a relative uncertainty of $10\%$ and $20\%$. 

\begin{figure*}
    \includegraphics[width=0.49\textwidth]{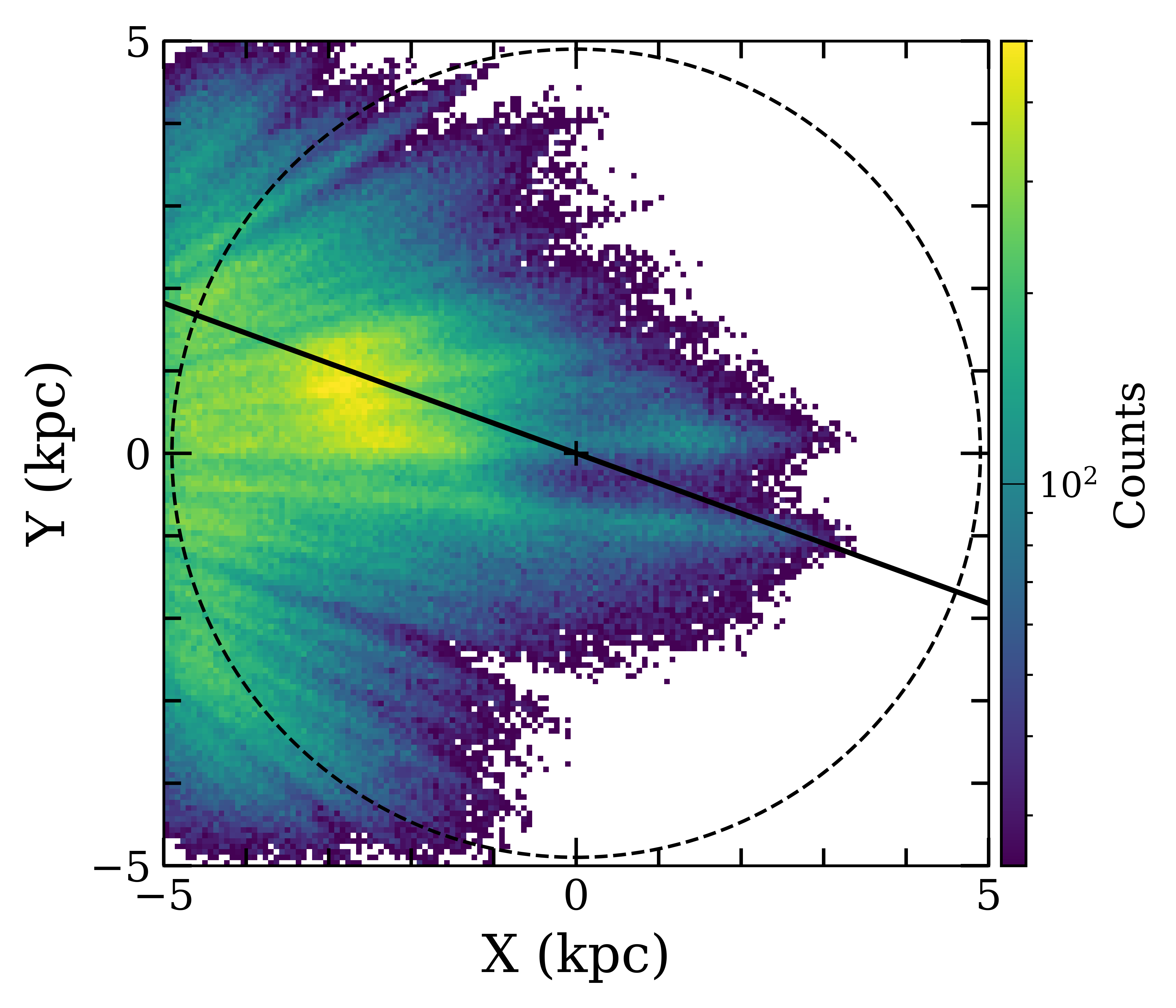}
    \includegraphics[width=0.49\textwidth]{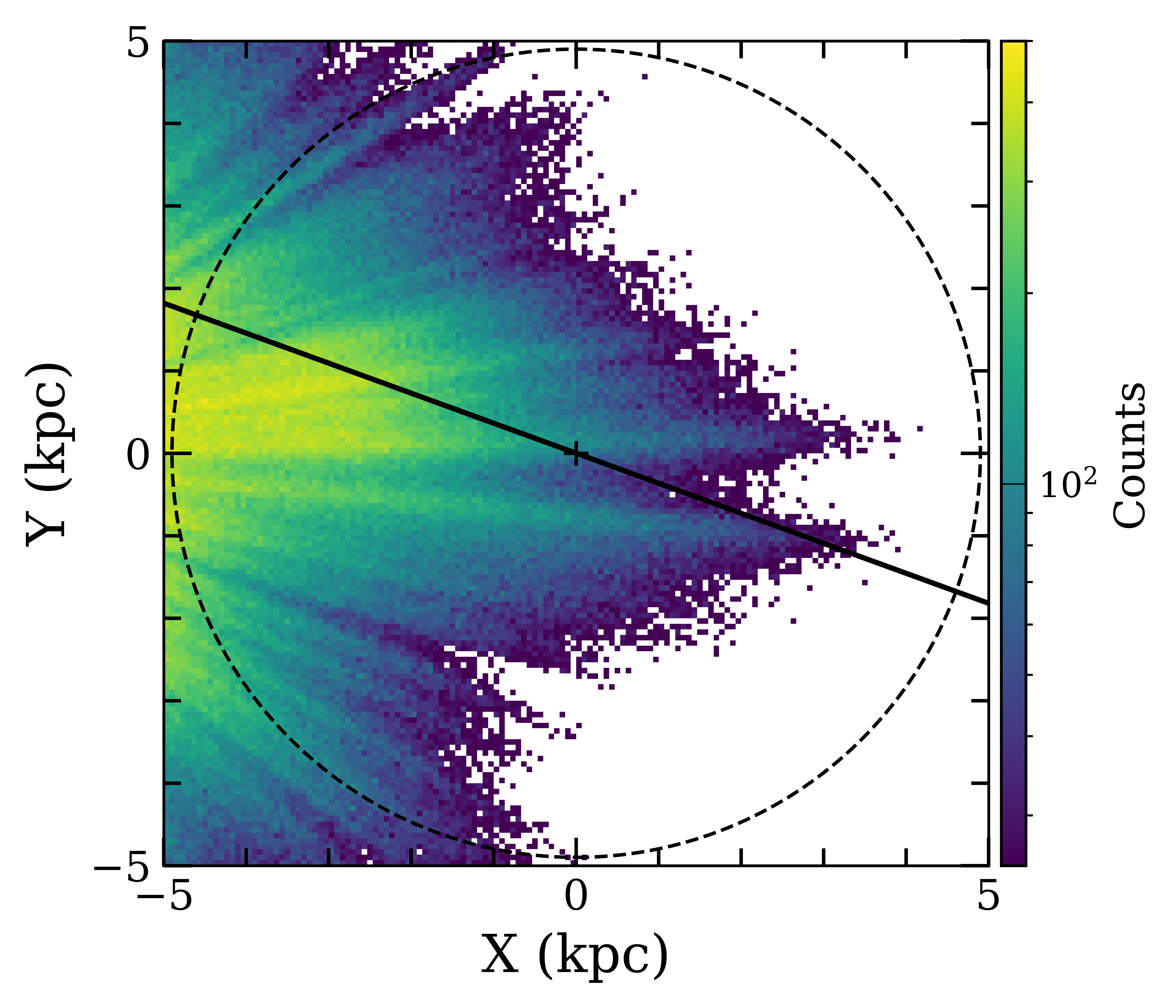}\\
    \includegraphics[width=0.49\textwidth]{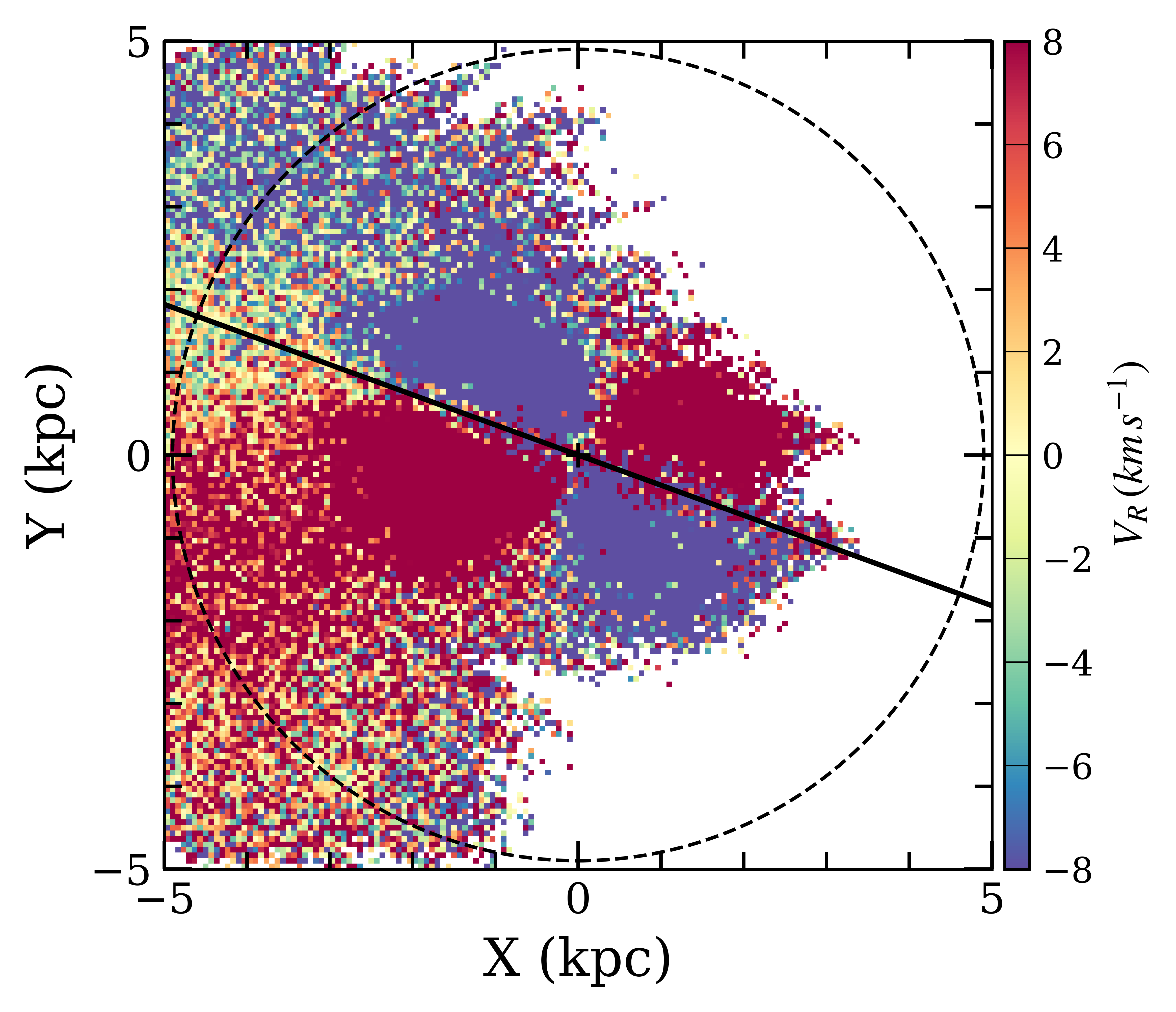}
    \includegraphics[width=0.49\textwidth]{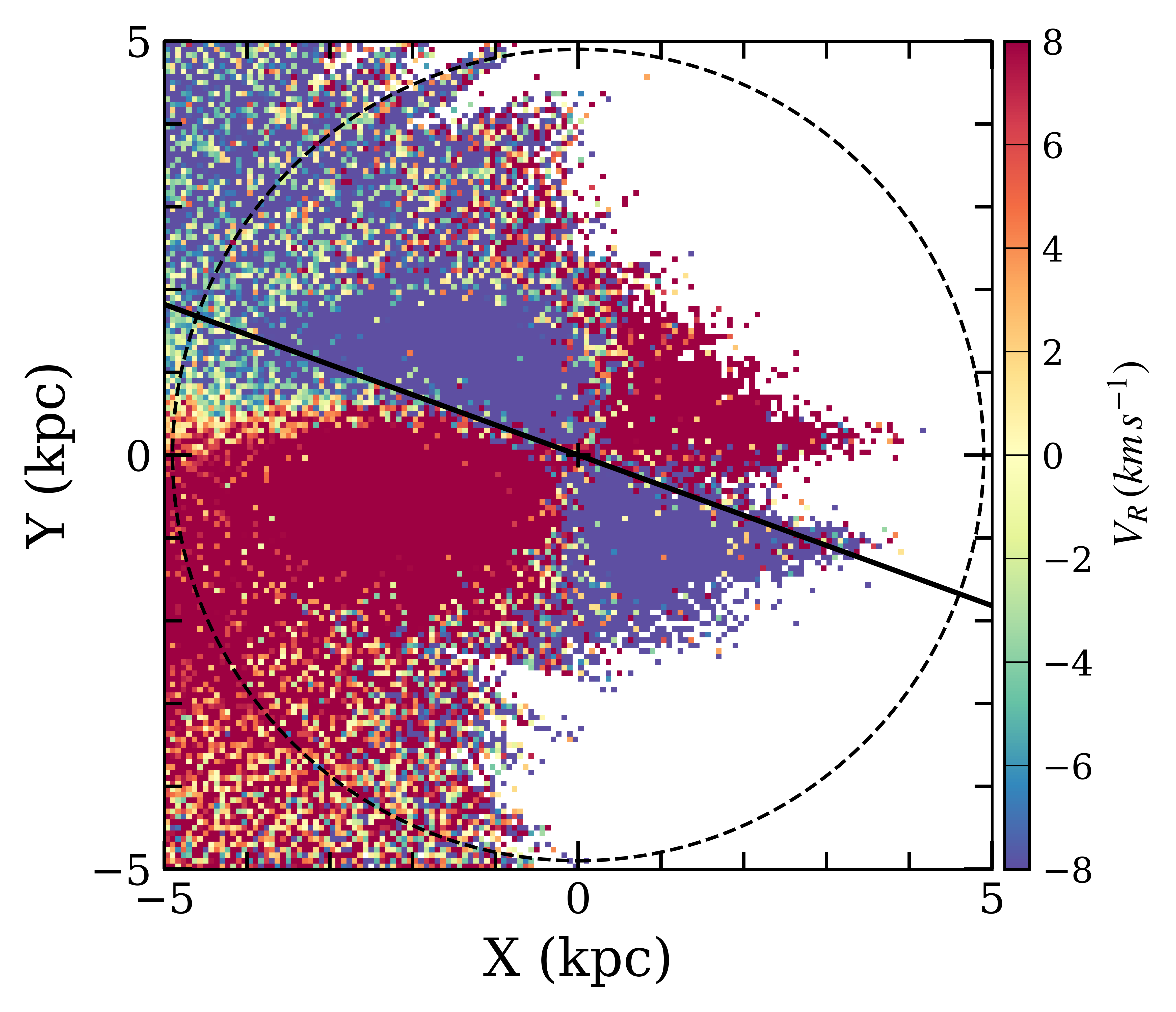}\\
    \includegraphics[width=0.49\textwidth]{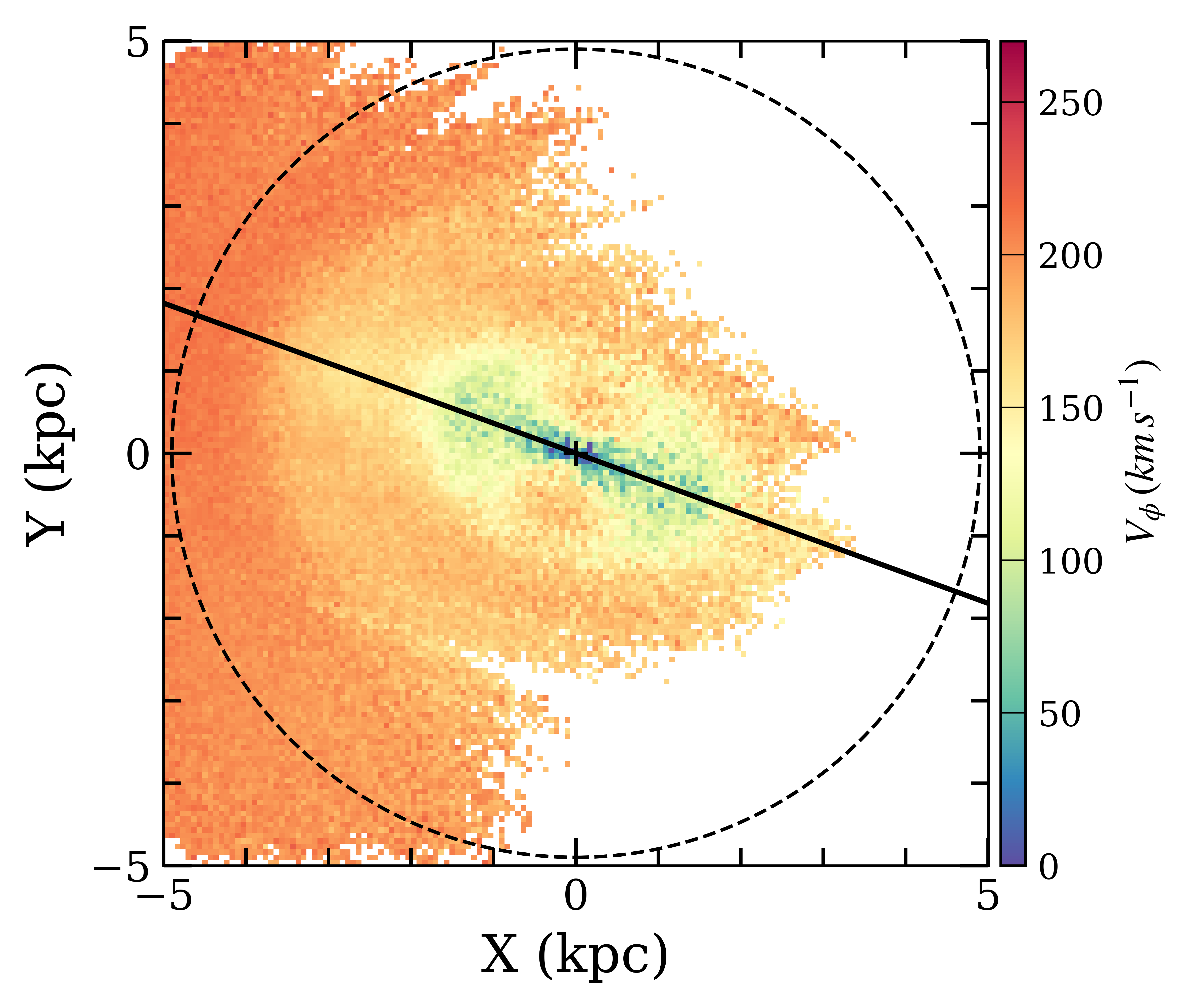}
    \includegraphics[width=0.49\textwidth]{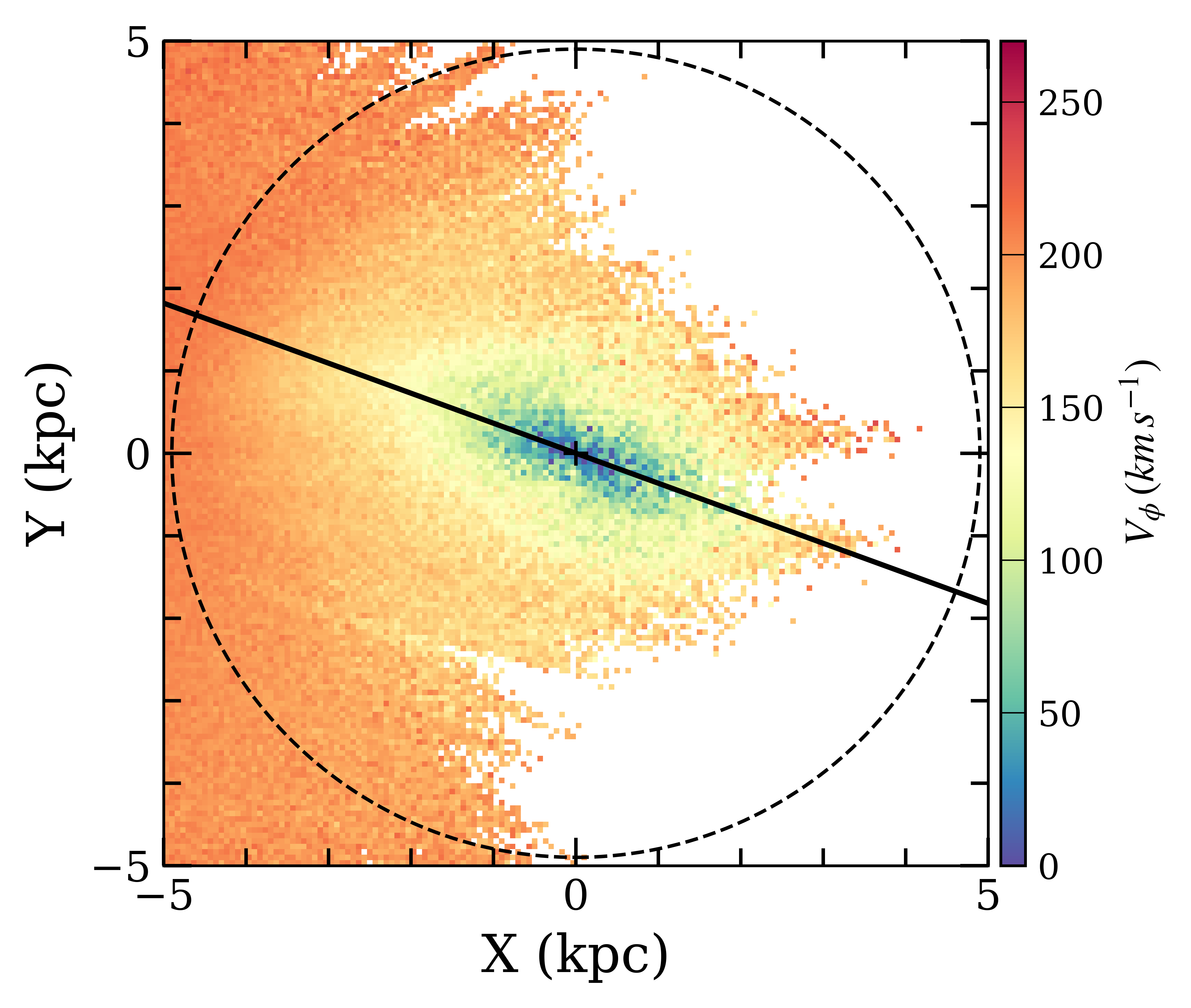}
    \caption{Spatial distribution (top), Median $V_r$ (middle) and median $V_{\phi}$ (bottom) from a barred test particle simulation. Left: variables without uncertainties. Right: variables with uncertainties. Dashed black circle shows the corotation radius, black solid line the bar orientation angle of $20^{\circ}$ and the black cross, the position of the Galactic Centre.
    }
    \label{fig:XY_VrVphi_Simu}
\end{figure*}

The spatial distribution (top), and radial (middle) and azimuthal (bottom) velocity maps corresponding to a density distribution with a Galactic bar oriented $20^{\circ}$ from the Sun -- Galactic centre line is shown in Fig.~\ref{fig:XY_VrVphi_Simu}, without (left) and with (right) uncertainties. As in the RGB sample, we only show cells with a minimum of 20 stars, and mask all others.  The solid black line shows the orientation of the bar major axis in the simulation. 

We fit the asymmetric velocity model of Eq.~\ref{eq:bisym} to the mock velocity fields in order to assess to which extent one can constrain the strength and phase angle of the bisymmetric perturbation. This simple model is motivated by the presence of   periodic features  seen in both \vrad\ and \vtan\ maps at low radius, as caused by the dynamics of the bar. Furthermore, comparable forms are regularly used to model the kinematics in barred galaxies \citep[see e.g.][]{Spekkens2007}, although the axisymmetric radial component $\overline{V}_R$ is rarely derived in the literature. For a disc in relative equilibrium, this component should indeed cancel out as a function of radius. However, with the current \gaia\ dataset is evident that this is not the case in the azimuthal range covered by the RGB sample. This is the reason why we decided to fit  $\overline{V}_R$ as well with both mock and real kinematics. 

As seen in Fig.~\ref{fig:XY_VrVphi_Simu}, the angle $\phi_{\rm b}$ that the main axis of the bisymmetry does with $Y=0$, the pseudo direction Sun $-$ Galactic centre, can be measured independently using \vrad\ or \vtan. Of course, this angle is that of the bar at low radius.
With \vrad, this angle is the direction where the right term of  Eq.~\ref{eq:bisym} changes its sign, that is $\phi_{\rm b} = \phi_{R} - \pi/4$. With \vtan\ it is the angle where this term is minimum, that is $\phi_{\rm b} = \phi_{\phi} - \pi/2$. An interesting feature seen in the middle right panel of Fig~\ref{fig:XY_VrVphi_Simu} (mock data with uncertainties) is that the direction of the change of sign of \vrad\ is no more perfectly aligned with the bar major axis, but is closer to the Sun $-$ Galactic centre line. We thus expect from the derivation of the simple asymmetric model that $\phi_{\rm bisym}$  is smaller than the true bar orientation.

\begin{figure*}
    \includegraphics[width=0.49\textwidth,trim={0 0.8cm 0 0},clip]{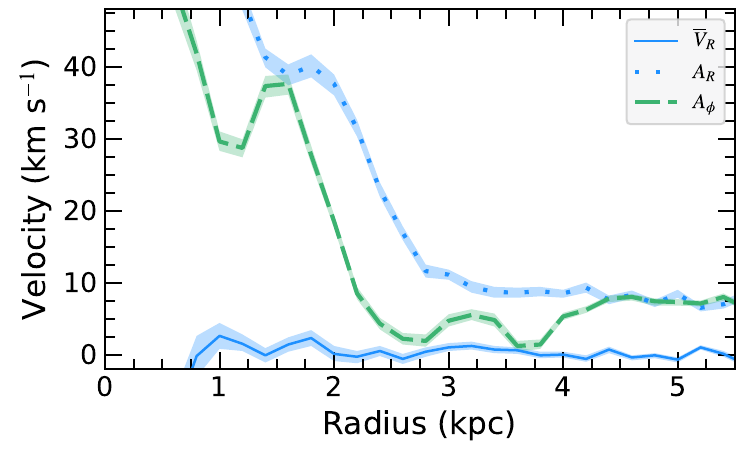}
    \includegraphics[width=0.49\textwidth,trim={0 0.8cm 0 0},clip]{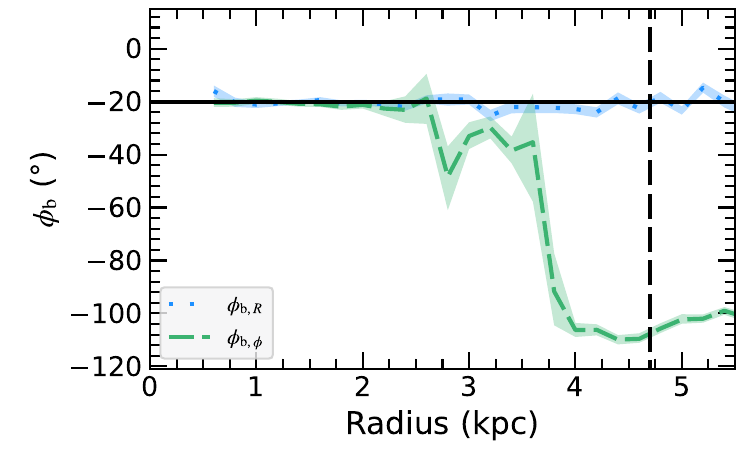}\\
    \includegraphics[width=0.49\textwidth,trim={0 0 0 0},clip]{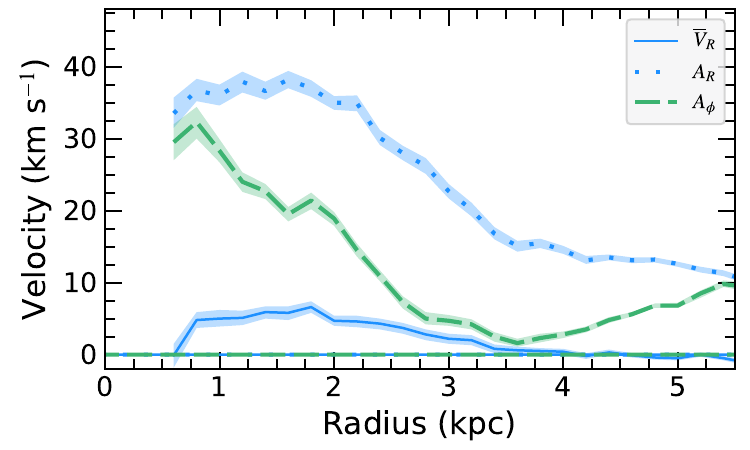}
    \includegraphics[width=0.49\textwidth,trim={0 0 0 0},clip]{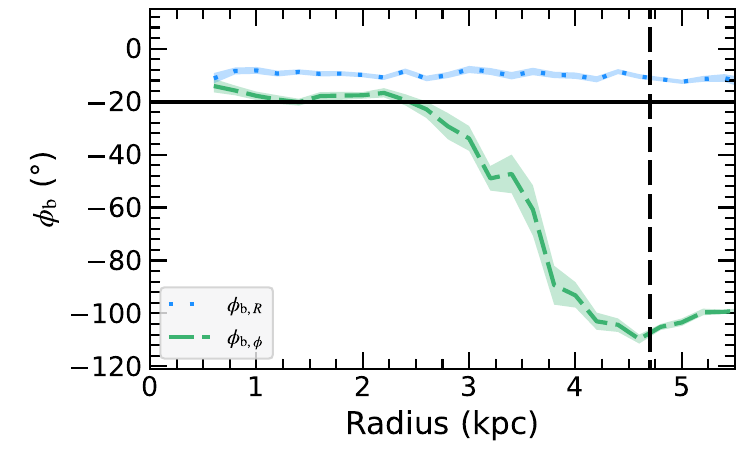}
    \caption{As in Fig.~\ref{fig:vradasym} applied to the simulation. Results for the model with and without  errors are shown in the top and bottom rows, respectively.   }
    \label{fig:modelVrasymsimu}
\end{figure*}

Figure~\ref{fig:modelVrasymsimu}  shows the amplitudes and phase angles of the bisymmetric model of \vrad\ and \vtan\ applied to the simulation with a bar phase angle of $-20\degr$, where we mask bins with $X>0$ to have the spatial distribution almost similar to that of the \gaia\ data. The bottom and top panels respectively show the results obtained with and without uncertainties added to the mock variables. The asymmetric model applied to \vtan\ recovers well the bar orientation within $R=2.2$ kpc, irrespective of the uncertainties, while for \vrad,  the phase angle of the bar is only recovered when uncertainties are not taken into account, and overestimates it in the other case. 

\begin{figure}
    \includegraphics[width=0.49\textwidth,trim={0 0 0 0},clip]{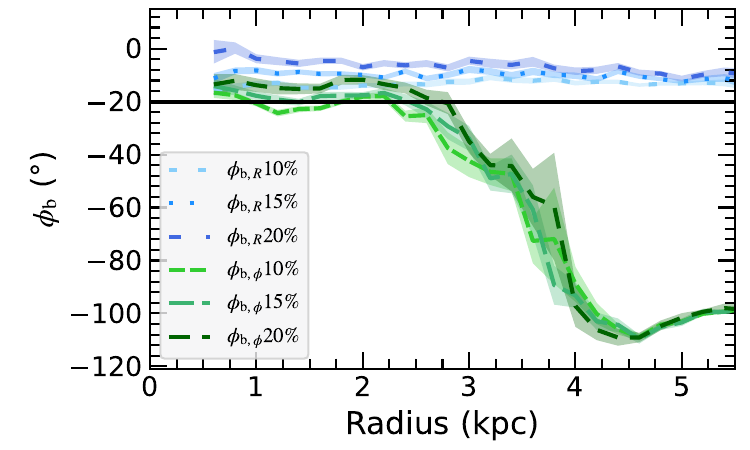}
    \caption{Phase angle of the bisymmetric model of \vrad\ (blues) and \vtan\ (greens) applied to the simulation with a bar phase angle of $20^{\circ}$ (shown as the black solid line). Different curves show a different relative uncertainty in distance applied in the simulation, namely $10\%$ (light color), $15\%$ (normal color) and $20\%$ (dark color).  }
    \label{fig:modelVrVphiSimuThreeErrors}
\end{figure}

To understand why the bisymmetric model works better with \vtan\ than with \vrad\ when uncertainties are added to the variables, we show in Fig.~\ref{fig:modelVrVphiSimuThreeErrors} the phase angle of \vrad\ (blue curves) and \vtan\ (green curves) obtained from a set of  three simulations where the only difference between them is the relative uncertainty in distance used, namely $10\%$, $15\%$ and $20\%$. Unsurprisingly, it shows that the larger the uncertainty in distance, the larger the difference between the recovered and the real phase angle. Interestingly, we also see that for \vtan, in the case of the largest relative uncertainty in distance of $20\%$, the difference between the estimated bar angle and the real orientation starts to be significant. The reason for the larger discrepancy in the radial case is that geometrically,  an error on the heliocentric distance translates into an incorrect  azimuthal angle, and a small change in azimuth affects the radial velocity more strongly than the azimuthal component.

\FloatBarrier

\section{Gaia funding acknowledgements}\label{appendixAck}
The \gaia\ mission and data processing have financially been supported by, in alphabetical order by country:
\begin{itemize}
\item the Algerian Centre de Recherche en Astronomie, Astrophysique et G\'{e}ophysique of Bouzareah Observatory;
\item the Austrian Fonds zur F\"{o}rderung der wissenschaftlichen Forschung (FWF) Hertha Firnberg Programme through grants T359, P20046, and P23737;
\item the BELgian federal Science Policy Office (BELSPO) through various PROgramme de D\'{e}veloppement d'Exp\'{e}riences scientifiques (PRODEX) grants and the Polish Academy of Sciences - Fonds Wetenschappelijk Onderzoek through grant VS.091.16N, and the Fonds de la Recherche Scientifique (FNRS), and the Research Council of Katholieke Universiteit (KU) Leuven through grant C16/18/005 (Pushing AsteRoseismology to the next level with TESS, GaiA, and the Sloan DIgital Sky SurvEy -- PARADISE);  
\item the Brazil-France exchange programmes Funda\c{c}\~{a}o de Amparo \`{a} Pesquisa do Estado de S\~{a}o Paulo (FAPESP) and Coordena\c{c}\~{a}o de Aperfeicoamento de Pessoal de N\'{\i}vel Superior (CAPES) - Comit\'{e} Fran\c{c}ais d'Evaluation de la Coop\'{e}ration Universitaire et Scientifique avec le Br\'{e}sil (COFECUB);
\item the Chilean Agencia Nacional de Investigaci\'{o}n y Desarrollo (ANID) through Fondo Nacional de Desarrollo Cient\'{\i}fico y Tecnol\'{o}gico (FONDECYT) Regular Project 1210992 (L.~Chemin);
\item the National Natural Science Foundation of China (NSFC) through grants 11573054, 11703065, and 12173069, the China Scholarship Council through grant 201806040200, and the Natural Science Foundation of Shanghai through grant 21ZR1474100;  
\item the Tenure Track Pilot Programme of the Croatian Science Foundation and the \'{E}cole Polytechnique F\'{e}d\'{e}rale de Lausanne and the project TTP-2018-07-1171 `Mining the Variable Sky', with the funds of the Croatian-Swiss Research Programme;
\item the Czech-Republic Ministry of Education, Youth, and Sports through grant LG 15010 and INTER-EXCELLENCE grant LTAUSA18093, and the Czech Space Office through ESA PECS contract 98058;
\item the Danish Ministry of Science;
\item the Estonian Ministry of Education and Research through grant IUT40-1;
\item the European Commission’s Sixth Framework Programme through the European Leadership in Space Astrometry (\href{https://www.cosmos.esa.int/web/gaia/elsa-rtn-programme}{ELSA}) Marie Curie Research Training Network (MRTN-CT-2006-033481), through Marie Curie project PIOF-GA-2009-255267 (Space AsteroSeismology \& RR Lyrae stars, SAS-RRL), and through a Marie Curie Transfer-of-Knowledge (ToK) fellowship (MTKD-CT-2004-014188); the European Commission's Seventh Framework Programme through grant FP7-606740 (FP7-SPACE-2013-1) for the \gaia\ European Network for Improved data User Services (\href{https://gaia.ub.edu/twiki/do/view/GENIUS/}{GENIUS}) and through grant 264895 for the \gaia\ Research for European Astronomy Training (\href{https://www.cosmos.esa.int/web/gaia/great-programme}{GREAT-ITN}) network;
\item the European Cooperation in Science and Technology (COST) through COST Action CA18104 `Revealing the Milky Way with \gaia (MW-Gaia)';
\item the European Research Council (ERC) through grants 320360, 647208, and 834148 and through the European Union’s Horizon 2020 research and innovation and excellent science programmes through Marie Sk{\l}odowska-Curie grant 745617 (Our Galaxy at full HD -- Gal-HD) and 895174 (The build-up and fate of self-gravitating systems in the Universe) as well as grants 687378 (Small Bodies: Near and Far), 682115 (Using the Magellanic Clouds to Understand the Interaction of Galaxies), 695099 (A sub-percent distance scale from binaries and Cepheids -- CepBin), 716155 (Structured ACCREtion Disks -- SACCRED), 951549 (Sub-percent calibration of the extragalactic distance scale in the era of big surveys -- UniverScale), and 101004214 (Innovative Scientific Data Exploration and Exploitation Applications for Space Sciences -- EXPLORE);
\item the European Science Foundation (ESF), in the framework of the \gaia\ Research for European Astronomy Training Research Network Programme (\href{https://www.cosmos.esa.int/web/gaia/great-programme}{GREAT-ESF});
\item the European Space Agency (ESA) in the framework of the \gaia\ project, through the Plan for European Cooperating States (PECS) programme through contracts C98090 and 4000106398/12/NL/KML for Hungary, through contract 4000115263/15/NL/IB for Germany, and through PROgramme de D\'{e}veloppement d'Exp\'{e}riences scientifiques (PRODEX) grant 4000127986 for Slovenia;  
\item the Academy of Finland through grants 299543, 307157, 325805, 328654, 336546, and 345115 and the Magnus Ehrnrooth Foundation;
\item the French Centre National d’\'{E}tudes Spatiales (CNES), the Agence Nationale de la Recherche (ANR) through grant ANR-10-IDEX-0001-02 for the `Investissements d'avenir' programme, through grant ANR-15-CE31-0007 for project `Modelling the Milky Way in the \gaia era’ (MOD4Gaia), through grant ANR-14-CE33-0014-01 for project `The Milky Way disc formation in the \gaia era’ (ARCHEOGAL), through grant ANR-15-CE31-0012-01 for project `Unlocking the potential of Cepheids as primary distance calibrators’ (UnlockCepheids), through grant ANR-19-CE31-0017 for project `Secular evolution of galxies' (SEGAL), and through grant ANR-18-CE31-0006 for project `Galactic Dark Matter' (GaDaMa), the Centre National de la Recherche Scientifique (CNRS) and its SNO \gaia of the Institut des Sciences de l’Univers (INSU), its Programmes Nationaux: Cosmologie et Galaxies (PNCG), Gravitation R\'{e}f\'{e}rences Astronomie M\'{e}trologie (PNGRAM), Plan\'{e}tologie (PNP), Physique et Chimie du Milieu Interstellaire (PCMI), and Physique Stellaire (PNPS), the `Action F\'{e}d\'{e}ratrice \gaia' of the Observatoire de Paris, the R\'{e}gion de Franche-Comt\'{e}, the Institut National Polytechnique (INP) and the Institut National de Physique nucl\'{e}aire et de Physique des Particules (IN2P3) co-funded by CNES;
\item the German Aerospace Agency (Deutsches Zentrum f\"{u}r Luft- und Raumfahrt e.V., DLR) through grants 50QG0501, 50QG0601, 50QG0602, 50QG0701, 50QG0901, 50QG1001, 50QG1101, 50\-QG1401, 50QG1402, 50QG1403, 50QG1404, 50QG1904, 50QG2101, 50QG2102, and 50QG2202, and the Centre for Information Services and High Performance Computing (ZIH) at the Technische Universit\"{a}t Dresden for generous allocations of computer time;
\item the Hungarian Academy of Sciences through the Lend\"{u}let Programme grants LP2014-17 and LP2018-7 and the Hungarian National Research, Development, and Innovation Office (NKFIH) through grant KKP-137523 (`SeismoLab');
\item the Science Foundation Ireland (SFI) through a Royal Society - SFI University Research Fellowship (M.~Fraser);
\item the Israel Ministry of Science and Technology through grant 3-18143 and the Tel Aviv University Center for Artificial Intelligence and Data Science (TAD) through a grant;
\item the Agenzia Spaziale Italiana (ASI) through contracts I/037/08/0, I/058/10/0, 2014-025-R.0, 2014-025-R.1.2015, and 2018-24-HH.0 to the Italian Istituto Nazionale di Astrofisica (INAF), contract 2014-049-R.0/1/2 to INAF for the Space Science Data Centre (SSDC, formerly known as the ASI Science Data Center, ASDC), contracts I/008/10/0, 2013/030/I.0, 2013-030-I.0.1-2015, and 2016-17-I.0 to the Aerospace Logistics Technology Engineering Company (ALTEC S.p.A.), INAF, and the Italian Ministry of Education, University, and Research (Ministero dell'Istruzione, dell'Universit\`{a} e della Ricerca) through the Premiale project `MIning The Cosmos Big Data and Innovative Italian Technology for Frontier Astrophysics and Cosmology' (MITiC);
\item the Netherlands Organisation for Scientific Research (NWO) through grant NWO-M-614.061.414, through a VICI grant (A.~Helmi), and through a Spinoza prize (A.~Helmi), and the Netherlands Research School for Astronomy (NOVA);
\item the Polish National Science Centre through HARMONIA grant 2018/30/M/ST9/00311 and DAINA grant 2017/27/L/ST9/03221 and the Ministry of Science and Higher Education (MNiSW) through grant DIR/WK/2018/12;
\item the Portuguese Funda\c{c}\~{a}o para a Ci\^{e}ncia e a Tecnologia (FCT) through national funds, grants SFRH/\-BD/128840/2017 and PTDC/FIS-AST/30389/2017, and work contract DL 57/2016/CP1364/CT0006, the Fundo Europeu de Desenvolvimento Regional (FEDER) through grant POCI-01-0145-FEDER-030389 and its Programa Operacional Competitividade e Internacionaliza\c{c}\~{a}o (COMPETE2020) through grants UIDB/04434/2020 and UIDP/04434/2020, and the Strategic Programme UIDB/\-00099/2020 for the Centro de Astrof\'{\i}sica e Gravita\c{c}\~{a}o (CENTRA);  
\item the Slovenian Research Agency through grant P1-0188;
\item the Spanish Ministry of Economy (MINECO/FEDER, UE), the Spanish Ministry of Science and Innovation (MICIN), the Spanish Ministry of Education, Culture, and Sports, and the Spanish Government through grants BES-2016-078499, BES-2017-083126, BES-C-2017-0085, ESP2016-80079-C2-1-R, ESP2016-80079-C2-2-R, FPU16/03827, PDC2021-121059-C22, RTI2018-095076-B-C22, and TIN2015-65316-P (`Computaci\'{o}n de Altas Prestaciones VII'), the Juan de la Cierva Incorporaci\'{o}n Programme (FJCI-2015-2671 and IJC2019-04862-I for F.~Anders), the Severo Ochoa Centre of Excellence Programme (SEV2015-0493), and MICIN/AEI/10.13039/501100011033 (and the European Union through European Regional Development Fund `A way of making Europe') through grant RTI2018-095076-B-C21, the Institute of Cosmos Sciences University of Barcelona (ICCUB, Unidad de Excelencia `Mar\'{\i}a de Maeztu’) through grant CEX2019-000918-M, the University of Barcelona's official doctoral programme for the development of an R+D+i project through an Ajuts de Personal Investigador en Formaci\'{o} (APIF) grant, the Spanish Virtual Observatory through project AyA2017-84089, the Galician Regional Government, Xunta de Galicia, through grants ED431B-2021/36, ED481A-2019/155, and ED481A-2021/296, the Centro de Investigaci\'{o}n en Tecnolog\'{\i}as de la Informaci\'{o}n y las Comunicaciones (CITIC), funded by the Xunta de Galicia and the European Union (European Regional Development Fund -- Galicia 2014-2020 Programme), through grant ED431G-2019/01, the Red Espa\~{n}ola de Supercomputaci\'{o}n (RES) computer resources at MareNostrum, the Barcelona Supercomputing Centre - Centro Nacional de Supercomputaci\'{o}n (BSC-CNS) through activities AECT-2017-2-0002, AECT-2017-3-0006, AECT-2018-1-0017, AECT-2018-2-0013, AECT-2018-3-0011, AECT-2019-1-0010, AECT-2019-2-0014, AECT-2019-3-0003, AECT-2020-1-0004, and DATA-2020-1-0010, the Departament d'Innovaci\'{o}, Universitats i Empresa de la Generalitat de Catalunya through grant 2014-SGR-1051 for project `Models de Programaci\'{o} i Entorns d'Execuci\'{o} Parallels' (MPEXPAR), and Ramon y Cajal Fellowship RYC2018-025968-I funded by MICIN/AEI/10.13039/501100011033 and the European Science Foundation (`Investing in your future');
\item the Swedish National Space Agency (SNSA/Rymdstyrelsen);
\item the Swiss State Secretariat for Education, Research, and Innovation through the Swiss Activit\'{e}s Nationales Compl\'{e}mentaires and the Swiss National Science Foundation through an Eccellenza Professorial Fellowship (award PCEFP2\_194638 for R.~Anderson);
\item the United Kingdom Particle Physics and Astronomy Research Council (PPARC), the United Kingdom Science and Technology Facilities Council (STFC), and the United Kingdom Space Agency (UKSA) through the following grants to the University of Bristol, the University of Cambridge, the University of Edinburgh, the University of Leicester, the Mullard Space Sciences Laboratory of University College London, and the United Kingdom Rutherford Appleton Laboratory (RAL): PP/D006511/1, PP/D006546/1, PP/D006570/1, ST/I000852/1, ST/J005045/1, ST/K00056X/1, ST/\-K000209/1, ST/K000756/1, ST/L006561/1, ST/N000595/1, ST/N000641/1, ST/N000978/1, ST/\-N001117/1, ST/S000089/1, ST/S000976/1, ST/S000984/1, ST/S001123/1, ST/S001948/1, ST/\-S001980/1, ST/S002103/1, ST/V000969/1, ST/W002469/1, ST/W002493/1, ST/W002671/1, ST/W002809/1, and EP/V520342/1.
\end{itemize}

The GBOT programme  uses observations collected at (i) the European Organisation for Astronomical Research in the Southern Hemisphere (ESO) with the VLT Survey Telescope (VST), under ESO programmes
092.B-0165,
093.B-0236,
094.B-0181,
095.B-0046,
096.B-0162,
097.B-0304,
098.B-0030,
099.B-0034,
0100.B-0131,
0101.B-0156,
0102.B-0174, and
0103.B-0165;
%
%
and (ii) the Liverpool Telescope, which is operated on the island of La Palma by Liverpool John Moores University in the Spanish Observatorio del Roque de los Muchachos of the Instituto de Astrof\'{\i}sica de Canarias with financial support from the United Kingdom Science and Technology Facilities Council, and (iii) telescopes of the Las Cumbres Observatory Global Telescope Network.

\end{appendix}

\end{document}